\documentclass[10pt,english,journal]{IEEEtran}
\usepackage[utf8]{inputenc}
\usepackage[table,xcdraw]{xcolor}
\usepackage{multirow}
\usepackage{multicol}
\usepackage{graphics}
\usepackage{hyperref}
\usepackage{booktabs}
\usepackage{rotating}
\usepackage{comment}
\usepackage{amssymb}
\usepackage{pifont}% http://ctan.org/pkg/pifont
\newcommand{\cmark}{\ding{51}}%
\newcommand{\xmark}{\ding{55}}%

\usepackage{float}
\usepackage{caption}
\usepackage{enumitem}
\usepackage{listings}
\usepackage{amsmath}
\usepackage{subfigure}
\usepackage{epsfig} 

\usepackage{algorithm,algorithmicx}
\usepackage{algpseudocode}

\usepackage{array}

\usepackage{url}
\usepackage{graphicx}

\usepackage{color}
\usepackage{soul}

\begin{document}
% To change the default using IEEEtran.bst style
\bstctlcite{IEEEexample:BSTcontrol}

\title{Split Federated Learning for 6G Enabled-Networks: Requirements, Challenges and Future Directions}

\author{
Houda Hafi,
Bouziane Brik,~\IEEEmembership{Senior~Member,~IEEE},
Pantelis A. Frangoudis,
and~Adlen Ksentini,~\IEEEmembership{Senior Member,~IEEE.} 

\IEEEcompsocitemizethanks{\IEEEcompsocthanksitem Houda Hafi is with Abdelhamid Mehri University, Constantine, Algeria (e-mail: houda.hafi@univ-constantine2.dz).}

\IEEEcompsocitemizethanks{
\IEEEcompsocthanksitem  Bouziane Brik is with the University of Burgundy, France (e-mail: bouziane.brik@u-bourgogne.fr).}

\IEEEcompsocitemizethanks{\IEEEcompsocthanksitem Pantelis A. Frangoudis is with the Distributed Systems Group, TU Wien, Vienna, Austria (e-mail: pantelis.frangoudis@dsg.tuwien.ac.at).}

\IEEEcompsocitemizethanks{\IEEEcompsocthanksitem Adlen Ksentini is with EURECOM, France (e-mail: adlen.ksentini@eurecom.fr).}

}

\maketitle

\begin{abstract}
Sixth-generation (6G) networks anticipate intelligently supporting a wide range of smart services and innovative applications. Such a context urges a heavy usage of Machine Learning (ML) techniques, particularly Deep Learning (DL), to foster innovation and ease the deployment of intelligent network functions/operations, which are able to fulfill the various requirements of the envisioned 6G services. The revolution of 6G networks is driven by massive data availability, moving from centralized and big data towards small and distributed data. This trend has motivated the adoption of distributed and collaborative ML/DL techniques. Specifically, collaborative ML/DL consists of deploying a set of distributed agents that collaboratively train learning models without sharing their data, thus improving data privacy and reducing the time/communication overhead. This work provides a comprehensive study on how collaborative learning can be effectively deployed over 6G wireless networks. In particular, our study focuses on Split Federated Learning (SFL), a technique recently emerged promising better performance compared with existing collaborative learning approaches. We first provide an overview of three emerging collaborative learning paradigms, including federated learning, split learning, and split federated learning, as well as of 6G networks along with their main vision and timeline of key developments. We then highlight the need for split federated learning towards the upcoming 6G networks in every aspect, including 6G technologies (e.g., intelligent physical layer, intelligent edge computing, zero-touch network management, intelligent resource management) and 6G use cases (e.g., smart grid 2.0, Industry 5.0, connected and autonomous systems). Furthermore, we review existing datasets along with frameworks that can help in implementing SFL for 6G networks. We finally identify key technical challenges, open issues, and future research directions related to SFL-enabled 6G networks.
\end{abstract}


\section{Introduction}
\label{sec:Introduction}
\subsection{Context and Motivation}
\IEEEPARstart{6}{G} wireless networks are growing to take a substantially more holistic approach, catalyzing smart services and innovative applications~\cite{chowdhury20206g}\cite{xie20216g}\cite{6G}\cite{6G1}. 6G is expected to ensure highly efficient and timely data collection, transfer, learning, and synthesizing at any time and anywhere. Applications such as Smart Grid 2.0, Extended Reality (XR), Holographic Tele-presence (HT), space and deep-sea tourism, and Industry 5.0 represent the mainstream applications of next-generation 6G systems~\cite{6G4}\cite{6G5}\cite{6G7, GE_IoT}.  
6G will be driven by a vision towards \textit{ubiquitous} intelligence integrated into every aspect of mobile networks~\cite{AI_6G2}\cite{GE_TNSE}, from network management and operations, to the specifics of intelligent vertical services powered by 6G. From a network infrastructure perspective, AI tools will play an integral role in automating multiple operations/functions in 6G wireless communication networks, and enabling a closed-loop optimization to support the emerging 6G services~\cite{ali20206g}\cite{6G10}\cite{AI_6G2}. From a service provision perspective, heavily data-driven applications will pervade, that are featuring Machine/Deep Learning (ML/DL) workflows spanning heterogeneous and potentially massive-scale networks; these applications need to be efficiently supported by the 6G infrastructure substrate, and this gives rise to interesting communication-computation co-design problems~\cite{Talwar21}.

These trends are currently being reflected in the activities of standardization bodies. The ITU Telecommunication Standardization Sector (ITU-T) has already established many focus groups (FGs), to promote data-driven AI applications in next-generation networks, such as FG-ML5G about ML for 5G networks, and FG-DPM for data processing and management coming from IoT and smart cities. The European Telecommunications Standards Institute (ETSI), on the other hand, has initiated the Experiential Networked  Intelligence Industry Specification Group (ENI ISG), which targets beyond-5G networks with the aim of introducing AI-driven facilities for cognitive network management. In addition, both academia and industry sectors have initiated the development of AI-based schemes to improve the performance of next-generation networks.
Notable examples include (i) HEXA-X\footnote{\url{https://hexa-x.eu/}} and its successor, HEXA-X-II\footnote{\url{https://hexa-x-ii.eu/}}, two EU-funded 6G flagship projects where AI-based 6G technology enablers are a core theme,  (ii) DETERMINISTIC6G,\footnote{\url{https://deterministic6g.eu/}} another EU-funded project that leverages advanced DL techniques for network performance awareness, in order to provide deterministic networking capabilities to 6G networks, and (iii) Nokia's strategy to lead the 6G development in the US.\footnote{\url{https://www.bell-labs.com/institute/blog/nokia-is-leading-the-6g-\\
conversation-in-the-us/}}

Developments in 6G are driven by the trend towards exploiting the massive availability of data, which in turn calls for moving from centralized and big data management to small and distributed data~\cite{xiao2020toward}\cite{6G}. Thus, 6G networks should leverage both small and distributed data sets at their infrastructures to optimize network performance. This trend will be manifested in  the heavy use of distributed and collaborative ML/DL techniques, which go beyond traditional and centralized ones~\cite{liu2020federated}\cite{elbir2021federated}. Specifically, collaborative ML/DL consists in deploying a set of distributed agents, that collaborate with each other to train learning models, without sharing their local data~\cite{AI_6G2}. In this context, Federated Learning (FL), proposed by Google in 2017~\cite{mcmahan2017communication}, has emerged to build cooperative learning models among a set of learners, while protecting the privacy of learners' data. However, implementing FL on top of 6G networks is still challenging mainly due to the heterogeneity of 6G-connected entities (cars, drones, sensors, haptic devices, flying vehicles, etc.) in terms of resource capabilities, as well as because of concerns from a ML \emph{model} privacy perspective, since, by design, the continuously updated versions of a learning model are shared with learners during the training process~\cite{Challenge_FL}.

To address such challenges, another DL technique was recently proposed by MIT Media Lab's Camera Culture group called Split Learning or Split Neural Networks (SplitNN)~\cite{SPLIT_MIT}.  
As its name indicates, it consists in splitting a global neural network into multiple sections, and training each section on an independent device (learner), by using the local device data. Thus, the training of the learning model is performed by transferring the output of the last layer of each section (\emph{smashed data}) as input to the next section, from one involved learner to another. Hence, compared with FL, SplitNN improves model privacy since no single learner is in possession of the global model, and reduces the computation required by the different learners to build a global learning model. Nevertheless, the main challenge of SplitNN is related to the time overhead needed to build a learning model, due to its sequential way of training. This may be very challenging in 6G settings, particularly for massive networks and when training time matters~\cite{thapa2020splitfed}.

Finally, split federated learning (SplitFed or SFL) comes to merge the two distributed DL solutions (FL and SplitNN), to design an enhanced hybrid collaborative learning algorithm~\cite{thapa2020splitfed}. In particular, SplitFed splits the neural network among the involved learners and server, as in SplitNN, to optimize both data/model privacy and compute resource usage. Moreover, SplitFed improves on training time as compared to SplitNN, by adopting the parallel model update paradigm of FL.

It is clear that SplitFed offers advantages over both FL and SplitNN by optimizing model privacy, learners' computation resources, and training time overhead. This motivates us to focus on SplitFed and show its main benefits when leveraging it over 6G wireless networks.  

\begin{table*}[!h]
\caption{Existing surveys on 6G, AI for 6G, and distributed learning for 6G. \textbf{H: High, M: Medium, and L: Low}.}
\label{Surveys}
\centering
\begin{tabular}{|l|
>{\columncolor[HTML]{96FFFB}}c |c|
>{\columncolor[HTML]{96FFFB}}c |
>{\columncolor[HTML]{96FFFB}}c |
>{\columncolor[HTML]{F8A102}}c |
>{\columncolor[HTML]{F8A102}}c |
>{\columncolor[HTML]{96FFFB}}c |
>{\columncolor[HTML]{96FFFB}}c |l|}
\hline
\multicolumn{1}{|c|}{\cellcolor[HTML]{EFEFEF}\textbf{Works.}} &
  \cellcolor[HTML]{EFEFEF}\textbf{\rotatebox{90}{Centralized Learning}} &
  \cellcolor[HTML]{EFEFEF}\textbf{\rotatebox{90}{Federated Learning}} &
  \cellcolor[HTML]{EFEFEF}\textbf{\rotatebox{90}{Split Learning}} &
  \cellcolor[HTML]{EFEFEF}\textbf{\rotatebox{90}{Split Federated Learning}} &
  \cellcolor[HTML]{EFEFEF}\textbf{\rotatebox{90}{6G Technical Aspects}} &
  \cellcolor[HTML]{EFEFEF}\textbf{\rotatebox{90}{6G Use Cases}} &
  \cellcolor[HTML]{EFEFEF}\textbf{\rotatebox{90}{\begin{tabular}[c]{@{}c@{}}Existing Datasets \& Tools\\ (Implementation)\end{tabular}}} &
  \cellcolor[HTML]{EFEFEF}\textbf{\rotatebox{90}{\begin{tabular}[c]{@{}c@{}}Open Challenges \&\\ Future Directions\end{tabular}}} &
  \multicolumn{1}{c|}{\cellcolor[HTML]{EFEFEF}\textbf{Main Achievement}} \\ \hline \hline
\cite{6G}\cite{6G7} &
  L &
  \cellcolor[HTML]{96FFFB}L &
  L &
  L &
  H &
  H &
  L &
  L &
  \begin{tabular}[c]{@{}l@{}}A holistic vision about 6G systems along with their main drivers, promising\\  applications, enabling technologies, and performance requirements.\end{tabular} \\ \hline
\cite{6G2} &
  L &
  \cellcolor[HTML]{96FFFB}L &
  L &
  L &
  H &
  H &
  L &
  \cellcolor[HTML]{F8A102}H &
    \begin{tabular}[c]{@{}l@{}}Different 6G-enabled scenarios with their challenges\\  and open issues.\end{tabular} 
  \\ \hline
\cite{Timeline1} &
  L &
  \cellcolor[HTML]{96FFFB}L &
  L &
  L &
  H &
  H &
  L &
  L &
  \begin{tabular}[c]{@{}l@{}}Main enablers of potential 6G applications in addition to a fullstack \\ perspective on 6G requirements and scenarios.\end{tabular} \\ \hline
\cite{Timeline2} &
  L &
  \cellcolor[HTML]{96FFFB}L &
  L &
  L &
  H &
  H &
  L &
  L &
  \begin{tabular}[c]{@{}l@{}}Potential 6G applications in addition to the key features of 6G such as \\ privacy, high security, and secrecy.\end{tabular} \\ \hline
\cite{6G4} &
  L &
  \cellcolor[HTML]{96FFFB}L &
  L &
  L &
  H &
  H &
  L &
  L &
  \begin{tabular}[c]{@{}l@{}}Main 6G technological trends with their potential enabled applications\\  and requirements.\end{tabular} \\ \hline
\cite{6G5} &
  \cellcolor[HTML]{67FD9A}M &
  \cellcolor[HTML]{96FFFB}L &
  L &
  L &
  H &
  H &
  L &
  \cellcolor[HTML]{F8A102}H &
  \begin{tabular}[c]{@{}l@{}}A taxonomy including use-cases, enabling computing/communication/\\ networking technologies, and promising AI techniques.\end{tabular} \\ \hline
\cite{6G6} &
  \cellcolor[HTML]{67FD9A}M &
  \cellcolor[HTML]{67FD9A}M &
  L &
  L &
  H &
  H &
  L &
  \cellcolor[HTML]{F8A102}H &
  \begin{tabular}[c]{@{}l@{}}Emerging technologies that can assist 6G architecture development in\\ meeting use-cases' requirements.\end{tabular} \\ \hline
\cite{6G8} &
  \cellcolor[HTML]{67FD9A}M &
  \cellcolor[HTML]{96FFFB}L &
  L &
  L &
  H &
  H &
  L &
  L &
  \begin{tabular}[c]{@{}l@{}}6G architectural updates featuring by the introduction of AI, \\ ubiquitous 3D coverage, and improved network stack.\end{tabular} \\ \hline
\cite{6G9} &
  L &
  \cellcolor[HTML]{96FFFB}L &
  L &
  L &
  H &
  H &
  L &
  L &
  \begin{tabular}[c]{@{}l@{}}A novel network architecture to facilitate the migration \\ from 5G to 6G.\end{tabular} \\ \hline
\cite{6G10} &
  \cellcolor[HTML]{67FD9A}M &
  \cellcolor[HTML]{96FFFB}L &
  L &
  L &
  H &
  H &
  L &
  L &
  \begin{tabular}[c]{@{}l@{}}5G limits and main motivations toward the need to move\\  to 6G wireless communication.\end{tabular} \\ \hline
\cite{6G11} &
  L &
  \cellcolor[HTML]{96FFFB}L &
  L &
  L &
  H &
  H &
  L &
  L &
  \begin{tabular}[c]{@{}l@{}}Key technologies of 6G in terms of critical requirements, \\ main drivers, enabling technologies, and system architectures.\end{tabular} \\ \hline
\cite{6G12} &
  L &
  \cellcolor[HTML]{96FFFB}L &
  L &
  L &
  H &
  H &
  L &
  \cellcolor[HTML]{F8A102}H &
  \begin{tabular}[c]{@{}l@{}}Different use cases for 6G enabling techniques, recent developments, \\ and open challenges with potential solutions.\end{tabular} \\ \hline
\cite{Timeline} &
  \cellcolor[HTML]{F8A102}H &
  \cellcolor[HTML]{F8A102}H &
  L &
  L &
  H &
  H &
  L &
  L &
  \begin{tabular}[c]{@{}l@{}}6G-enabled AI applications and AI-enabled 6G performance and \\ design optimization.\end{tabular} \\ \hline
\cite{AI_6G4} &
  \cellcolor[HTML]{F8A102}H &
  \cellcolor[HTML]{F8A102}H &
  L &
  L &
  H &
  H &
  L &
  L &
 \begin{tabular}[c]{@{}l@{}}How AI is revolutionized the 6G \\  communication technology.\end{tabular} 
   \\ \hline
\cite{AI_6G6} &
  \cellcolor[HTML]{F8A102}H &
  \cellcolor[HTML]{F8A102}H &
  L &
  L &
  H &
  \cellcolor[HTML]{34FF34}M &
  L &
  L &
  \begin{tabular}[c]{@{}l@{}}Various ML techniques applied to networking, communication, \\ and security aspects in 6G vehicular networks.\end{tabular} \\ \hline
\cite{AI_6G3} &
  \cellcolor[HTML]{F8A102}H &
  \cellcolor[HTML]{F8A102}H &
  L &
  L &
  \cellcolor[HTML]{34FF34}M &
  H &
  L &
  \cellcolor[HTML]{F8A102}H &
  \begin{tabular}[c]{@{}l@{}}What AI can bring at both the physical and link layers in 6G networks, \\ Major challenges when using AI, and future directions to mitigate them.\end{tabular} \\ \hline
\cite{AI_6G5} &
  \cellcolor[HTML]{F8A102}H &
  \cellcolor[HTML]{F8A102}H &
  L &
  L &
  H &
  \cellcolor[HTML]{34FF34}M &
  L &
  L &
  \begin{tabular}[c]{@{}l@{}}An AI-based architecture for 6G networks to mainly enable resource \\ management, knowledge discovery, and automatic service adjustment.\end{tabular} \\ \hline
\cite{liu2020federated} &
  L &
  \cellcolor[HTML]{F8A102}H &
  L &
  L &
  H &
  H &
  L &
  \cellcolor[HTML]{34FF34}M &
  \begin{tabular}[c]{@{}l@{}}Combination between FL and 6G, and FL-enabling applications in \\ 6G networks.\end{tabular} \\ \hline
\cite{AI_6G2} &
  L &
  \cellcolor[HTML]{F8A102}H &
  L &
  L &
  H &
  H &
  L &
  \cellcolor[HTML]{F8A102}H &

 \begin{tabular}[c]{@{}l@{}}  Main requirements that are driving convergence \\  between FL and 6G.\end{tabular}
  
 \\ \hline
\cite{FL_6G} &
  L &
  \cellcolor[HTML]{F8A102}H &
  L &
  L &
  H &
  H &
  L &
  L &
   \begin{tabular}[c]{@{}l@{}} How FL distributed learning can be deployed  \\ over wireless networks.\end{tabular}
   \\ \hline
\cite{SL_6G} &
  L &
  \cellcolor[HTML]{34FF34}M &
  \cellcolor[HTML]{34FF34}M &
  \cellcolor[HTML]{F8A102}H &
  \cellcolor[HTML]{34FF34}M &
  \cellcolor[HTML]{34FF34}M &
  L &
  \cellcolor[HTML]{F8A102}H &
  \begin{tabular}[c]{@{}l@{}}Different technologies enabling to combine FL and SL, at the edge \\ computing in IoT networks.\end{tabular} \\ \hline
\cite{SFL_UAV} &
  L &
  \cellcolor[HTML]{34FF34}M &
  \cellcolor[HTML]{34FF34}M &
  \cellcolor[HTML]{F8A102}H &
  \cellcolor[HTML]{96FFFB}L &
  \cellcolor[HTML]{96FFFB}L &
  L &
  L &
  \begin{tabular}[c]{@{}l@{}}FL and SL convergence on top of identically and non-independent \\ data related to 6G drone networks.\end{tabular} \\ \hline
\multicolumn{1}{|c|}{\textbf{Our Survey}} &
  \textbf{L} &
  \cellcolor[HTML]{34FF34}\textbf{M} &
  \cellcolor[HTML]{34FF34}\textbf{M} &
  \cellcolor[HTML]{F8A102}\textbf{H} &
  \textbf{H} &
  \textbf{H} &
  \cellcolor[HTML]{F8A102}\textbf{H} &
  \cellcolor[HTML]{F8A102}\textbf{H} &
  \textbf{\begin{tabular}[c]{@{}l@{}}A comprehensive survey on the use of combined FL and SL over the \\ 6G networks, covering technical aspects, use-cases, existing datasets\\  and tools, open challenges, and future research directions.\end{tabular}} \\ \hline
\end{tabular}
\end{table*}

\subsection{Review of Existing Related Surveys}
So far, many survey papers about the emerging 6G networks have been proposed to review 6G technologies, development, and enablers~\cite{6G}\cite{6G2}\cite{6G4}\cite{Timeline}\cite{Timeline1}\cite{Timeline2}. In~\cite{6G}\cite{6G7}, the authors presented a holistic vision about 6G systems and their main tenets. The primary drivers of 6G are also identified along with their promising applications, enabling technologies, and performance requirements. Another survey work was proposed in~\cite{6G2}. The authors focused more on the different 6G-enabled scenarios with their challenges and open issues. A 6G framework describing the main 6G actors/components was also designed. In~\cite{Timeline1}, the authors discussed technologies that may help to transform wireless networks toward 6G, and be considered as main enablers for many potential 6G applications. A full-stack perspective on 6G requirements and scenarios is also provided. Similarly, the authors describe a human-centric vision about 6G networks in~\cite{Timeline2}. A systematic framework, including potential 6G applications in addition to key features of 6G such as privacy, improved security, and secrecy, is also provided. An exhaustive survey about the current developments towards 6G was proposed in~\cite{6G4}. The authors also highlight the main technological trends with their potential enabled applications and requirements. Ongoing research projects and standardization efforts are also outlined. In~\cite{6G5}, recent advances toward developing 6G systems have been explored. The authors propose a taxonomy including use-cases, enabling computing/communication/networking technologies, and promising AI techniques. Open research challenges are identified and discussed. The emerging technologies that can assist 6G architecture development in meeting use-case requirements are identified and described in~\cite{6G6}. These technologies include blockchain, AI, Terahertz communications, quantum communications, cell-free communications, dynamic network slicing, and integrated sensing and communication, among others. Potential challenges and research directions are also discussed. In~\cite{6G8}, the authors give a comprehensive survey on mobile network evolution towards 6G, while focusing more on the architectural updates featuring by the introduction of AI, ubiquitous 3D coverage, and improved network stack. Potential technologies that may help in forming green and sustainable networks, such as Terahertz and visible light communication, are also discussed. Besides, a novel next-generation network architecture was designed to facilitate the migration from 5G to 6G in~\cite{6G9}. This architecture also provides various applications and technologies of 6G networks. In~\cite{6G10}, the authors study the main motivations toward the need to move to 6G wireless communication. The authors analyse the limits of 5G networks, providing a new synthesis of emerging services, including high precision manufacturing, holographic communications, and AI, while the key technologies of 6G in terms of critical requirements, different drivers, enabling technologies, and system architectures, were studied in~\cite{6G11}. In~\cite{6G12}, the authors provide a roadmap study about the different use cases for 6G enabling techniques, discussing recent developments on 6G and open challenges with potential solutions, followed by a development timeline of 6G.

Besides, a wide range of survey and review papers have discussed the application of AI and its benefits for 6G networks~\cite{Timeline}\cite{AI_6G3}\cite{AI_6G4}\cite{AI_6G5}\cite{AI_6G6}. These works focus mostly on ML/DL algorithms. In~\cite{Timeline}, the authors discussed both 6G-enabled AI applications and AI-enabled 6G performance and design optimization, while how AI is revolutionizing 6G communication technology was described in~\cite{AI_6G4}. Another comprehensive survey about various ML techniques applied to networking, communication, and security aspects in 6G vehicular networks is proposed in~\cite{AI_6G6}. In~\cite{AI_6G3}, the authors focused on what AI can bring at both the physical layer and link layer in 6G networks, they also present major challenges when using AI, and provide some future directions to mitigate them in 6G networks. An AI-based architecture for 6G networks to mainly enable smart resource management, knowledge discovery, and automatic service adjustment, is designed in~\cite{AI_6G5}.

However, few survey works have addressed distributed/collaborative learning techniques for 6G networks. Most of them are focused on federated learning (FL)~\cite{liu2020federated}\cite{AI_6G2}\cite{FL_6G}. In~\cite{liu2020federated}, the authors introduced the combination between FL and 6G, and provide enabling applications in 6G networks. Critical issues, suitable FL techniques, and future research directions leveraging FL for 6G are also described. Similarly, the main requirements which are driving convergence between FL and 6G are identified in~\cite{AI_6G2}. In addition, the authors designed a novel FL-based architecture, and showed its benefits in dealing with the emerging challenges of 6G. Moreover, future research directions and critical open challenges in FL-enabled 6G are also reviewed. In~\cite{FL_6G}, the authors gave a comprehensive study on how distributed learning can be deployed over wireless networks, while focusing more on FL, distributed inference, federated distillation, and multi-agent reinforcement learning.

On split learning (SL), only one survey work~\cite{SL_6G} has been presented, to the best of our knowledge. In this work, the authors reviewed both FL and SL, and provided a survey on different technologies enabling to combine them in an Internet of Things (IoT) context. In~\cite{SFL_UAV}, the authors first proposed a combined architecture of both FL and SL to leverage their advantages. Then, they studied their convergence under non-independent and identically distributed (non-IID) data related to 6G drone networks. Numerical results showed the efficiency of the generated learning model when combining both FL and SL.

Table~\ref{Surveys} compares existing survey studies. As we show, even though survey articles addressing 6G, and AI for 6G systems exist, there is a lack of comprehensive surveys that combine split learning and 6G to explore the potential of split learning for developing efficient, reliable, privacy-preserving AI-powered 6G systems. The relevant studies on the integration of AI with 6G networks~\cite{Timeline}\cite{AI_6G3}\cite{AI_6G4}\cite{AI_6G5}\cite{AI_6G6}\cite{SL_6G}\cite{SFL_UAV} focus more on FL rather than the recent split learning approach. In~\cite{SL_6G} and~\cite{SFL_UAV}, the authors studied the combination of FL and SL, but the focus was specifically on IoT networks and Unmanned Aerial Vehicles (UAV), and the technical aspects of 6G as well as 6G-enabled use-cases remained largely unexplored. Therefore, the related literature is missing a comprehensive survey of split learning (SL and SFL) and its potential in designing the upcoming 6G systems, which could be valuable in guiding practitioners and researchers. This is the gap our work aims to fill.

\begin{figure*}[h!]
	\centering
	\includegraphics[height=5.8in,width=7.4in]{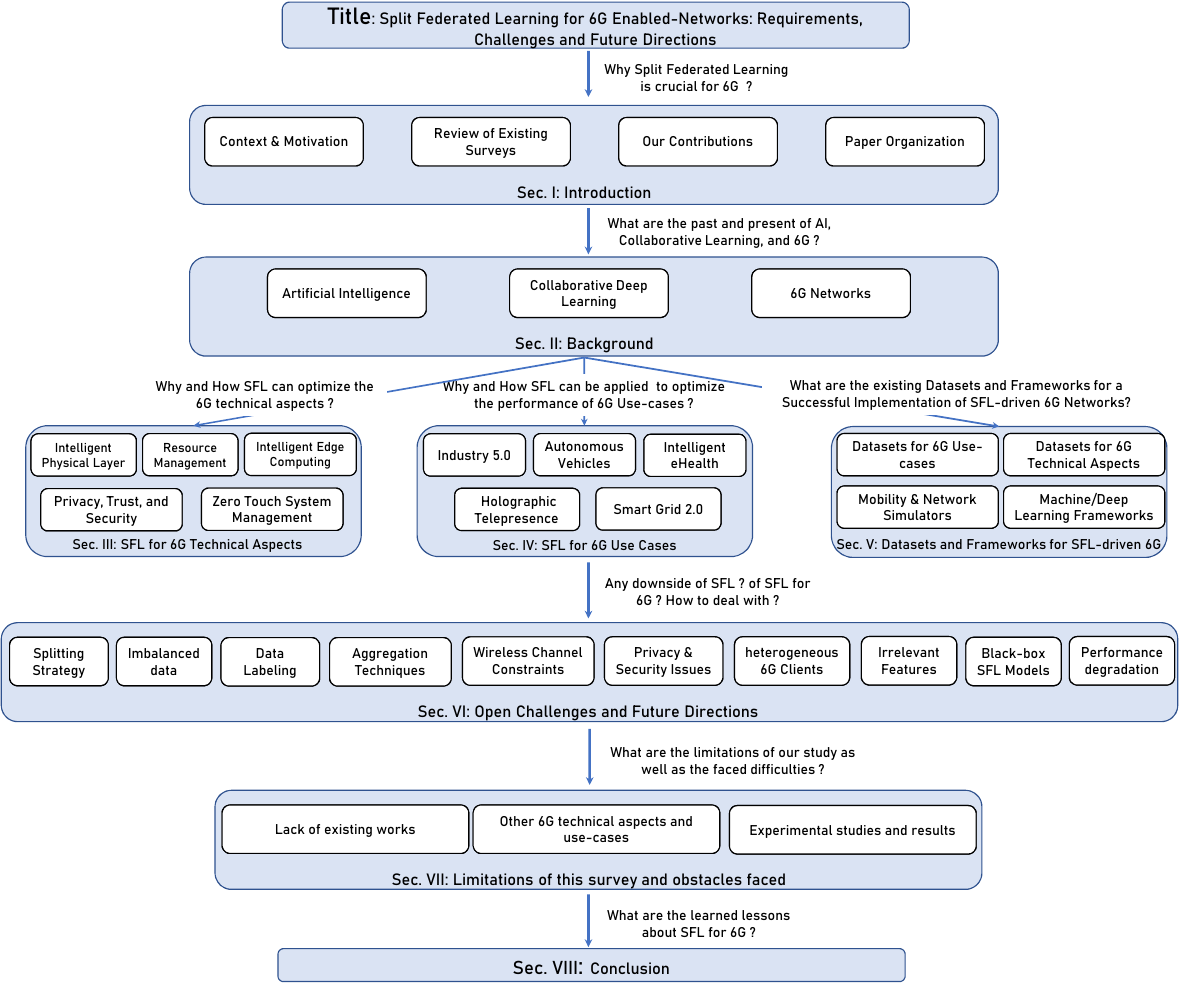}
	\caption{The structure of the article.}
	\label{stc_paper}
\end{figure*}

\subsection{Our Contributions}
The main contributions of this article are summarized below.
\begin{itemize}
    
    \item \textbf{Bridging the gap between SplitFed and 6G Networks:} Understanding the integration of SplitFed within 6G networks requires viewing what this new distributed learning algorithm represents. To do this, this survey examines first the algorithms that precede SplitFed learning to guide the reader to recognize the different existing techniques before the emergence of SplitFed. Additionally, the reader will gain a comprehension of the principles of SplitFed and how 6G can benefit from it. There are myriads of papers on the applications of AI, such as Deep Learning and Federated Learning in 6G networks~\cite{liu2020federated}~\cite{kaur2021machine}. However, the interplay between SplitFed and 6G is absent from the discussion. This comprehensive survey fills this gap by analyzing the important contributions that the connection between SplitFed and 6G can lead to.
   
    \item \textbf{A comprehensive survey of SplitFed for the most important technical aspects and use-cases of 6G:} A deeper dive in the chief 6G technical aspects (e.g., intelligent physical layer, intelligent edge computing, resource management) and 6G use cases (e.g., holographic telepresence, digital twin, and intelligent e-health) is taken to investigate how SplitFed can help in enhancing the functionalities of all 6G stakeholders. 
    
    \item \textbf{Towards a successful implementation of SplitFed over 6G networks:} This work describes various tools that can support the development, evaluation, and validation of SFL-based solutions for 6G networks. We first list multiple existing datasets for different 6G technical aspects and use-cases. Then, we overview multiple existing frameworks related to B5G networks as well as collaborative AI techniques. Finally, this article discusses multiple open implementation challenges along with their potential solutions, as future directions, in applying SFL to 6G systems.
\end{itemize}

\begin{table*}[h!]
\caption{List of Acronyms.}
\label{Acro}
\centering 
\begin{tabular}{|l|l|l|l|}
\hline
\textbf{Acronym}               & \textbf{Definition}&\textbf{Acronym} & \textbf{Definition} \\ \hline \hline
\textbf{1G}                    & First Generation  & \textbf{LIME}                  & Local Interpretable Model-Agnostic Explanations \\ \hline
\textbf{2G}                    & Second Generation & \textbf{LIS}                   & Large Intelligent Surfaces \\ \hline
\textbf{3G}                    & Third Generation  & \textbf{LSTM}                  & Long Short Term Memory \\ \hline
\textbf{3GPP}                  & 3rd Generation Partnership Project  & \textbf{LTE-A}                 & Long-Term Evolution Advanced \\ \hline
\textbf{4G}                    & Fourth Generation  & \textbf{MAC}                   & Medium Access Control \\ \hline
\textbf{5G}                    & Fifth Generation   & \textbf{MBBLL}                 & Mobile BroadBand and Low-Latency \\ \hline
\textbf{6G}                    & Sixth Generation  & \textbf{MEC}                   & Multi-access Edge Computing  \\ \hline
\textbf{A2C}                   & Advantage Actor Critic  & \textbf{MIMIC-III}             & Medical Information Mart for Intensive Care III  \\ \hline
\textbf{AE}                    & Auto Encoder & \textbf{MIMO}                  & Multiple-Input Multiple-Output  \\ \hline
\textbf{AI}                    & Artificial Intelligence  & \textbf{MIoT}                  & Massive Internet of Things  \\ \hline
\textbf{AMC}                   & Automatic Modulation Classification & \textbf{MIT}                   & Massachusetts Institute of Technology  \\ \hline
\textbf{AMF}                   & Access and Mobility Management Function & \textbf{ML}                    & Machine Learning \\ \hline
\textbf{AV}                    & Autonomous Vehicle & \textbf{mLLMT}                 & massive Low-Latency Machine Type communication \\ \hline
\textbf{ANN}                   & Artificial Neutral Networks & \textbf{MLOps}                 & ML system operations \\ \hline
\textbf{AR}                    & Augmented Reality & \textbf{MLP}                   & Multi-Layer Perceptron \\ \hline
\textbf{AWID}                  & Aegean WiFi Intrusion Dataset & \textbf{mMTC}                  & massive Machine Type Communications \\ \hline
\textbf{B5G}                   & Beyond Fifth-Generation  & \textbf{MR}                    & Mixed Reality \\ \hline
\textbf{BBU}                   & Baseband Unit & \textbf{MSE}                   & Mean Sequared Error \\ \hline
\textbf{BPSK}                  & Binary Phase Shift Keying & \textbf{NFV}                   & Network Function Virtualization \\ \hline
\textbf{BS}                    & Base Station & \textbf{NG RAN}                & New Generation RAN \\ \hline
\textbf{CAPEX}                 & CAPital EXpenditures & \textbf{NOMA}                  & Non-Orthogonal Multiple Access \\ \hline
\textbf{CAE}                   & Convolutional Auto-Encoder & \textbf{NR-MAC}                & New Radio Medium Access Control\\ \hline
\textbf{CAV}                   & Connected Autonomous Vehicles & \textbf{NS}                    & Network Slicing\\ \hline
\textbf{CD}                    & Continuous Delivery & \textbf{NS3}                   & Network Simulator 3 \\ \hline
\textbf{CNN}                   & Convolutional Neural Network & \textbf{PSK}                   & Phase Shift Keying\\ \hline
\textbf{CPS}                   & Cyber-Physical Systems & \textbf{QoE}                   & Quality of Experience\\ \hline
\textbf{CS}                    & Compressive Sensing & \textbf{QoS}                   & Quality of Service\\ \hline
\textbf{CT}                    & Continuous Training & \textbf{RAN}                   & Radio Access Network\\ \hline
\textbf{CV}                    & Connected Vehicles & \textbf{RAT}                   & Radio Access Technologies\\ \hline
\textbf{DARPA}                 & Defense Advanced Research Projects Agency &\textbf{RL}                    & Reinforcement Learning\\ \hline
\textbf{DDoS}                  & Distributed Denial of Service & \textbf{RM}                    & Resource Management\\ \hline
\textbf{DevOps}                & DEVelopment and IT Operations & \textbf{RNN}                   & Recurrent Neural Network\\ \hline
\textbf{DL}                    & Deep Learning & \textbf{RSPQ}                  & Reference Signals Received Quality\\ \hline
\textbf{DLT}                   & Distributed Ledger Technologies & \textbf{RSPR}                  & Reference Signals Received Power\\ \hline
\textbf{DNN}                   & Deep Neural Network & \textbf{RSU}                   & Road Side Unit\\ \hline
\textbf{DP}                    & Differential Privacy & \textbf{RT}                    & Real Time\\ \hline
\textbf{DQN}                   & Deep Q-Network & \textbf{SCM}                   & Structural Causal Models\\ \hline
\textbf{DRL}                   & Deep Reinforcement Learning & \textbf{SDN}                   & Software Defined Network\\ \hline
\textbf{DSRC}                  & Dedicated Short Range Communications & \textbf{SFL/SplitFed}          & Split Federated Learning  \\ \hline
\textbf{E2E}                   & End-to-End  & \textbf{SG}                    & Smart Grid  \\ \hline
\textbf{EDOs}                  & Energy Data Owners  & \textbf{SHAP}                  & SHapley Additive exPlanations\\ \hline
\textbf{EI}                    & Edge Intelligence  & \textbf{SIC}                   & Successive Interference Cancellation\\ \hline
\textbf{eMBB}                  & enhanced Mobile Broadband  & \textbf{SL/SplitNN}            & Split Learning/Split Neural Network \\ \hline 
\textbf{ENI}                   & Experiential Networked Intelligence  & \textbf{SLA}                   & Service Level Agreement\\ \hline
\textbf{ETSI}                  & European Telecommunications Standards Institute & \textbf{SM}                    & Spectrum Management\\ \hline
\textbf{FDMA}                  & Frequency Division Multiple Access& \textbf{SMO}                   & Service Management and Orchestration\\ \hline
\textbf{FedAvg}                & Federated Averaging & \textbf{SNR}                   & Signal-to-Noise Ratio\\ \hline
\textbf{FeMBB}                 & Further enhanced Mobile Broadband & \textbf{SSN}                   & Self-Sustaining Networks\\ \hline
\textbf{FG}                    & Focus Group & \textbf{SUMO}                  & Simulation of Urban Mobility\\ \hline
\textbf{FG-DPM}                & Focus Group for Data Processing and Management & \textbf{SVM}                   & Support Vector Machines\\ \hline
\textbf{FL}                    & Federated Learning & \textbf{TDMA}                  & Time Division Multiple Access\\ \hline
\textbf{GAN}                   & Generative Adversarial Network  & \textbf{TFF}                   & TensorFlow Federated\\ \hline
\textbf{gNB}                   & gNodeB  & \textbf{THz}                   & Tera Hertz\\ \hline
\textbf{GPS}                   & Global Positioning System  & \textbf{TL}                    & Transfer Learning\\ \hline
\textbf{HE}                    & Horizon Europe & \textbf{TTI}                   & Transmission Time Interval\\ \hline
\textbf{HO}                    & Handover & \textbf{UAV}                   & Unmanned Aerial Vehicles\\ \hline
\textbf{HT}                    & Holographic Tele-presence & \textbf{UE}                    & User Equipment\\ \hline
\textbf{IDS}                   & Intrusion Detection Systems & \textbf{umMTC}                 & ultra-massive Machine-Type Communication \\ \hline
\textbf{IEC}                   & Intelligent Edge Computing & \textbf{uRLLC}                 & ultra-Reliable Low Latency Communications\\ \hline
\textbf{IEEE}                  & Institute of Electrical and Electronics Engineers & \textbf{V2X}                   & Vehicle-To-Everything\\ \hline
\textbf{IID}                   & Independent and Identically Distributed  & \textbf{VNF}                   & Virtual Network Function \\ \hline
\textbf{IoE}                   & Internet of Everything & \textbf{VoIP}                  & Voice Over IP \\ \hline
\textbf{IoMT}                  & Internet of Medical Things & \textbf{VR}                    & Virtual Reality\\ \hline
\textbf{IoT}                   & Internet of Things & \textbf{WBANs}                 & Wireless Body Area Networks\\ \hline
\textbf{IoV}                   & Internet of Vehicles & \textbf{WG}                    & Working Group\\ \hline
\textbf{IP}                    & Internet Protocol &\textbf{Wifi}                  & Wireless Fidelity\\ \hline
\textbf{IRS}                   & Intelligent Reflecting Surfaces &\textbf{XAI}                   & eXplainable AI\\ \hline
\textbf{ITS}                   & Intelligent Transportation Systems & \textbf{XR}                    & eXtended Reality\\ \hline
\textbf{ITU}                   & International Telecommunication Union & \textbf{ZSM}                   & Zero-touch Network and Service Management\\ \hline
\textbf{KPI}                   & Key Performance Indicators \\ \hline
\end{tabular}
\end{table*}

\subsection{Paper Outline}
As shown in Fig.~\ref{stc_paper}, the organization of this article is as follows: Section~\ref{sec:Introduction} delineates the significant role of AI in 6G, highlights our motivation, and provides an in-depth review on existing related surveys in addition to the main contributions of this paper. Section~\ref{sec:Background} gives a background on AI, collaborative learning, and 6G networks, which are required for describing the potentials of SplitFed learning for 6G networks. In Section~\ref{sec:SplitFed for 6G Technical Aspects}, we describe the main challenges and requirements pertinent to different key 6G technical aspects, and show how SFL can help in optimizing operations and performance through a realistic scenario for each such aspect. Section~\ref{sec:SplitFed for 6G Use Case} discusses five emerging 6G use cases with a particular focus on how SFL can help in optimizing some of their performance characteristics. To help towards a successful implementation of SFL on top of 6G networks, we list several existing datasets and frameworks (tools) that can support the development, evaluation, and validation of SFL-based solutions for 6G networks. As with any new technology, SFL has its limitations and challenges which are presented in Section~\ref{sec:Future Directions and Discussions}. This section gives not only the main limitations of SFL, but also the still-open challenges of applying SFL to 6G networks along with future directions. We also provide the main limitations of our study as well as the faced difficulties, that may be considered to stimulate further research in Section~\ref{sec:Limitations of the Survey}. Finally, we conclude the article in Section~\ref{sec:Conclusion}. For the ease of reference, the acronyms used in this article are summarized in Table~\ref{Acro}, in alphabetical order. We should finally note that, in the course of this article, we use the terms SplitFed and SFL interchangeably.

\section{Background}
\label{sec:Background}
This section gives a background on AI, collaborative learning, and 6G networks, which are required for describing the potential of SplitFed learning for 6G networks. We start by introducing AI and machine/deep learning as a main branch of AI; this may help non-expert readers get a better understanding of the concepts we introduce next, namely the main collaborative learning mechanisms of interest in this article: federated, split, and SplitFed learning. We finally provide an overview of 6G networks, along with their main vision and development timeline.

\subsection{Artificial Intelligence fundamentals} 
Artificial intelligence is a computer science field that aims to mimic human mind capabilities in solving problems and making decisions. It enables machines, e.g., computer systems, to simulate human intelligence process through algorithms and rules, by combining various fields, such as reasoning, planning, learning, communicating, perception, and interaction. In our study, we focus more on machine/deep learning as an AI branch.
\subsubsection{Machine Learning Algorithms}
Machine learning (ML) is a branch of AI that leverages statistical and mathematical models to process and generate inferences from robust dataset patterns. ML consists in building learning models based on training data, typically to obtain accurate predictions, as results. ML algorithms are usually classified into three different categories:
\begin{itemize}

    \item \textbf{Supervised learning}: It consists in mapping specific inputs to an output considering structured and labeled data. For instance, to train a learning model to recognize pictures of dogs and cats, the model should consider pictures labeled as dogs and cats. This category of ML models can deal with two main problems: regression to predict a continuous (real) value, e.g., temperature and velocity, and classification to predict the class to which a particular input data example may belong, e.g., picture of cat or dog. Various supervised learning algorithms have been developed. For example, \textit{linear} and \textit{logistic regression} aim to learn the correlation between input and output data by estimating the parameters of a linear or logistic model fit to them~\cite{LLR}. \textit{Random decision forests} or \textit{random forests} build a set of decisions trees in order to make both regression and classification. \textit{Random forests} also belong to another category of learning algorithms, called ensemble learning, which further includes algorithms such as boosting machines and AdaBoost~\cite{RF}. \textit{Support Vector Machines (SVM)}, on the other hand, create classification and regression learning models building on a statistical learning theory framework. Particularly, they aim to learn the optimal hyperplane that separates the data instances~\cite{SVM}. \textit{Artificial Neural Networks (ANN)} mimic the human brain, by linking a high number of artificial neurons (perceptrons) with each other via edges and their associated weights. The purpose of an ANN algorithm is to learn the optimal edge weights so that a loss function is minimized and inference accuracy is increased. Deep learning is based on ANNs with a large number of hidden layers in the neural network~\cite{ANN}.

    \item \textbf{Unsupervised learning:} In this family of approaches, unlabeled data are processed to deduce common patterns/information. 
    Unlike supervised learning, data labels are not known ahead of time. The ML algorithm processes the whole dataset, classifying it into groups based on common attributes. This category of ML includes three different sub-classes: clustering, dimensionality reduction, and association rules. \textit{K-means} is a clustering algorithm that divides data into $k$ clusters, according to the distance to each cluster's centroid. Different such distance metrics exist, such as dynamic time warping, Euclidean, or Manhattan distance~\cite{kmeans}. \textit{Dimensionality Reduction} is used to reduce the data dimension, while keeping its main attributes. \textit{Principal Component Analysis (PCA)} is one of the main dimensionality reduction algorithms, which operates by projecting data geometrically onto new components, called Principal Components~\cite{kmeans}. \textit{Association rule} mining learns the different associations between the input data, which will then help to determine the correlation/relationship among them. For example, associations between shoppers can be established based on their purchasing or browsing histories~\cite{Ass}.

    It is worth noting that there is also a mixed category named semi-supervised learning, where only some data are labeled~\cite{SSLM}. 
    In this category, a final output is known, and the learning algorithm should figure out how to structure and organize the data to achieve the desired outputs.
    
    \item \textbf{Reinforcement learning:} The basic idea of this class of learning is ``learning by doing.'' An \textit{agent} learns to perform particular tasks in a feedback loop by trial and error, until achieving a desirable performance. The agent receives either positive or negative reward when it performs the task either well or poorly, respectively. Hence, the agent aims to learn an optimal policy enabling to maximize the reward. A typical example of a reinforcement learning application is when teaching a robotic hand to pick up a ball. Popular reinforcement learning algorithms include \textit{Q-learning}, \textit{Deep Q-learning}, and \textit{Advantage Actor Critic (A2C)}~\cite{RL}. \textit{Q-learning} consists in finding the optimal policy to transit from a state of the system to another.
    It aims to learn a so-called \textit{Q-table}, where ``Q'' stands for quality. Each value of the table encodes the quality of picking a specific action when the agent is at a given state. \textit{Deep Q-learning} replaces the Q-table with an ANN. This allows the algorithm to handle problems with a continuous state space and cases where the state space is prohibitively large. Finally, \textit{A2C} consists in building two different learning networks, named actor and critic. The actor is in charge of making optimal actions, while the critic network assesses their qualities~\cite{A2C}. 
\end{itemize}    
\begin{figure*}[!t]
\centering
  \includegraphics[height=9cm,width=17cm]{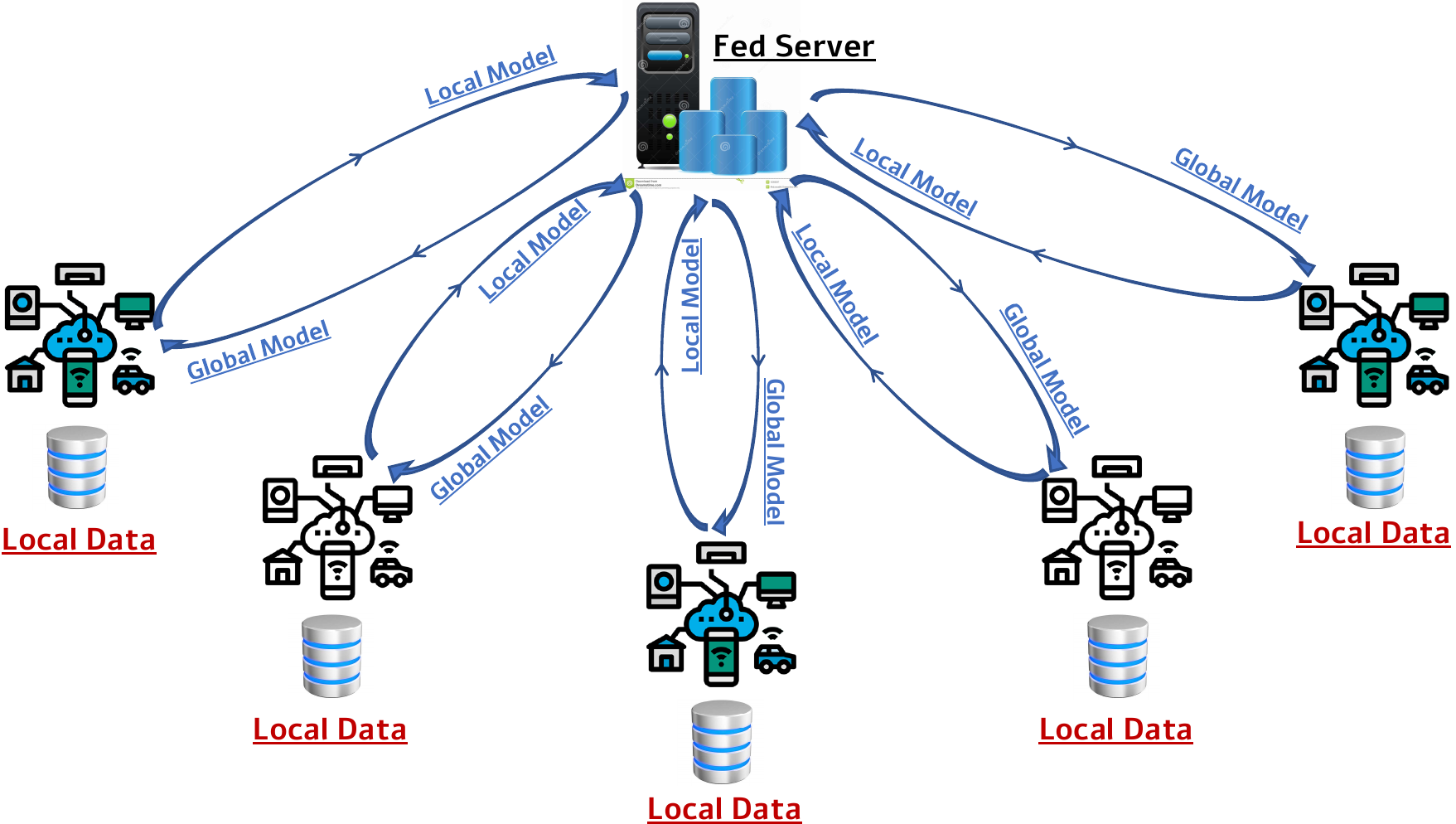}
  \caption{Federated Learning Training.}
   \label{FL}
\end{figure*}
\subsubsection{Deep Learning Algorithms}
An Artificial Neural Network (ANN) comprises a set of artificial neurons, named also perceptrons, which are computational components used to process and analyze data. ANNs are generally organized in layers of neurons, where neurons of a specific layer receive input from the layers before, operate on this input, and pass the output of their computations on to the layers that follow (though variations of this organization exist). Deep Learning (DL) is a sub-field of ML that is characterized by the application of ANNs with many layers. ANNs with more than three layers are usually referred to as Deep Neural Networks (DNN), and some modern neural architectures may include hundreds of layers. DL alleviates the need for manual feature engineering, allowing instead the system to automatically \emph{learn} features of the input data that are important, by hierarchically synthesizing higher-level features (e.g., shapes, in an image classification problem) from lower-level ones (e.g., edges or contours) as ANN layers progress. This is in contrast with other ML designs where humans need to explicitly define the representation of data in terms of input features. DL The main deep learning algorithms are as follows:
\begin{itemize}
    \item \textit{Feed-forward artificial neural networks} are one of the most used deep learning forms, where data are fed from the first to the last (output) layer, through multiple hidden layers, and thus multiple computational neurons~\cite{ANN}. A feed-forward ANN is usually coupled to a back-propagation algorithm, that works back from the results (output layer) towards the first layer in order to correct errors and improve prediction accuracy of the neural network. 

    \item \textit{Sequence Algorithms} enable to deal with sequential data-related problems (time series), such as speech recognition and language translation. \textit{Recurrent Neural Networks (RNN)}, such as \textit{Long Short-Term Memory (LSTM)} networks, are typical examples of such algorithms. They are mainly based on an internal memory to save what happened in the previous layer, to decide about the output of the current layer. For example, if we saved the first two words of the well-known sentence, ``Deep Learning Algorithms,'' it would be much easier to predict the third word ``Algorithms.'' Thus, recurrent architectures decide about their future outputs based on both the historical and actual states~\cite{LSTM}.

    \item \textit{Convolutional Neural Networks (CNN)} is another deep learning form that is mostly used for image recognition. As in the typical ANN structure, CNNs include input, hidden, and output layers. However, intermediate layers can include distinct layer types, such as convolutional, pooling, full-connected, and normalization layers~\cite{CNN}. These different types of layers can learn about both simple and more complex image features like colors and edges.

   \item \textit{Generative Adversarial Network (GAN)}: It is based on CNN to deal with unlabeled data (unsupervised learning) to extract common attributes from data~\cite{GAN}. In particular, GAN includes two competitive neural networks. The first one generates new data examples (generator), while the second one evaluates the quality of such new data (discriminator). For instance, GAN has been widely used to generate new realistic images.

   \item \textit{Auto-encoder}: It is another unsupervised deep learning algorithm used to learn efficient encodings of unlabeled data. Auto-encoder comprises two main steps: encoder to code the input data, and decoder to reconstruct the input data from the code~\cite{Auto_Enco}. Applications include detecting anomalies and reducing the noise of images.    
\end{itemize}

DL models may be trained either in a centralized or a distributed fashion. Centralized learning can be performed by uploading all required data from all connected data sources, such as remote devices, to a central node, e.g., a cloud server, to train a global model. The latter can then be deployed to all involved entities, or it can be accessed remotely as a service delivered from the cloud. In a mobile network context, centralized training enables to optimize the energy consumption of connected devices which are typically battery-powered. At the same time, though, it is associated with other critical challenges related to device privacy threats due to sharing data, and increasing communication overhead and costs.
To overcome such challenges, distributed learning emerges as a potential solution, due to its potential for network cost savings and its inherent privacy preservation. Enabling a set of learners to collaboratively build machine learning models without sharing their private data enhances data sovereignty and is a form of user empowerment. This is a step towards democratizing deep learning processes, while at the same time bearing the potential for resource savings at the cloud end by distributing training load.

\begin{figure*}[!h]
\centering
  \includegraphics[height=9cm,width=18cm]{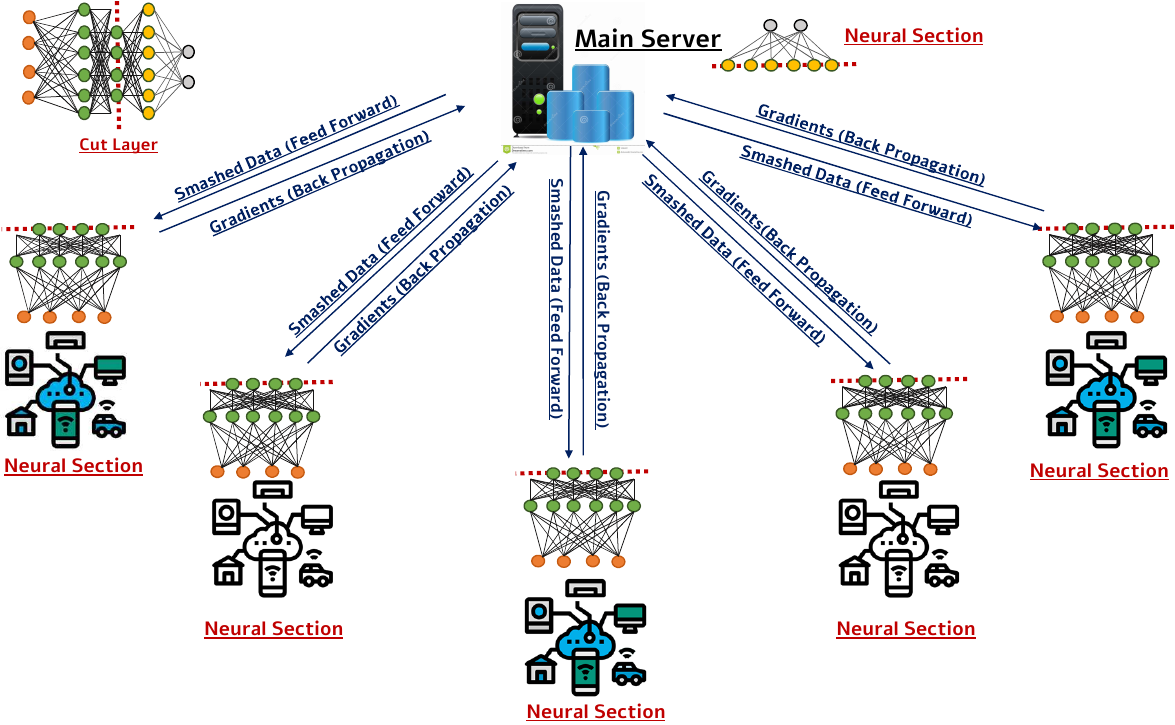}
  \caption{Split Learning Training.}
   \label{SplitNN}
\end{figure*}
\subsection{Background on Collaborative Deep Learning}
In this section, we give an overview of the main collaborative deep learning schemes, ranging from federated learning, to the recent split learning and SplitFed learning paradigms.
\subsubsection{Federated Deep Learning}
Federated Learning (FL) enables a set of nodes (clients) to collaboratively learn a prediction model, without sharing their own data~\cite{FL}\cite{FL1}. Thus, FL aims to build cooperative learning models, while protecting the privacy of learners' data. The FL process comprises three main steps (see Fig~\ref{FL}):
\begin{itemize}
    \item Local learning initialization: During this first step, a central node, e.g., cloud server, specifies learning hyper-parameters in terms of the neural architecture and deep learning algorithm to be used, along with its configuration (number of layers and neurons, activation functions, learning rate, optimizer, dataset features, number of iterations, minimum required accuracy, etc.). Such parameters are then transferred to all the involved learners.
    \item Training of local models: Once receiving the learning parameters from the central node, each learner starts to build its local learning model leveraging its own data. The local models, i.e., neural network weights, are then communicated back to the central node after either the specified number of iterations is reached, or the needed accuracy has been achieved. 
    \item Local models aggregation: During this step, the central node aggregates all the received local models to generate a global model, before sharing it with all learners. At this stage, different aggregation algorithms can be used such Federated Averaging (FedAvg)~\cite{mcmahan2017communication}, FedProx~\cite{fedprox}, FedPer~\cite{fedper}, and SCAFFOLD~\cite{scaffold}.    
\end{itemize}
Even though FL enables to train neural networks in a distributed, collaborative way, it still presents critical challenges that should be addressed. One of the main challenges related to FL is \textit{systems heterogeneity}, where the storage, communication and computing capabilities of involved learners may differ mainly due to the heterogeneity in hardware (memory and processing units), network connectivity (Wi-Fi, 3G, 4G, 5G), and power source and state (e.g., battery level). Thus, it is possible that not every learner is able to train a neural network, nor to periodically share it with a central node~\cite{Challenge_FL}. Moreover, while FL avoids sharing learners' data by design, sharing model updates during the training process can reveal private information, either to the central server, or to a third-party~\cite{Challenge_FL}.  

\subsubsection{Split Deep Learning}
To overcome FL's limits, another collaborative deep learning technique called ``Split Learning'' or ``Split Neural Learning'' (SplitNN) has recently been developed~\cite{SL}. As illustrated in Fig.~\ref{SplitNN}, it is used to build deep neural networks over multiple learners, while avoiding to share their labeled data. In SplitNN, a deep neural network is split into multiple sections (sub-layers) and each section is locally trained on a different learner (user or server). Thus, the training of the learning model is performed by transferring the weights of the last layer of each section (\emph{smashed data}), also named the \emph{cut layer}, to the next section. By this, SplitNN mitigates to share learners' data, and only the weights of the last layer are shared with the next learners.
Specifically, the neural networks in SplitNN are trained through two main steps:
\begin{itemize}
  \item Forward propagation: Each learner (for example an IoT device) trains a partial deep neural network up to the cut layer. The outputs of the cut layer are then transferred to the next learner (server), that continues the training without access to the data of the other learners (IoT devices).
    \item Backward propagation: This consists in back-propagating the gradients from the last section, to the first section of the neural network. Only the gradients of the cut layer are sent back from the server to the IoT devices.
\end{itemize}

\begin{figure*}[!h]
\centering
  \includegraphics[height=9cm,width=18cm]{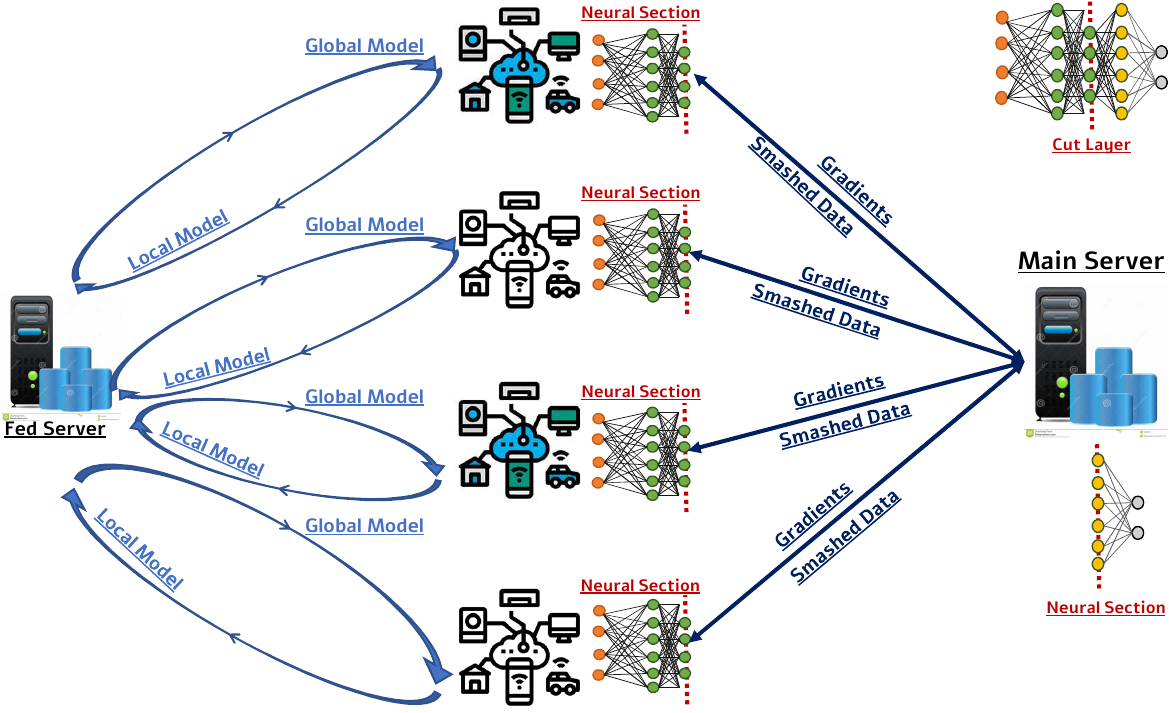}
  \caption{SplitFed Learning Training.}
   \label{SplitFed}
\end{figure*}
This process is repeated until training the whole neural network and reaching the required accuracy. In practice, SplitNN can be configured in three different ways:
\begin{itemize}
    \item Vanilla split learning: It is the simplest configuration, where the deep neural network is split between a set of learners and a server, where the last section of layers is located at the server. Learners start to train their neural networks until the cut layer. The weights of the cut layers are then sent by the learners to the server to complete the rest of training. For backward propagation at the server, the gradients are back propagated from the last layer towards the cut layer. The gradients at the cut layer are then communicated back to the different learners, to continue the backward propagation step.
    
    \item Split learning without label sharing: This configuration also consists in splitting the neural network among learners and a server. 
    However, the label of each example is located at the learner's side. The neural network is partitioned in a way that learners maintain the first and the last layers of the neural network, so (i) the output of the last cut layer in the forward pass is sent back by the server to learners, which in turn (ii) start back propagation by building the gradients from the last section of the neural network without sharing the corresponding labels, and passing the gradients on to the server, which eventually (iii) sends back its output and the back propagation process is finalized at the learners' end. This configuration is ideal for applications where labels incorporate very sensitive information like patients' disease status.
    
    \item Vertically partitioned split learning: It is suitable when multiple institutions, for example different network operators, aim to train a common network over vertically partitioned data (i.e., where each institution holds a different set of features of the dataset) through a central server and without sharing their data. The deep neural network is split in a way that the institutions share the same neural network sections, albeit with different input features each. The last layers are located at the server. Institutions train their neural networks up to the cut layer, and the institutions' outputs at the cut layer are then aggregated and sent to the central server that continues the training process. 
\end{itemize}

Compared to FL, SplitNN improves data privacy by sharing only the weights of a sub-section of the neural network, up to the cut layer, rather than sharing the model updates during the training process. In addition, SplitNN strongly reduces the computation required by the different learners to generate a global learning model, since each learner is in charge of only a part of the neural network. 

Despite the main advantages of SplitNN, however, it presents a critical performance issue: The sequential nature of training in SplitNN makes client resources idle, since only one learner can be engaged with the server at one instance. In particular, in settings with multiple learners, after one learner finishes back propagation -- thus a new version of the global model, partitioned between the learner and the server, is available -- the next one needs to begin its forward pass on the most up-to-date model. Synchronizing the learner-side part of the model can take place either in a centralized (the learner-side model portion is uploaded by the last learner to a central server accessible by other learners) or in a peer-to-peer manner. In any case, new versions of the global model are created one learner at a time, while the rest remain idle. This may generate a considerable training time overhead, especially when the number of learners is large~\cite{thapa2020splitfed}.   

\subsubsection{SplitFed Deep Learning}
SplitFed (or SFL) merges the two distributed ML solutions, FL and SplitNN, to construct an enhanced hybrid collaborative learning algorithm, that always follows the learner-server model~\cite{thapa2020splitfed}. It inherits the dual advantages of both FL and SplitNN. It partitions the model into learner/client and server sides, but all the sub-models are trained in parallel. In addition to the main server that existed in the earlier SplitNN design, a novel component is introduced in the architecture called \emph{Fed Server} (see Fig.~\ref{SplitFed}). The working process follows a series of steps: At first, the Fed Server starts the procedure by sending the global learner-side model portion to all participating clients. Next, all learners run at the same time the forward propagation function using their own local data, till the cut layer of learners, where they pass the smashed data to the cut layer of the main server. Following the SplitNN principle explained so far, the server takes over the rest of the forward propagation, calculates the cost function and back-propagates up to the cut layer of the server. Notably, as we shall describe next, it is possible to execute this server-side process in parallel for multiple clients. Then, each client continues the back propagation on its learner-side model portion. The cycle of forward and backward propagation between the learners and the server is carried out for some rounds without the Fed Server. Then, the learners communicate their updates to the Fed Server, that aggregates them and creates a global learner-side model, which is sent back to all the involved learners.

\begin{table}[!h]
\caption{Comparison of Collaborative Learning Algorithms.\\ \textbf{H: High, M: Medium, and L: Low}.}
\label{Comparative_Study_2}
\centering
\begin{tabular}{|l|
>{\columncolor[HTML]{FFCE93}}c |
>{\columncolor[HTML]{FFCE93}}c |c|
>{\columncolor[HTML]{96FFFB}}c |
>{\columncolor[HTML]{FFCE93}}c |
>{\columncolor[HTML]{96FFFB}}c |}
\hline
\multicolumn{1}{|c|}{\cellcolor[HTML]{EFEFEF}\textbf{\begin{tabular}[c]{@{}c@{}}Collaborative\\ Learning\end{tabular}}} &
  \cellcolor[HTML]{EFEFEF}\textbf{\rotatebox{90}{\begin{tabular}[c]{@{}c@{}}Model\\ Privacy\end{tabular}}} &
  \cellcolor[HTML]{EFEFEF}\textbf{\rotatebox{90}{\begin{tabular}[c]{@{}c@{}}Model\\ Aggregation\end{tabular}}} &
  \cellcolor[HTML]{EFEFEF}\textbf{\rotatebox{90}{\begin{tabular}[c]{@{}c@{}}Training Time\\ Overhead\end{tabular}}} &
  \cellcolor[HTML]{EFEFEF}\textbf{\rotatebox{90}{\begin{tabular}[c]{@{}c@{}} Computation \\ Resources\end{tabular}}} &
  \cellcolor[HTML]{EFEFEF}\textbf{\rotatebox{90}{\begin{tabular}[c]{@{}c@{}}Distributed \\ Computing\end{tabular}}} &
  \cellcolor[HTML]{EFEFEF}\textbf{\rotatebox{90}{\begin{tabular}[c]{@{}c@{}}Access to\\ Raw Data\end{tabular}}} \\ \hline \hline
\textbf{\begin{tabular}[c]{@{}l@{}}Federated \\ Learning\end{tabular}} &
  \cellcolor[HTML]{96FFFB}L &
  H &
  \cellcolor[HTML]{96FFFB}L &
  \cellcolor[HTML]{FFCE93}H &
  H &
  L \\ \hline
\textbf{\begin{tabular}[c]{@{}l@{}}Split \\ Learning\end{tabular}} &
  H &
  \cellcolor[HTML]{96FFFB}L &
  \cellcolor[HTML]{FFCE93}H &
  L &
  H &
  L \\ \hline
\textbf{\begin{tabular}[c]{@{}l@{}}Split Federated \\ Learning\end{tabular}} &
  \textbf{H} &
  \textbf{H} &
  \cellcolor[HTML]{9AFF99}\textbf{M} &
  \textbf{L} &
  \textbf{H} &
  \textbf{L} \\ \hline
\end{tabular}
\end{table}

\begin{figure*}[!h]
\centering
  \includegraphics[height=8cm,width=18cm]{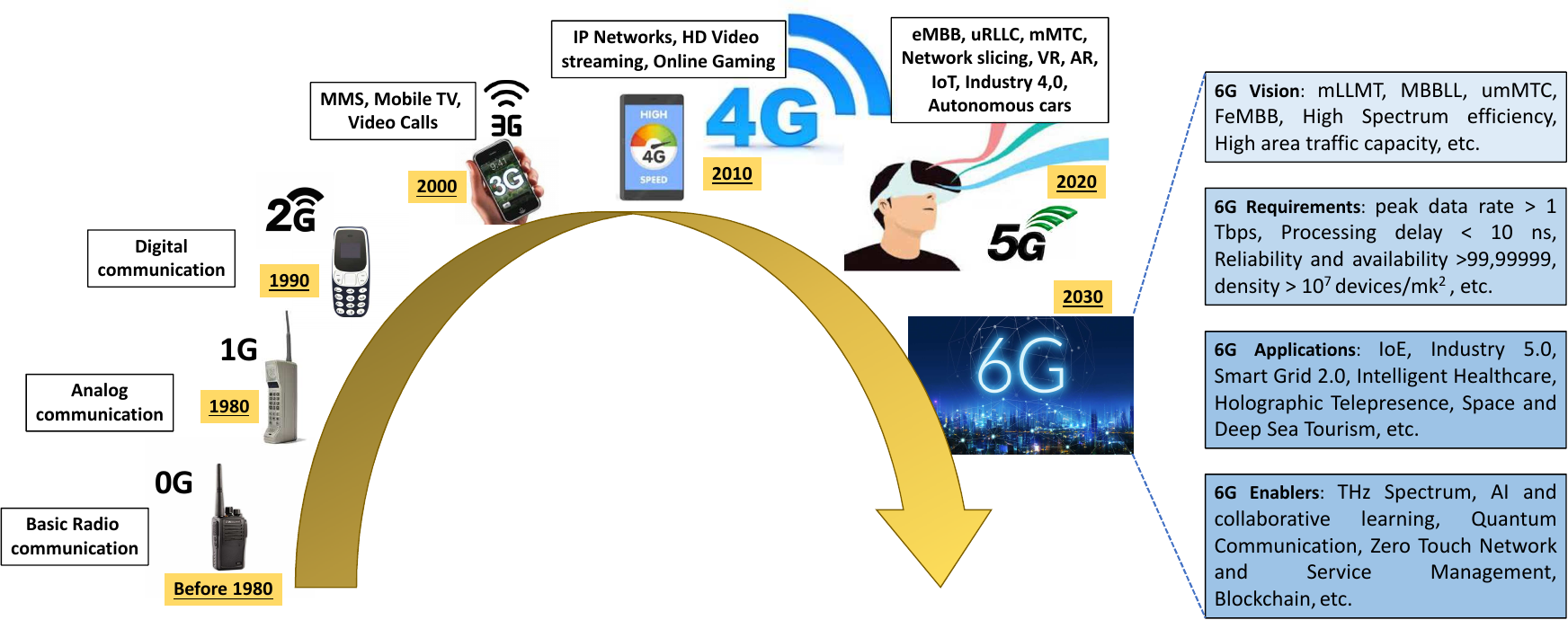}
  \caption{Mobile Networks Evolution.}
   \label{Evo}
\end{figure*}

There are two ways to perform server-side synchronization:
\begin{itemize}

    \item With Aggregation: The main server is responsible not only of training its part of the model over the smashed data received from clients, but also to aggregate the back propagation results corresponding to each client's data into a single global server-side model at each learning epoch. This takes place by executing federated averaging over the values it computed during the backward pass on each individual learner's smashed data. It should be noted that the main server processes these smashed data \emph{in parallel}.  

    \item Without Aggregation: There is no aggregation at the main server, and the server-side model is updated in every single forward-backward pass, processing smashed data from the various clients \emph{sequentially}.  The smashed data themselves are received by clients synchronously. Thus, at each instance, the main server selects one client at random, in order to perform forward-back propagation. The clients' operations remain the same (forward-back propagation), where their models are sent periodically to the Fed Server to aggregate them and generate a global (learner-side) model.
     
\end{itemize}
To summarize, SplitFed splits the neural network among the involving learners and server, as in SplitNN, to optimize both data/model privacy and compute resource use. Moreover, SplitFed improves on training time, as compared to SplitNN, by integrating the parallel model update paradigm of FL. Table~\ref{Comparative_Study_2} compares the three collaborative learning forms, FL, SplitNN, and SplitFed, according to six main criteria: privacy of the learning model, model aggregation, training time overhead, needed computation resources, distributed computing, and access to the raw data.

It is clear that the three collaborative learning approaches enable a high degree of computation distribution, without any access to the raw data. However, SplitFed offers more advantages as compared to both FL and SplitNN by optimizing model privacy, learners' computation resources, and training time overhead. This represents the main motivation of our work to focus on SplitFed and demonstrate its benefits when leveraging it over B5G/6G wireless networks.  

\subsection{Background on Sixth-Generation (6G) Networks}
6G mobile networks are expected to evolve towards connected intelligence with the support of a wide range of services with diverse and stringent requirements. In this section, we provide an overview of the emerging 6G mobile networks, ranging from mobile network evolution, to today's 6G vision and development timeline of 6G-enabled networks. 

\subsubsection{Evolution of Mobile Networks}

During the last four decades, mobile networks have transformed through five different generations. Each new generation integrates more capabilities and technologies to enhance and empower our lifestyle and work. Before 1980s, pre-cellular mobile generation was referred as the zeroth-generation (0G) of mobile networks. It offered basic voice communication using devices such as walkie-talkies~\cite{0G}. In the 1980s, the first-generation (1G) of cellular networks was launched, to support analog cellular telephony~\cite{1G}. Second-Generation (2G) cellular telephony was introduced in the early 1990's. 2G featured a transition from analog to digital technology, to provide new services such as MMS, picture messages, and text messages in addition to voice communication~\cite{2G}. The International Telecommunication Union (ITU) then launched initiatives to unify a frequency band in the 2000 MHZ, supporting a single wireless communication standard for all countries. The third-generation (3G) was introduced based on such standard to enable new and advanced services, while optimizing network capacity~\cite{3G}. These services include video calls, multimedia messages, mobile TV, GPS (global positioning system), etc. The fourth-generation (4G) succeeds 3G cellular networks, to introduce further improved mobile services, including Voice Over IP (VoIP), online gaming, High-definition mobile TV, mobile web access, and 3D television~\cite{4G}.

Currently, multiple network operators are deploying 5G mobile communication worldwide to support further advanced services, such as ultra Reliable Low Latency Communication (uRLLC) to ensure a communication latency down to 1 ms, enhanced Mobile Broadband (eMBB) to achieve data throughput up to 10 Gbps, and massive Machine Type Communication (mMTC) to support a massive deployment of devices, in particular over 100x more devices per unit area as compared to 4G. In 5G, network availability and reliability are expected to reach 99.999\%~\cite{5G}. Indeed, network slicing and network softwarization are the main technology enablers of 5G that introduce more programmability, dynamicity, and abstraction of networks~\cite{5G1}. These capabilities have enabled promising applications including Augmented Reality (AR), Virtual Reality (VR), Mixed Reality (MR), Internet of Things (IoT), autonomous vehicles, and Industry 4.0~\cite{5G_app}\cite{5G_app1}. 
\begin{figure*}[!h]
\centering
  \includegraphics[height=8cm,width=15cm]{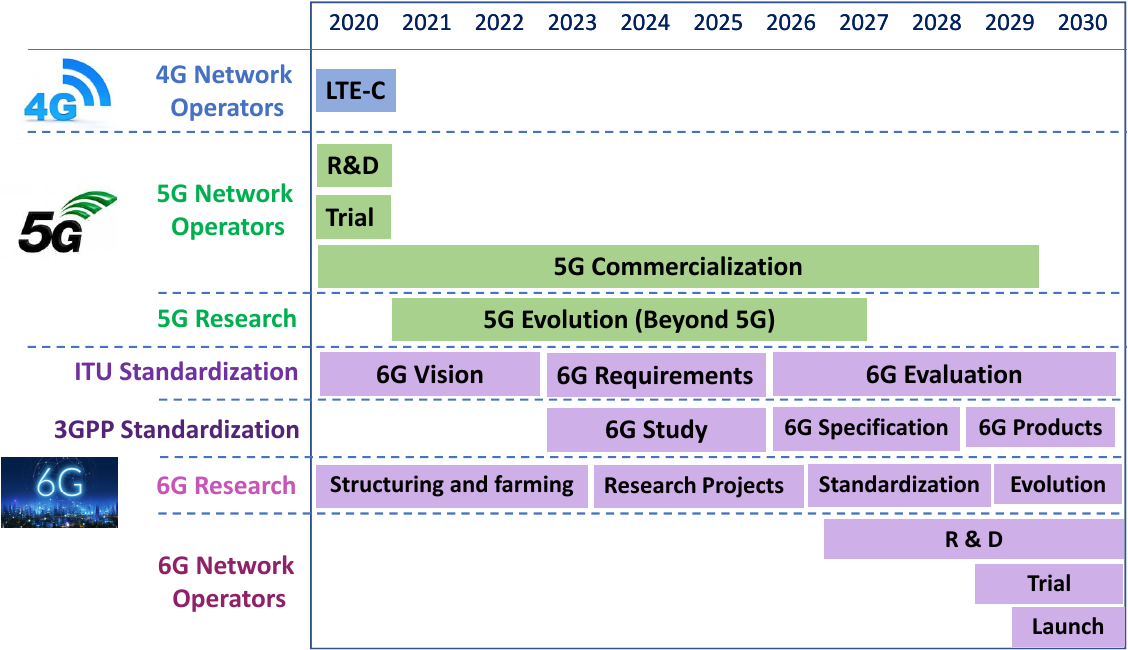}
  \caption{Expected Timeline of 6G development~\cite{Timeline}.}
   \label{Time}
\end{figure*}
Recent developments have brought about several new concepts, such as Non-Orthogonal Multiple Access (NOMA), beyond sub 6GHz to THz communication, Edge Intelligence (EI), Self-Sustaining Networks (SSN), swarm networks, and Large Intelligent Surfaces (LIS)~\cite{6G}\cite{6G1}. These concepts are expected to play a vital role in empowering the next generations of wireless networks. In addition, these concepts are also expected to be the main enablers of many new applications such as Unmanned Aerial Vehicles (UAV), smart grid 2.0, Extended Reality (XR), Holographic Telepresence (HT), space and deep-sea tourism, and Industry 5.0. However, the requirements of these applications, including accurate sensing and localization, availability of powerful computing resources, ultra-high data rates, extremely low latency, very high reliability and availability, surpass the 5G network capabilities~\cite{6G2}\cite{liu2020federated}. This has motivated the research and industrial communities to envision 6G communication networks, which are expected to consider the emerging communication concepts and applications.

Fig.~\ref{Evo} gives the evolution of cellular networks, while showing the key features of each generation. It also illustrates the main envisaged 6G enablers, vision, requirements, and applications.  

\subsubsection{Today's 6G Vision}
As envisioned today, extreme peak data rates over 1 Tbps and a very low end-to-end delay of 0.1 ms are expected to be provided by 6G networks. To attain such goals, processing delays at the sub-microsecond range for specific tasks may be required, and this will be facilitated by the pervasive use of edge intelligence, a core theme in the 6G vision. Both network reliability and availability are expected to go beyond 99.99999\%. 6G networks are expected to provide high connection density of over 107 devices/km2, and thus support Internet of Everything (IoE), which connects massive numbers of Cyber-Physical Systems (CPS), devices, actuators, and sensors. Furthermore, extreme mobility up to 1000 kmph is expected to be supported by 6G in addition to a spectrum efficiency estimated to 5$\times$ that of 5G~\cite{6G}.

To enable emerging applications, 6G networks are expected to meet multiple new requirements including massive Low-Latency Machine Type communication (mLLMT), Mobile BroadBand and Low-Latency (MBBLL), ultra-massive Machine-Type Communication (umMTC), and Further enhanced Mobile Broadband (FeMBB). These new requirements are enabled thanks to the new emerging technologies in terms of Compressive Sensing (CS), distributed/collaborative learning, Edge AI, THz spectrum, 3D networking, and blockchain/Distributed Ledger Technologies (DLT). To this end, considerable efforts are made by research community to developing and specifying 6G technologies, applications, services, vision, and standards~\cite{6G_spe}\cite{6G_spe1}.
\subsubsection{Expected 6G Development Timeline}
Developments of 6G networks progress along with the finalization of 4G LTE-C, that followed LTE-B and LTE-Advanced as well as the commercialization and deployment of 5G networks~\cite{Timeline}. By 2023, the definition of the 6G vision in terms of requirements and development evaluation, standards, technologies, etc. is expected. The technical specifications of 6G are expected to be developed by standardization bodies like 3rd Generation Partnership Project (3GPP) and ITU by 2026-2027~\cite{Timeline}. Network operators are also expected to initiate 6G research and development by 2026-2027, in order to perform 6G network trials by 2028-2029, and to start deploying 6G communication networks by 2030~\cite{6G2},~\cite{Timeline}~\cite{Timeline1}. Fig~\ref{Time} illustrates the expected timeline of 6G standardization, development, and deployment. 

\section{SFL for 6G: Technical Aspects} 
\label{sec:SplitFed for 6G Technical Aspects}

\subsection{Intelligent Physical Layer}

\subsubsection{Introduction, requirements, and existing solutions}
One of the major features of 6G is the ability to connect a massive number of intelligent devices (IoE)~\cite{chowdhury20206g}. Undoubtedly, this would lead to a dramatic growth in the number of users and emerging applications that impose diverse performance requirements: extremely high data rate, extremely high reliability and ultra low latency. As previous cellular networks, the issue of spectrum's scarcity is still ongoing, thus achieving these goals is a challenging task. Notwithstanding the wide spectrum offered by 5G, it is insufficient to cover the future 6G needs. For that, additional frequency bands are a necessity. Multi-band spectrum that combines sub-6 GHz, millimeter wave (30 – 300 GHz), Terahertz (0.06 – 10 THz) and non-radio frequencies (RFs) (visible and optical bands such as Li-Fi) will be a central solution for 6G networks~\cite{xie20216g}. To usefully exploit the limited resources, smart techniques for sharing and managing are mandatory. The intervention of AI in the physical layer will be a further addition to deal with problems that are cumbersome to model accurately with conventional mathematical methods. 

In the literature, many works are devoted to the utilization of AI paradigms (ML, DL, FL) in the physical layer~\cite{tanveer2021machine}~\cite{restuccia2020deep}. AI will render the transmission more reliable by improving different physical layer aspects, e.g., signal modulation, channel estimation, and error control. In~\cite{elbir2021federated}, the authors proposed a channel estimation model based on Federated Learning, and their results show that their approach offers 16$\times$ lower overhead than centralized learning.

Similarly, Automatic Modulation Classification (AMC) is an attractive solution widely used for intelligent radio systems. In a typical communication environment, the modulation scheme is shared between both the transmitter and receiver. However, this would increase the signalling overhead, while a sniffer can interpret and identify the modulation scheme used for transmission. AMC consists in the identification of the modulation type of the received signal without prior knowledge of the transmitter modulation. A reliable modulation classifier needs to sustain a high accuracy and low loss under various channel conditions and SNR (Signal-to-Noise Ratio) rate. Of late, many studies address the integration of deep learning algorithms to replace conventional classifiers. For example, in~\cite{8891763} the authors proposed a new AMC multi-class model with four possible outputs (BPSK, QPSK, 8-PSK, 16-QAM) based on a Recurrent Neural Network (RNN), which is the most suitable for sequential data. The classifier exhibits noteworthy results under different noise conditions. A CNN architecture has also been successfully employed by~\cite{zeng2019spectrum} for the processing of graphical representations of signals in a spectrogram form. Recently, a distributed classification method for AMC based on federated learning has been introduced in~\cite{wang2021federated}. The proposed solution, so-called FedeAMC, shows a slight performance gap with a centralized solution, i.e., CentAMC (less than 2\%). 
Another solution to set up a reliable communication link at the physical layer in 6G systems is to use a huge number of antennas at the transmitter and receiver sides. This technology is called Massive MIMO (Multiple-Input, Multiple-Output)~\cite{de2022overview}. It enables the transmission/reception of signals from/to multiple users simultaneously. Beamforming is a thriving technique used in Massive MIMO. It is based on smart small antennas that increase the transmitted energy to a specific direction, in order to create narrow beams destined for particular users. This method can boost the link's capacity, by reducing interference, providing more signal paths and higher throughput. On the other hand, the orchestration of the massive number of antennas would be complex and hardly manageable. In this regard, many researchers leverage AI capabilities, by introducing new antenna design processes through ML. For instance, in~\cite{elbir2020federated}, a FL-CNN based model was designed for analog beamformers, where simulation results demonstrated that the framework minimizes the overhead of channel state information (CSI) collection and transmission, and it is more tolerant to channel changes and imperfections.

\subsubsection{Challenges and how SFL can help}
Despite the promising performance of traditional ML and FL in physical layer design, they present some drawbacks. Most studies apply centralized ML that entails the availability of datasets at a central node, e.g., the base station (BS), wherein a transmission of local data from user equipment (UE) is prerequisite. Thus, the BS starts the training process after the collection of the required datasets from the respective sources. Further, in existing FL-based approaches, the generated datasets, e.g., the received pilots are kept intact at the client side, and the base station forwards a replica of the same model to all clients (e.g., mobile phones). Accordingly, each client trains independently the whole model which is prohibitive in terms of computing resources, that may not always be available, due to the heterogeneous hardware constraints. Another problem of FL stems from the migration of the total parameters of the physical model, which causes privacy and security issues. To cope with these limitations, SplitFed (SFL) could be a better alternative. Contrary to FL, the SFL technique does not share the entire model. As a matter of fact, the model is cleaved into two parts, one part for the clients and the other for the main server. Then, only the dedicated sub-model is migrated towards each client. Therefore, this will enhance the computing and energy consumption and raise the level of the model's privacy. In contrast to SplitNN, the SFL client-side model is trained by all the wireless devices at once, using their own raw data (CSI, received signal, beamformer information, etc.), which accelerates the learning stage. As mentioned earlier, SFL can be applied for various practical physical layer applications, ranging from channel estimation to error control. The base station can act as a bridge between the clients and the Fed Server to send the model updates for aggregation purposes. Indeed, SFL can be seen as a hybrid solution that combines the advantages of both centralized and distributed ML paradigms, in order to confront the performance parameters expected to surge in 6G networks.

\subsubsection{Realistic Scenario}
\begin{figure*}[!h]
    \centering
  \includegraphics[height=10cm,width=16cm]{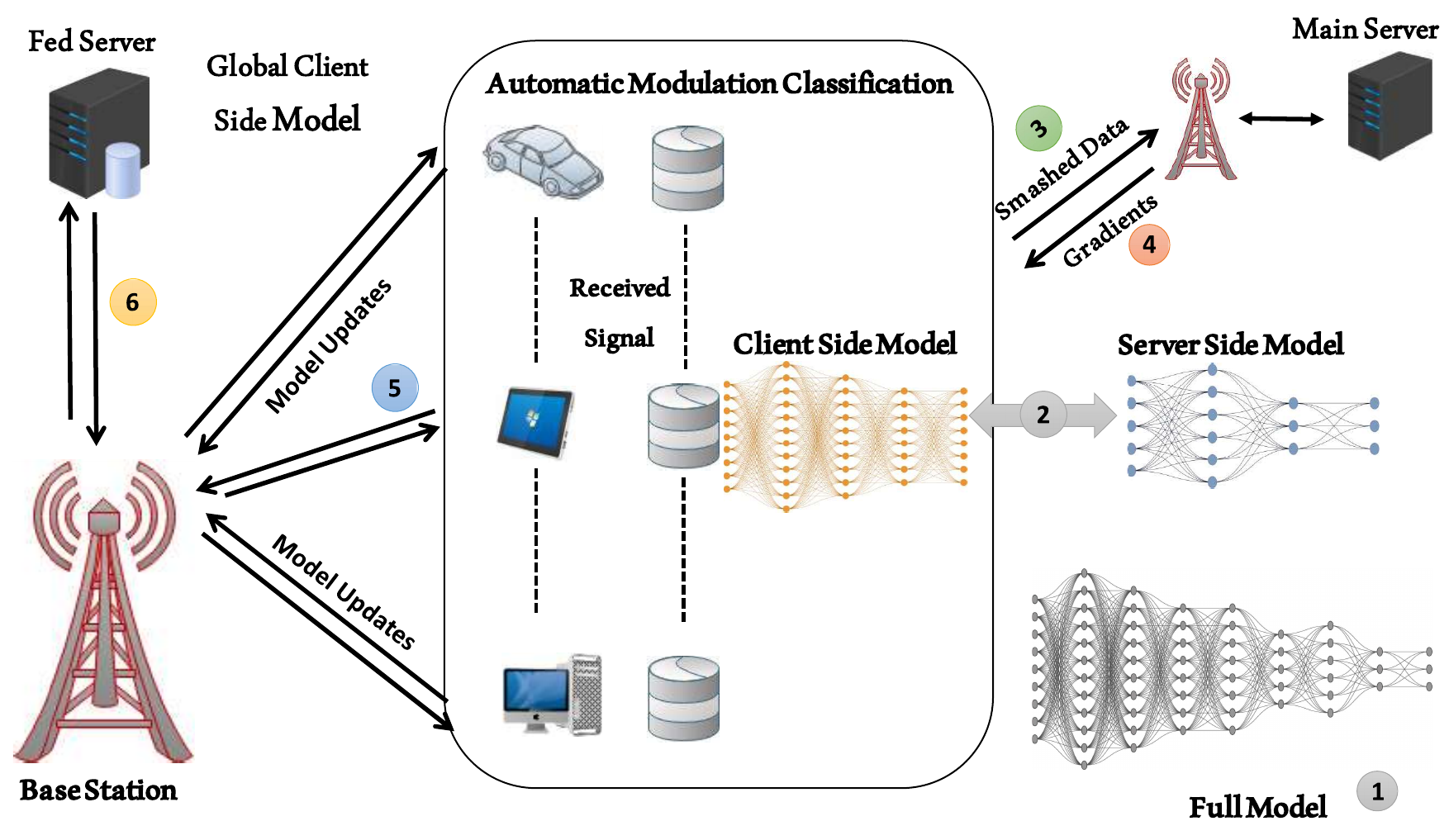}
    \caption{SFL-based modulation recognition application.}
    \label{fig:SFL_PHY}
\end{figure*}

In order to elucidate the use of SFL at the PHY layer, we propose an illustration of a realistic scenario for modulation recognition based on SFL (see \figurename~\ref{fig:SFL_PHY}). The model proposed in~\cite{zeng2019spectrum} used a CNN architecture, where the input is the corresponding spectrograms of the signals with a dimension of $100 \times 100 \times 3$. The model comprises four convolutional layers with a kernel size of $3 \times 3$ and different number of filters from 64, 32, to 12 and 8. The size of both zero-padding and stride are set to 1. The pooling size of the max-pooling layer is (2, 2). The fully connected layer consists of 128 neurons.  
To apply SFL, we consider a system with $k$ clients [$c_{1}$, $c_{2}$, \dots, $c_{k}$] and the client with the highest computing resources will represent the main server; we assume it to be the last client $c_{k}$. First, we divide the global model $W$ (see step 1 in \figurename~\ref{fig:SFL_PHY}) into $W_{C}$ (client side) and $W_{S}$ (server side) (step 2 in the figure). Then, we allocate the first five layers to the clients (Conv, Max Pooling, Conv, Max Pooling and Conv) and the remaining layers to the main server (Max Pooling, Conv, Fully Connected and Softmax). The used dataset is RadioML2016.10a~\cite{o2016radio}, which considers 11 modulation methods, 20 different signal-to-noise ratios (SNRs) and 1000 signals per modulation mode per SNR. $700$ random signals, per modulation mode per SNR, are chosen as training data, and the remaining are divided into validation and test data. We distribute the dataset between the $(k-1)$ clients for training, So, each client will have [$ 700 \times 11 \times 20 \mathbin{/} (k-1)$] signals. In the first iteration, the clients from $c_{1}$ to $c_{k-1}$ train the $W_{C}$ until the third Conv layer. Then, each of them applies the ReLU activation function and transmits the output to the main server (client $c_{k}$) (step 3 in the figure). The rest of the forward operation is performed by the main server and the recognition accuracy is measured. Next, the client $c_{k}$ back-propagates the model until its Max Pooling layer and sends the activation gradients for the clients to continue the back-propagation on their client-side local model (step 4 in the figure). After some iterations, the $k-1$ clients forward their local weights to the Fed Server to reconstruct the new global $W_{C}$. 
The resulting averaged model is then re-forwarded to all clients and the process restarts (step 6 in the figure). The training is stopped once the validation loss is not decreased (has stable values). Using SplitFed in this scenario will be beneficial on several levels. First, it would be time and energy saving for the client-side, as each learner would just train five layers instead of nine layers. It is also advantageous in terms of storage requirements because of the small number of trainable parameters implicated, compared to FL that typically require a large number parameters due to the learnable layers. All these previous points make SplitFed more suitable especially when no computational capabilities are available. 

\subsection{Resource Management}
\subsubsection{Introduction, requirements, and existing solutions}
Resource Management (RM) in 6G would face significant challenges. Owning to the uncommon number of expected connected devices, network resources are projected to be under a high strain. Accordingly, to meet the diverse needs of each device and/or service, RM algorithms should be able no only to allocate the resources but also to optimize, adapt, prioritize and secure the allocation process. Typically, RM problems are resolved using optimization and heuristics-based methods. However, the foreseeable services and their performance in 6G will hinder the utilization of traditional solutions. For instance, one of the key capabilities of 5G and beyond is Network Slicing. The general rule that stands behind this technology is to build multiple virtual networks (aka slices) on the top of a single physical infrastructure~\cite{khalili2019network}. A slice is a set of resources (memory, computing, and network) and functions (virtual network functions) customized to support a specific service, and deployed at different levels, such as RAN, core network, edge and cloud computing facilities~\cite{BOUTIBA}. The resource allocation to different running slices should be managed in an automatic and flexible way. For that, a range of research works tackled AI-assisted network-slicing along the complete lifecycle of a slice~\cite{shen2020ai,bega20}, from admission control~\cite{Ojijo20}, resource orchestration~\cite{Bakri21, Rezazadeh23} to radio scheduling~\cite{Setayesh22}. It should be noted that while significant efforts have been put in these directions in the context of 5G, emerging 6G use cases will challenge Network Slice customization, management and orchestration in multiple ways: Application requirements will cut across the traditional 5G service classes (eMBB, URLLC, and MIoT)~\cite{Timeline1}, which implies that slice resource management mechanisms will need to address new throughput-reliability-latency (and potentially privacy) trade-offs. Furthermore, handling massive numbers of potentially short-lived slices can strain the control and management planes.

Another aspect of RM is the power allocation problem, where the transmitted power should conserve the signal's quality without causing interference. In~\cite{sanguinetti2018deep}, the authors have developed a DL model for MIMO power control to predict the power allocation profile of any UE based on its position. The model has been trained to learn the mapping between the UE's position and the optimal power allocation. The limited radio resources and the massive channel access expected in 6G make the orthogonal multiple access schemes (e.g., TDMA, FDMA and CDMA), used in the previous generations of cellular networks, unable to fulfill the needs of users. To handle this struggle, NOMA (Non-Orthogonal Multiple Access) is a good candidate for 6G. It covers multiple users using the same resource block (same time and frequency) which brings about inter-user interference~\cite{islam2019non}. To mitigate the latter, successive interference cancellation (SIC) process is applied. In~\cite{9348107}, a DNN-aided SIC architecture was studied, where all the users' signals are successively decoded by the base station from the strongest to the weakest. The input of each DNN is the composite signal that contains all received signals, and the decoded signals of all previous users (the input for the first user is only the composite signal), and the output is the decoded signal of the corresponding user. On the side, the high-speed and high-mobility of some nodes (e.g, drones and air-taxis) and the use of mm-waves in 6G may lead to many handover (HO) events~\cite{angjo2021handover}. Cell selection and handover management are major issues that must be handled in 6G to allow users to continue communicating smoothly without interruption, by moving from one AP/BS to another. DL-based HO techniques prove their power in literature. In~\cite{hu2019trajectory}, Hu et al. proposed a novel intelligent HO control method where the handover decision-making is based on the results of a deep learning-based model for trajectory prediction. According to the test results, the method achieves higher accuracy than the other traditional systems (on average, the accuracy value is 8\% better). 

\subsubsection{Challenges and how SFL can help}
The frequent changes in network conditions and configurations (fluctuating number of users,  dynamic network status) cause a degradation in traditional resource allocation methods, which assume static networks and depend on fixed network models. Within the last decade, a considerable number of works leveraging data-driven techniques to solve issues related to resource management in B5G/6G networks have been proposed~\cite{guan2021customized}~\cite{thantharate2023adaptive6g}. Data-driven approaches permit the training of a model on a large amount of data, associate the relationship between input and output, and forecast an optimal resource allocation. Among the manifold data-driven algorithms that can be used for resource allocation, we find traditional DL schemes, reinforcement learning with its DL-based variants, such as Deep Q-Learning, and FL. 
However, the proposed techniques cannot deal with all the challenges, especially when executed over resource-constrained client devices.
The training of a complex FL-based resource management model with low-resource devices can lead to many problems, starting from the lack of storage resources, where the whole model cannot be loaded into the small memory of the device and also cannot be trained due to the lack of compute resources. In addition, the training of the model will consume a lot of energy and will be time consuming. In this scenario, to ensure the training process in resource-constrained devices, we need to split the model into shards between the different entities, which will optimize the resource utilization with a fast convergence time. So, SFL is a recommended technique for the 6G network resource allocation optimization, especially when device resources are scarce.

\subsubsection{Realistic Scenario} 
\begin{figure*}[!h]
    \centering
  \includegraphics[height=10cm,width=16cm]{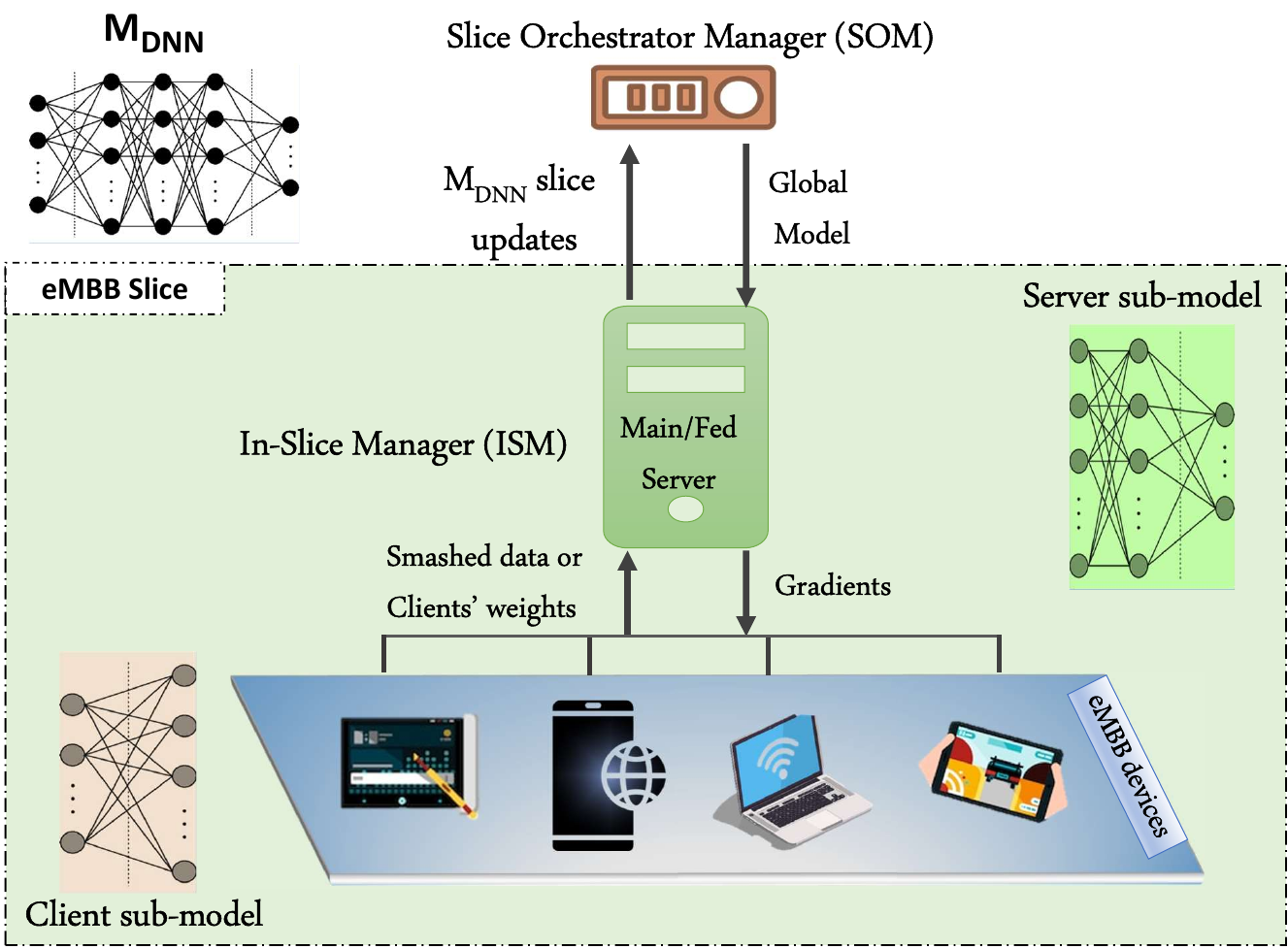}
    \caption{An illustration of SFL for Resource Management (First Scenario).}
    \label{fig:RM-SFL}
\end{figure*}

To demonstrate the efficiency of SFL in solving resource management issues, we project its application on Network Slicing architectures in 5G and beyond. In~\cite{thantharate2023adaptive6g} the authors propose a new framework called ADAPTIVE6G based on the Transfer Learning (TL) paradigm that offers incontrovertible advantages by reusing an already trained neural network model, instead of developing a fresh model from scratch, to resolve related problems~\cite{nguyen2022transfer}. ADAPTIVE6G considers three slices A (eMBB), B (mIoT), and C (URLLC) respectively. Founded on the data collected from all slices, it builds up a traditional Deep Neural Model $M_{DNN}$ to predict the total network loads. $M_{DNN}$ contains five neural layers: the input, output, and three hidden layers. 
After optimizing $M_{DNN}$, the learned weights are held as TL parameters to train a new model called $M_{ADAPTIVE6G}$ using the dataset of each slice individually, releasing three models $M_{eMBB}$, $M_{mIoT}$ and $M_{URLLC}$ (one for each slice). Performing training on the entire model and dataset may spend more time and energy for devices that know tiny size and limited resources. In this scenario, SFL may be applied in two manners: \begin{itemize}[label=\textbullet, font=\LARGE]
    \item First, with the $M_{DNN}$ model by splitting the five layers among the clients within each slice and the main server. As depicted in Fig.~\ref{fig:RM-SFL}, the first two layers are running on the eMBB devices whereas the last three layers are running on the main server, represented by the in-slice manager entity (ISM). Given that context, the in-slice manager plays twofold roles, namely the main server and fed server for all slice devices. After some iterations, each in-slice manager forwards the obtained model towards the Slice Orchestrator Manager (SOM) to create the global model.
    
    \item Second, applying SFL on the $M_{ADAPTIVE6G}$ model inside each slice would be immensely beneficial in terms of efficient use of network resources. With this aim, we suggest a resource-aware split strategy where the number of layers running on each slice is not fixed but varying per slice based on its device capabilities. Using this approach, we propose to split the $M_{ADAPTIVE6G}$ learning model into just one layer as a client segment inside the mIoT slice due to its low-power devices, while the remaining four layers are sent to the main server. Following the same concept, eMBB and URLLC devices can have more layers as they have more computing resources and large memory compared with mIoT slice devices. To achieve that, we suggest preserving two layers for eMBB and URLLC clients and running the other three layers on the main server. In this scenario, ISM acts as the main server in close proximity to slice clients whilst the SOM entity acts as a fed server that aggregates the sub-models updates contributed by the participating clients.
\end{itemize}

\subsection{Intelligent Edge Computing}
\subsubsection{Introduction, requirements, and existing solutions} 
Copious volumes of data are generated incessantly by the various kinds of ubiquitous devices. The transmission of this amount of data to remote cloud servers puts a strain on the network infrastructure, while their centralized storage and processing requires massive cloud resources. In the reverse direction, massive content and service delivery, as well as bounded latency requirements of real-time applications, call for resource disaggregation and the delivery of services from locations closer to end users. 
The Multi-Access Edge Computing (MEC) paradigm offers answers to these challenges~\cite{Taleb17}. It consists in moving computing resources close to the data source and executing network operator or third-party services at telco-operated edge data centers close to the radio access network~\cite{Zaki_2, Zaki_3}. ETSI provides a comprehensive set of standards specifying various aspects of the MEC architecture~\cite{MEC003}. The integration of AI algorithms into the edge enabled the emergence of a new form called Intelligent Edge Computing (IEC), which is a missing element in 5G networks~\cite{peltonen20206g}. IEC will well be an important component in future 6G networks, by making services more intelligent, secure, autonomous, reliable and scalable. The appearance of distributed learning such as FL and SplitNN has supported its progress by the utilization of the edge capabilities to train and share models providing added value and optimized services. A wave of applications stimulates the deployment of an edge-native solutions, such as live video-based facial recognition in smart spaces and air pollution monitoring, to name a few, that call for real-time data processing. IEC starts to draw a keen interest among specialists and research communities, wherein series of works study the intersection between AI and edge computing. A recent study in~\cite{al2021survey} provides a comprehensive survey on the IEC technologies in 6G. This article describes the necessary IEC's concepts and raises new open challenges and future directions in IEC within 6G networks. In~\cite{xiao2020toward}, the authors propose a self-learning architecture based on self-supervised Generative Adversarial Nets (GANs), to illustrate the performance improvement that can be achieved by automatic data learning and synthesizing at the edge of the network. Self-learning is a prominent field in ML, that allows automatic data collection, label generation, feature extraction, and model construction without human involvement. Self-supervised learning is one of the principal axes of self-learning that permits the generation of labeled data set from unlabeled data~\cite{liu2021self}.

\subsubsection{Challenges and how SFL can help}
Although intelligent edge computing is an attractive technology to cover the limitations of the cloud, there are some challenges that need to be addressed, such as security and privacy-associated ones, wherein different types of attacks may be launched, such as data poisoning, data evasion, and privacy attacks. At the same time, there is a persistent need of the cloud in particular for big data processing due to the limited computation and storage capabilities of edge servers. Therefore, lightweight AI algorithms must be utilized to provide smart applications for edge scenarios. In effect, SplitFed and IEC form a complete unit, where each of them will enhance and emphasize the qualities of the other. The integration of SplitFed will accommodate IEC's technology requirements. Because of its model segmentation feature, a large model can be optimally trained at the edge on massive data generated in smart spaces permitting a more efficient use of device resources. Also, with IEC, both the Fed Server and the main server would be deployed at the edge network instead of the cloud, which reduces the distance between servers and edge devices, leading to a low latency connection in the forward and backward propagation steps, higher training speed, and less network traffic. Furthermore, in addition to the protection of user data, SFL improves on model privacy and ensures a high reliability due to the high number of edge nodes (user devices and servers), that participate in the training process. Moreover, the proximity feature makes the prediction of the end-user's location easier, and helps in the training of SFL models for localization based services.

\subsubsection{Realistic Scenario} 
\begin{figure*}[!h]
    \centering
  \includegraphics[height=10cm,width=18cm]{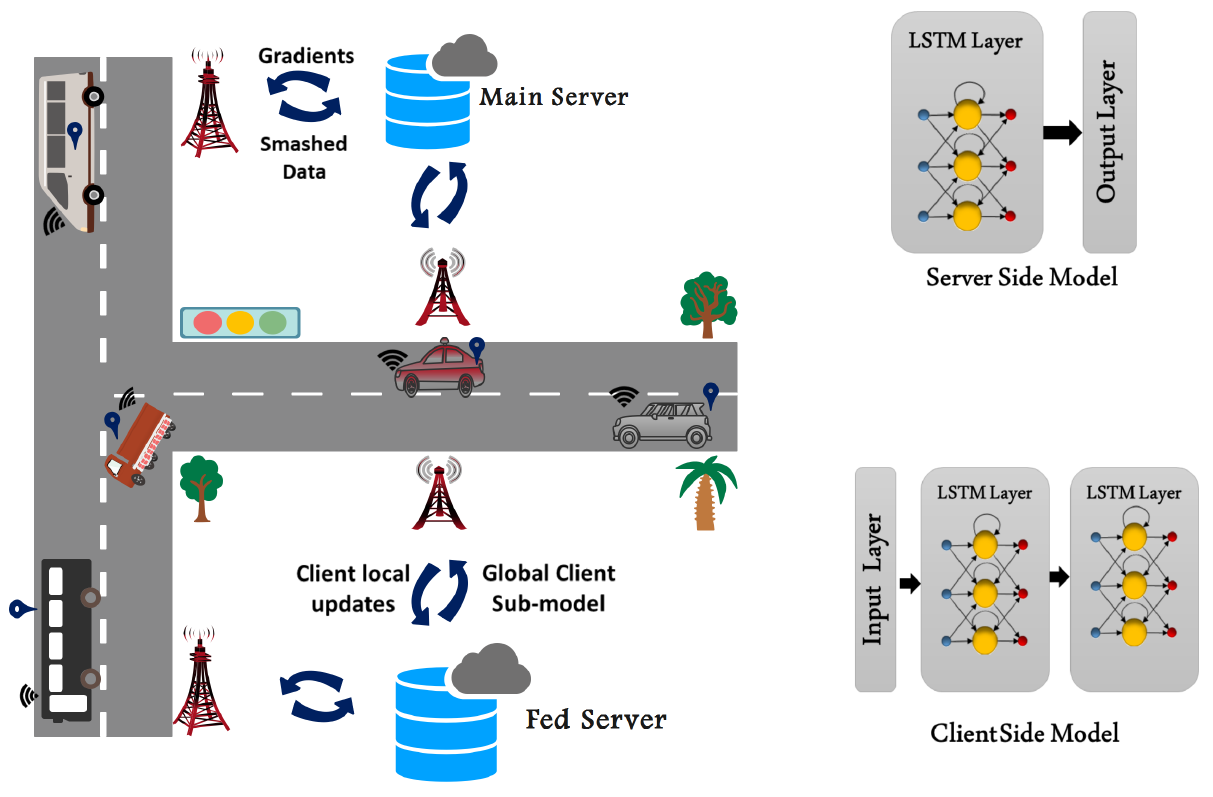}
    \caption{SFL-enabled vehicle mobility prediction for a collision avoidance system.}
    \label{fig:Mobility-SFL}
\end{figure*}

To depict how SFL can be applied in Intelligent Edge Computing, we examine a realistic scenario. In~\cite{brik2021toward}, the authors propose a MEC deep learning-powered framework for an optimal execution of vehicle collision detection and avoidance services. The proposed architecture involves predicting first the density of vehicles to be covered by a MEC host. Then, according to the observed vehicle density, the required MEC computing resources are deduced. To predict the vehicle mobility, a long short-term memory (LSTM) based model is used. It consists of five layers that start with a fully connected input layer of $56$ neurons, three stacked LSTM layers, each with $56$ neurons and an output layer of $45$ neurons. The evaluation process was made on real-world taxi GPS data, consisting of 464019 entries, recorded over 30 days in the city of San Francisco~\cite{epfl-mobility-20090224}. The system assumes that each taxi forwards periodically the vector (timestamp, ID, GPS coordinates) to a central server. In this regard, the incorporation of SFL could bring significant improvements. \figurename~\ref{fig:Mobility-SFL} illustrates how SFL can help
in enhancing mobility prediction in the proposed framework. First, the data remain on participating taxis and each taxi carries out a part of model training. As stated before, only model parameters would be transferred over the network. Moreover, both servers involved in the SFL paradigm would be at the edge, near the taxis, which will enhance learning performance. Once a vehicle completes its local training task, it sends the smashed data to the nearest base station which plays the role of a bridge between the taxis and the main server. Next, the main server continues with the forward/backward propagation procedures and returns, through the closest base station, the adjusted weights to the taxis' cut layers for the rest of training. The process is repeated until reaching an expected level of accuracy.

\subsection{Privacy, Trust, and Security}
\subsubsection{Introduction, requirements, and existing solutions}
6G is a hyper-connected network that allows communication over the ground, air, sea and space~\cite{ray2021review}, connecting the digital, virtual, and physical worlds~\cite{zhang2019technology}. Undoubtedly, this would open up a range of subject matters in terms of privacy, trust, and security. How to deal with threats, vulnerabilities, and attacks in an ultra-dense, heterogeneous, and complex network? How to assure the CIA triad (Confidentiality, Integrity, and Availability) and offer the best user experience? How to build a trustworthy network and defend users' personal data against bad practices? In the literature, a plethora of thematic solutions applying varied approaches have been proposed. 

In recent years, Distributed Ledger Technologies (DLT) such as Blockchain are one of the most explored techniques in this field due to their exclusive features such as decentralization, replication, immutability, transparency, and traceability~\cite{leema2021fundamentals}.
This is manifested in the ongoing discussion on the integration of Blockchain with networks beyond 5G~\cite{Khan22WC}\cite{Sabra_1}. In such networks, deployment cost challenges associated with operating smaller (and denser) cells due to higher radio frequencies, and accountability challenges due to supporting multi-vendor settings arise. These challenges call for efficient and trustworthy \emph{resource and infrastructure sharing} among different providers, and Blockchain can be seen as the fabric to support automated, transparent, and accountable SLA management mechanisms to implement it~\cite{Faisal22}. Accountability and transparency are critical to build trust among the involved stakeholders, such as network operators and device vendors. From a technical standpoint, smart contracts executing on the Blockchain can be used, among others, to encode rules to create and regulate a RAN resource marketplace~\cite{Le23IoTJ, 9524474}, or to implement automated negotiation mechanisms among infrastructure providers for offering end-to-end network slices to vertical service providers~\cite{Sabra22}. 

On the federated learning front, Issa et al.~\cite{IssaBlockFed23} and Zhu et al.~\cite{ZhuBlockFed23} survey recent works that integrate Blockchain into federated learning design to address specific security and performance issues of the latter. 
This body of works addresses such issues in two major ways: (i) Using Blockchain and the associated payment mechanisms to incentivize FL nodes to contribute computation resources in exchange for specific rewards. (ii) Replacing the traditional centralized FL design with a Blockchain-powered, decentralized one, where there is no single node that acts as the FL server; instead, this role is shared by multiple FL nodes in a peer-to-peer manner, where nodes exchange encrypted model updates and smart contracts execute secure model aggregation, with the global model being recorded in the Blockchain in an immutable, transparent, and reliable way.

Contemporary cybersecurity solutions leverage the capabilities of machine learning, since it enables the automatic detection of abnormal behaviors. Several ML algorithms confirmed their reliability in analyzing traffic data and detecting malicious attacks. For instance, in~\cite{gojic2022proposal}, authors proposed a new centralized solution combining RNN with an Autoencoder to detect DDOS attacks. In another work~\cite{d2021effective}, CNN with RNN have been used for Android Malware Detection. Recently, FL has also been actively introduced for malicious attacks detection. In~\cite{rey2022federated}, authors proposed a FL framework using MLP and AE, to detect malicious software in IoT devices. In another work~\cite{yang2022improved}, authors leverage FL to enhance privacy protections in 6G cybertwin networks.

\subsubsection{Challenges and how SFL can help}
 
With its vision towards connected intelligence, 6G is expected to both feature AI-driven operation, and support key use cases that make heavy use of AI (see Section~\ref{sec:SplitFed for 6G Use Case}). Therefore, it inherits AI security and privacy challenges that have received significant attention. Based on the model life cycle (training and testing), four types of relevant attacks can be distinguished: model extraction and model inversion attacks (privacy threats), and adversarial and poisoning attacks (security threats)~\cite{9294026}. 
The goal of model extraction attack is to generate a substitute model architecture as a close approximation to the original model based on a query dataset and the relation between the input and output pairs. The model inversion attack, proposed in 2015 by Fredrikson et al~\cite{fredrikson2015model}, consists of finding the input that has a high resemblance to the record that was used in the training set. In adversarial attacks, the attacker leads the target model to report false predictions with a high confidence. In poisoning attacks, the adversary focuses on polluting the model's training data. A variety of techniques have been integrated to solve privacy-preserving DL, for instance, differential privacy (DP) and homomorphic encryption (HE). The former shields from the inversion attack by applying noise on the input data whilst the latter is a cryptography technology that enables to perform computation and processing on encrypted data without revealing the original one. However, both mechanisms raise model privacy-accuracy trade-offs. The defenses for DL security threats are divisible into two main categories: Adversarial defenses, such as pre-processing and malware detection, and poisoning defenses that aim to remove the poisoning samples during the training phase. At present, there is no universal approach to address all deep learning privacy and security issues. A synergy between FL and SplitNN can be used to mitigate some of these problems, particularly for model privacy violations. In view of the fact that the model is split into sub-models and the duplicate instances of the client side copy, the model would be more robust and reliable regarding models' attacks. For instance, if one sub-model was invaded, the rest remain intact and the attacker cannot infer all the attributes and parameters of the model.

\subsubsection{Realistic Scenario} The core objective of an Intrusion Detection System (IDS) is guarding data, services and applications against threats and attacks. Standard data-driven IDS involves the presence of very large amounts of network traffic data in a central location for training purposes. However, gathering data in a single site abets intrusion attempts and facilitates data stealing. In addition, the transmission of data over a vulnerable environment endangers its security. Furthermore, the centralized training on huge network traffic can affect latency whereas IDS claims analysis responsiveness. Authors in~\cite{binbusayyis2021unsupervised} propose an unsupervised Deep Learning Approach for IDS that encompasses a one dimensional convolutional auto-encoder (1D CAE) and one-class classifier SVM (OCSVM). The former (CAE) is utilized as a feature representation learning method while the OCSVM is used for attack detection. Note that one class classification is considered because of the imbalanced IDS dataset (the model is trained only with the knowledge of normal traffic). SFL applied to Intrusion Detection Systems can provide an effective strategy by maximizing work division. For instance, in the features learning phase, each host in the network downloads and trains its client side CAE model (the encoder part) locally. Afterwards, the server side recuperates the compressed data (memorized in the Bottleneck layer) and decompresses them (the decoder layer). Then, it calculates the MSE reconstruction loss and back-propagates the model parameters. The cycle continues until convergence is reached. The application of SFL at this stage alleviates the computational complexity for the central processing server, preserves data privacy and enhances bandwidth utilization.

\subsection{Zero Touch System Management}
\subsubsection{Introduction, requirements, and existing solutions}
The distributed network architecture and multi-service support enabled by new technologies (SDN, MEC, network slicing, and NFV) involve increased complexity in the management of 5G and beyond networks where the existing solutions are inefficient in managing, monitoring and orchestrating all the operations and services of the network. In 2017, ETSI adopted a new framework called the Zero-touch Network and Service Management (ZSM)~\cite{etsi2019zero}. It aims to minimize human intervention by enabling the full automation of all processes and networking services. Emerging disciplines such as AI, ML, DL and Big Data play a significant role in the self-governing of the network (for instance: self-configuration, self-optimization, self-healing and self-protection). Note that the aforementioned attributes can be expanded to self-* to support more of the network's autonomic capabilities. This would be helpful in reducing typical causes of humans errors, improving network performance, and likewise shorten the time and operational costs. There have been many research works that analyze the benefits of integration of AI/DL algorithms in ZSM~\cite{liyanage2022survey}\cite{benzaid2020ai} in order to create an automated network for customers. However, the success of the full automation process depends on multiple parameters such as the learning algorithm being used and the quality of the input data. Besides, many relevant challenges would accompany this network transformation as we will see in the next section.

\subsubsection{Challenges and how SFL can help}

To deal with the increased network complexity of beyond-5G systems, full E2E automation is needed and AI-based ZSM offers a good answer. However it also has limitations. For instance, a higher accuracy with a short training time is one of the fundamental challenges in AI/ML model-based ZSM systems. Interface-level security issues (e.g., Open API security threats), E2E management, scalability, privacy, near real-time systems are among the dares that confront the application of ZSM. As we have seen in the earlier sections, SplitFed could be introduced at any mentioned point. For instance, it could be applied in the case of network multi-tenancy, where multiple slices have a variety of resource requirements (including radio access network, core network, and cloud computing resources) provided by different administrative domains. The latter possess relevant and pertinent data for global management and orchestration procedures, but are less willing to share it with the global orchestrator. Under these circumstances, the collaboration feature of SplitFed among the administrative domains preserves data privacy while enabling network learning.  

\subsubsection{Realistic Scenario} It would be tricky to disentangle ML/DL and ZSM. Both concepts are elemental parts for the automation of network operations. ML/DL could be integrated in the management of several network categories described under the label of the FCAPS model (Fault, Configuration, Accounting, Performance and Security). In this example, we focus on the work presented in~\cite{brik2020predicting} where the authors formulate a new FL-based solution for performance and security functions. The proposed model predicts slices' service-oriented Key Performance Indicators (KPIs) to act quickly to the decline of one or more QoS parameters of a running network slice and maintain the specification of a Service Level Agreement (SLA). The authors assume the presence of an in-slice manager to monitor the service-level KPI, and train the FL local model. SplitFed, as an intelligent, decentralized and cooperative solution, could be used to further enhance the performance of the model. Instead of training the entire deep neural network model, each in-slice manager trains a sub-model using its private data subset. After some forward-backward passes between the participating clients and the main server, the client-side sub-models are aggregated by the Fed Server. Considering the parallel and continuous aspects of SFL, it is coherent that the convergence rate of the proposed model would be improved.

As could be observed throughout this section, all the examined papers either apply a traditional centralized learning or federated algorithms as a distributed solution. In Table~\ref{tab:sum-technical-aspect}, we highlight, for a list of selected works, how SFL can enhance both techniques by identifying the corresponding SFL agents (clients and servers), as well as the benefits that would accrue from the application of SFL.

%comparative table
% \usepackage{multirow}
% \usepackage{graphicx}
%\begin{table*}[]
\begin{sidewaystable*}
\caption{SFL Contribution on a Selected Group of Studies Related to 6G Technical Aspects.}
\label{tab:sum-technical-aspect}
\setlength\tabcolsep{2.5pt}
\centering
\resizebox{\textwidth}{!}{%
\begin{tabular}{|c|l|c|c|c|l|l|cccl}
\hline
\rowcolor[HTML]{C0C0C0} 
\cellcolor[HTML]{C0C0C0} &
  \multicolumn{1}{c|}{\cellcolor[HTML]{C0C0C0}} &
  \cellcolor[HTML]{C0C0C0} &
  \cellcolor[HTML]{C0C0C0} &
  \cellcolor[HTML]{C0C0C0} &
  \multicolumn{1}{c|}{\cellcolor[HTML]{C0C0C0}} &
  \multicolumn{1}{c|}{\cellcolor[HTML]{C0C0C0}} &
  \multicolumn{4}{c}{\cellcolor[HTML]{C0C0C0}\textbf{SFL Contribution}} \\ \cline{8-11} 
\rowcolor[HTML]{C0C0C0} 
\cellcolor[HTML]{C0C0C0} &
  \multicolumn{1}{c|}{\cellcolor[HTML]{C0C0C0}} &
  \cellcolor[HTML]{C0C0C0} &
  \cellcolor[HTML]{C0C0C0} &
  \cellcolor[HTML]{C0C0C0} &
  \multicolumn{1}{c|}{\cellcolor[HTML]{C0C0C0}} &
  \multicolumn{1}{c|}{\cellcolor[HTML]{C0C0C0}} &
  \multicolumn{3}{c|}{\cellcolor[HTML]{C0C0C0}\textbf{SFL Agents}} &
  \multicolumn{1}{c|}{\cellcolor[HTML]{C0C0C0}} \\ \cline{8-10}
\rowcolor[HTML]{C0C0C0} 
\multirow{-3}{*}{\cellcolor[HTML]{C0C0C0}\textbf{\begin{tabular}[c]{@{}c@{}}Technical \\ Aspect\end{tabular}}} &
  \multicolumn{1}{c|}{\multirow{-3}{*}{\cellcolor[HTML]{C0C0C0}\textbf{Ref}}} &
  \multirow{-3}{*}{\cellcolor[HTML]{C0C0C0}\textbf{Focus}} &
  \multirow{-3}{*}{\cellcolor[HTML]{C0C0C0}\textbf{\begin{tabular}[c]{@{}c@{}}DL \\ Technique\end{tabular}}} &
  \multirow{-3}{*}{\cellcolor[HTML]{C0C0C0}\textbf{\begin{tabular}[c]{@{}c@{}}DL \\ Algorithm\end{tabular}}} &
  \multicolumn{1}{c|}{\multirow{-3}{*}{\cellcolor[HTML]{C0C0C0}\textbf{Performance Metrics}}} &
  \multicolumn{1}{c|}{\multirow{-3}{*}{\cellcolor[HTML]{C0C0C0}\textbf{Limitations}}} &
  \multicolumn{1}{c|}{\cellcolor[HTML]{C0C0C0}\textbf{Clients}} &
  \multicolumn{1}{c|}{\cellcolor[HTML]{C0C0C0}\textbf{Fed Server}} &
  \multicolumn{1}{c|}{\cellcolor[HTML]{C0C0C0}\textbf{Main Server}} &
  \multicolumn{1}{c|}{\multirow{-2}{*}{\cellcolor[HTML]{C0C0C0}\textbf{SFL Benefits}}} \\ \hline
 &
  ~\cite{elbir2021federated} &
  \begin{tabular}[c]{@{}c@{}}ChannelNet: \\ Channel Estimation\\ Scheme for Massive MIMO\end{tabular} &
  \begin{tabular}[c]{@{}c@{}}Supervised \\ Federated \\ Learning\end{tabular} &
  CNN &
  \begin{tabular}[c]{@{}l@{}}- Communication Overhead\\ - Time Computational Complexity\end{tabular} &
  \begin{tabular}[c]{@{}l@{}}- High Time Complexity\\ - Model Privacy Violation\end{tabular} &
  \multicolumn{1}{c|}{Cellular Users} &
  \multicolumn{1}{c|}{Edge Server} &
  \multicolumn{1}{c|}{Base Station} &
  \multicolumn{1}{l|}{\begin{tabular}[c]{@{}l@{}}- Preserving Model Privacy\\ - Reduce Time Complexity\end{tabular}} \\ \cline{2-11} 
 &
  ~\cite{wang2021federated} &
  \begin{tabular}[c]{@{}c@{}}FedeAMC:\\ Automatic Modulation Classifier\end{tabular} &
  \begin{tabular}[c]{@{}c@{}}Supervised \\ Federated \\ Learning\end{tabular} &
  CNN &
  \begin{tabular}[c]{@{}l@{}}- Confusion Matrix\\ - Classification Probability vs SNR\\ - Loss vs Epoch\end{tabular} &
  - High Model Leakage &
  \multicolumn{1}{c|}{\begin{tabular}[c]{@{}c@{}}- IoT devices\\ - 6G users\end{tabular}} &
  \multicolumn{1}{c|}{Edge Server} &
  \multicolumn{1}{c|}{\begin{tabular}[c]{@{}c@{}}Edge Server\\ Cloud Server\end{tabular}} &
  \multicolumn{1}{l|}{\begin{tabular}[c]{@{}l@{}}- Enhance Training Time\\ - Secure The Model\\ - Enhance the AMC Accuracy\end{tabular}} \\ \cline{2-11} 
 &
  ~\cite{butt2021ml} &
  Positioning &
  \begin{tabular}[c]{@{}c@{}}Supervised \\ Learning\end{tabular} &
  \begin{tabular}[c]{@{}c@{}}Decision \\ Tree\end{tabular} &
  - Positioning Accuracy &
  \begin{tabular}[c]{@{}l@{}}- Centralized Training\\ - Data and Model Disclosures\end{tabular} &
  \multicolumn{1}{c|}{- UEs} &
  \multicolumn{1}{c|}{- gNB/gNB-CU} &
  \multicolumn{1}{c|}{\begin{tabular}[c]{@{}c@{}}- gNB/gNB-CU\\ -OAM\end{tabular}} &
  \multicolumn{1}{l|}{- Enhance Users' Localization} \\ \cline{2-11} 
\multirow{-4}{*}{\begin{tabular}[c]{@{}c@{}}Intelligent \\ Physical\\  Layer\end{tabular}} &
  ~\cite{elbir2020federated} &
  \begin{tabular}[c]{@{}c@{}}HLHB: \\  Hybrid Beamforming \\ in mm-wave Massive MIMO\end{tabular} &
  \begin{tabular}[c]{@{}c@{}}Supervised \\ Federated \\ Learning\end{tabular} &
  MLP &
  \begin{tabular}[c]{@{}l@{}}- Accuracy\\ - Transmission Overhead\\ - Spectral Efficiency\end{tabular} &
  - Downlink Channel Congestion &
  \multicolumn{1}{c|}{Cellular Users} &
  \multicolumn{1}{c|}{- Base Station} &
  \multicolumn{1}{c|}{- Base Station} &
  \multicolumn{1}{l|}{\begin{tabular}[c]{@{}l@{}}- Reduce Training Time Complexity\\ - Enhance BW Utilization\\ - Reduce End Devices\\ Computational Resources Limitations\end{tabular}} \\ \hline
 &
  ~\cite{sanguinetti2018deep} &
  \begin{tabular}[c]{@{}c@{}}Power Allocation Strategy\\ in Massive MIMO\end{tabular} &
  \begin{tabular}[c]{@{}c@{}}Sequence\\ Algorithms\\    (RNN)\end{tabular} &
  LSTM &
  \begin{tabular}[c]{@{}l@{}}- Mean Squared Error (MSE)\\ - Cumulative Distribution Function \\    (CDF)\end{tabular} &
  \begin{tabular}[c]{@{}l@{}}- Centralized Solution\\ - Single Point of Failure\end{tabular} &
  \multicolumn{1}{c|}{User Equipment} &
  \multicolumn{1}{c|}{- Base Station} &
  \multicolumn{1}{c|}{Edge Server} &
  \multicolumn{1}{l|}{\begin{tabular}[c]{@{}l@{}}- Parallel Training\\ - Distributed and Decentralized\\ Learning\end{tabular}} \\ \cline{2-11} 
 &
  ~\cite{9348107} &
  \begin{tabular}[c]{@{}c@{}}Successive Interference \\ Cancellation (SIC) Architecture\end{tabular} &
  \begin{tabular}[c]{@{}c@{}}Supervised \\ Learning\end{tabular} &
  DNN &
  \begin{tabular}[c]{@{}l@{}}- Bit Error Rate (BER)\\ - Mean Squared Error (MSE)\end{tabular} &
  \begin{tabular}[c]{@{}l@{}}- Large Computational Burden\\ - High Transmission Overhead\\ - Large Transmission Delay\end{tabular} &
  \multicolumn{1}{c|}{\begin{tabular}[c]{@{}c@{}}Intelligent \\ Users\end{tabular}} &
  \multicolumn{1}{c|}{\begin{tabular}[c]{@{}c@{}}Compute Server \\ at the Edge\end{tabular}} &
  \multicolumn{1}{c|}{\begin{tabular}[c]{@{}c@{}}Compute Server \\ at the Edge\end{tabular}} &
  \multicolumn{1}{l|}{\begin{tabular}[c]{@{}l@{}}- Reduce the Computational load,\\ Training Overhead and Time\end{tabular}} \\ \cline{2-11} 
 &
  ~\cite{qi2020federated} &
  \begin{tabular}[c]{@{}c@{}}Proactive Handover \\ in mmWave Vehicular Networks\end{tabular} &
  \begin{tabular}[c]{@{}c@{}}Federated \\ Learning\end{tabular} &
  MLP &
  \begin{tabular}[c]{@{}l@{}}- Test Accuracy\\ - Communication Costs\\ - Number of Handover\\ - Average SNR\end{tabular} &
  - Downlink Congestion &
  \multicolumn{1}{c|}{Smart-cars} &
  \multicolumn{1}{l|}{\begin{tabular}[c]{@{}l@{}}- Road Side Units\\ - Compute Edge Server\end{tabular}} &
  \multicolumn{1}{l|}{\begin{tabular}[c]{@{}l@{}}- Road Side Units\\ - Edge Server\end{tabular}} &
  \multicolumn{1}{l|}{\begin{tabular}[c]{@{}l@{}}- Reduce the Uplink/Donwlink \\ Communication of the Learning Phase\\ - Enhance the Handover Accuracy\end{tabular}} \\ \cline{2-11} 
\multirow{-4}{*}{\begin{tabular}[c]{@{}c@{}}Resource\\  Management\end{tabular}} &
  ~\cite{thantharate2023adaptive6g} &
  \begin{tabular}[c]{@{}c@{}}ADAPTIVE6G:\\ Resource Management Framework\\ for Network Slicing in 5G and Future 5G\end{tabular} &
  \begin{tabular}[c]{@{}c@{}}Transfer\\  Learning\end{tabular} &
  DNN &
  \begin{tabular}[c]{@{}l@{}}- Predicted Network Load\\ - Correlation Coefficient 'R' \\ - Mean Squared Error MSE\end{tabular} &
  \begin{tabular}[c]{@{}l@{}}- Centralized Forecasting\\ - Single Point of Failure\end{tabular} &
  \multicolumn{1}{c|}{\begin{tabular}[c]{@{}c@{}}- Slice eMBB\\ - Slice mIoT\\ - Slice URLLC\end{tabular}} &
  \multicolumn{1}{c|}{- Edge Server} &
  \multicolumn{1}{l|}{\begin{tabular}[c]{@{}l@{}}- One of the slices\\ - Edge Server\end{tabular}} &
  \multicolumn{1}{l|}{\begin{tabular}[c]{@{}l@{}}- Share the Training Process\\ among the Different Slices\end{tabular}} \\ \hline
 &
  ~\cite{zeb2021edge} &
  \begin{tabular}[c]{@{}c@{}}Edge-native Framework\\ for Data Flow Prediction in 6G\end{tabular} &
  \begin{tabular}[c]{@{}c@{}}Sequence\\ Algorithms\\    (RNN)\end{tabular} &
  LSTM &
  \begin{tabular}[c]{@{}l@{}}- Root-Mean-Square Error (RMSE)\\ - Coefficient of Determination\end{tabular} &
  \begin{tabular}[c]{@{}l@{}}- Centralized Solution\\ - Heavy Computing Workload\\ during Training\end{tabular} &
  \multicolumn{1}{c|}{\begin{tabular}[c]{@{}c@{}}- End Users\\ (IoT Devices)\end{tabular}} &
  \multicolumn{1}{c|}{Edge Server} &
  \multicolumn{1}{c|}{Edge Server} &
  \multicolumn{1}{l|}{\begin{tabular}[c]{@{}l@{}}- Distributed, Decentralized and \\ Parallel Training\\ \end{tabular}} \\ \cline{2-11} 
 &
  ~\cite{brik2021toward} &
  \begin{tabular}[c]{@{}c@{}}AI-empowered MEC Resources\\  Management in IoV\end{tabular} &
  \begin{tabular}[c]{@{}c@{}}Sequence\\ Algorithms\\    (RNN)\end{tabular} &
  LSTM &
  - Mean Squared Error (MSE) &
  - Centralized Prediction &
  \multicolumn{1}{c|}{- Driving Cars} &
  \multicolumn{1}{c|}{RSU} &
  \multicolumn{1}{c|}{\begin{tabular}[c]{@{}c@{}}RSU\\ Edge Server\end{tabular}} &
  \multicolumn{1}{l|}{\begin{tabular}[c]{@{}l@{}}- Distributed Prediction\\ - Enhance Computation Latency\\ - Enhance the Mobility Prediction and \\  Computing Resource  Allocation\end{tabular}} \\ \cline{2-11} 
 &
  ~\cite{liu2021fedcpf} &
  \begin{tabular}[c]{@{}c@{}}FedCPF:\\ Communication Approach for Vehicular\\ Edge Computing (VEC) in 6G\end{tabular} &
  \begin{tabular}[c]{@{}c@{}}Federated \\ Learning\end{tabular} &
  MLR &
  \begin{tabular}[c]{@{}l@{}}- Test Accuracy\\ - Training Loss\\ - Communication Costs\end{tabular} &
  \begin{tabular}[c]{@{}l@{}}- Model Transmission Issues\\ - High Energy Consumption\end{tabular} &
  \multicolumn{1}{c|}{\begin{tabular}[c]{@{}c@{}}Vehicular Nodes\\ RSU\end{tabular}} &
  \multicolumn{1}{c|}{\begin{tabular}[c]{@{}c@{}}RSU\\ Base Station\\ Edge Server\end{tabular}} &
  \multicolumn{1}{c|}{\begin{tabular}[c]{@{}c@{}}RSU\\ Base Station\\ Edge Server\end{tabular}} &
  \multicolumn{1}{l|}{\begin{tabular}[c]{@{}l@{}}- Ameliorate the Convergence Speed\\ - Increase Model Protection\\ - Improve Energy Efficiency\end{tabular}} \\ \cline{2-11} 
\multirow{-4}{*}{\centering\begin{tabular}[c]{@{}c@{}}Intelligent\\ Edge \\ Computing\end{tabular}} &
  ~\cite{he2022acefl} &
  \begin{tabular}[c]{@{}c@{}}AceFL:\\ Federated Edge Learning Scheme\\ in 6G-enabled MEC Networks\end{tabular} &
  \begin{tabular}[c]{@{}c@{}}Federated \\ Learning\end{tabular} &
  \begin{tabular}[c]{@{}c@{}}CNN\\ $+$\\ MLCR\\ $+$ \\ LSTM\end{tabular} &
  \begin{tabular}[c]{@{}l@{}}- Training Performance\\  (Loss and Accuracy)\end{tabular} &
  \begin{tabular}[c]{@{}l@{}}- Model Transfer Constraints\\ - Intensive Resources Utilization\end{tabular} &
  \multicolumn{1}{c|}{\begin{tabular}[c]{@{}c@{}}Edge Users\\ (Mobile Devices)\end{tabular}} &
  \multicolumn{1}{c|}{\begin{tabular}[c]{@{}c@{}}Base Station\\ Edge Server\end{tabular}} &
  \multicolumn{1}{c|}{\begin{tabular}[c]{@{}c@{}}Edge Computing \\ Server\end{tabular}} &
  \multicolumn{1}{l|}{\begin{tabular}[c]{@{}l@{}}- Enhance Resource Consumption\\ - Enhance Training Efficiency\\ - Optimize Energy Usage\end{tabular}} \\ \hline
 &
  \cite{yang2022improved} &
  \begin{tabular}[c]{@{}c@{}}Privacy-Preserving in\\  Cybertwin-Driven\\ 6G System\end{tabular} &
  \begin{tabular}[c]{@{}c@{}}Federated \\ Learning\end{tabular} &
  DNN &
  \begin{tabular}[c]{@{}l@{}}-Test Accuracy\\ - KullbackLeibler (KL) divergence\end{tabular} &
  \begin{tabular}[c]{@{}l@{}}- Strong Storage and Processing\\ Costs to Train the Entire Model\end{tabular} &
  \multicolumn{1}{c|}{6G Terminals} &
  \multicolumn{1}{c|}{Base Station} &
  \multicolumn{1}{c|}{MEC Server} &
  \multicolumn{1}{l|}{\begin{tabular}[c]{@{}l@{}}- Enhance the Convergence Rate\\ - Adding a Model Security Layer  \\ - Model Parallel-Training\end{tabular}} \\ \cline{2-11} 
 &
  ~\cite{gojic2022proposal} &
  DDoS Attack Detection &
  \begin{tabular}[c]{@{}c@{}}Supervised\\ Learning\end{tabular} &
  \begin{tabular}[c]{@{}c@{}}RNN\\ $+$\\ AE\end{tabular} &
  {\begin{tabular}[c]{@{}l@{}}- Expected number of incorrectly \\ detected
attacks\\ - System's Reliability (R\_{s})\end{tabular}} &
  - Centralized Detection System &
  \multicolumn{1}{c|}{\begin{tabular}[c]{@{}c@{}}Connected\\ Devices\end{tabular}} &
  \multicolumn{1}{c|}{gNB} &
  \multicolumn{1}{c|}{MEC Server} &
  \multicolumn{1}{l|}{\begin{tabular}[c]{@{}l@{}}- Lightweight Training (Sub-Models)\end{tabular}} \\ \cline{2-11} 
 &
  ~\cite{d2021effective} &
  Android Malware Detection &
  \begin{tabular}[c]{@{}c@{}}Supervised\\ Learning\end{tabular} &
  \begin{tabular}[c]{@{}c@{}}CNN\\ $+$\\  RNN\end{tabular} &
  \begin{tabular}[c]{@{}l@{}}- Accuracy, Sensitivity, Specificity\\ - Precision, F-Score\\ - AUC\end{tabular} &
  \begin{tabular}[c]{@{}l@{}}- Model Inversion Attack\\ - Model Poisoning Attack\end{tabular} &
  \multicolumn{1}{l|}{- Medical Devices} &
  \multicolumn{1}{c|}{Edge Server} &
  \multicolumn{1}{c|}{Edge Server} &
  \multicolumn{1}{l|}{\begin{tabular}[c]{@{}l@{}}- Preserve Model Inversion Attack\\ - Preserve Model Poisoning Attack\end{tabular}} \\ \cline{2-11} 
 &
  ~\cite{shrestha2021machine} &
  \begin{tabular}[c]{@{}c@{}}Intrusion Detection System \\ for Cellular Connected UAV Networks\end{tabular} &
  \begin{tabular}[c]{@{}c@{}}Supervised \\ Learning\end{tabular} &
  \begin{tabular}[c]{@{}c@{}}LR, LDA,\\  KNN,  DT,\\  GN\end{tabular} &
  \begin{tabular}[c]{@{}l@{}}- Accuracy rate, Precision, Recall\\ - F1-score, False-Negative Rate\end{tabular} &
  \begin{tabular}[c]{@{}l@{}}- High Training Time\\ - Privacy Leakage of IDS Models\end{tabular} &
  \multicolumn{1}{c|}{UAVs} &
  \multicolumn{1}{c|}{MEC Server} &
  \multicolumn{1}{c|}{MEC Server} &
  \multicolumn{1}{l|}{\begin{tabular}[c]{@{}l@{}}- Distribute the Model Training among\\ Cellular Nodes\end{tabular}} \\ \cline{2-11} 
\multirow{-5}{*}{\begin{tabular}[c]{@{}c@{}}Privacy, \\ Trust, \\ and Security\end{tabular}} &
  \cite{rey2022federated} &
  \begin{tabular}[c]{@{}c@{}}Malware Detection\\ in IoT Devices\end{tabular} &
  \begin{tabular}[c]{@{}c@{}}Federated \\ Learning\\ (Supervised;\\ Unsupervised)\end{tabular} &
  \begin{tabular}[c]{@{}c@{}}MLP\\  $+$ \\ AE\end{tabular} &
  \begin{tabular}[c]{@{}l@{}}- Accuracy\\ - F1 Score\end{tabular} &
  \begin{tabular}[c]{@{}l@{}}- Training the Entire Model \\ - Low Battery Level \\ - Byzantine failure\end{tabular} &
  \multicolumn{1}{c|}{- IoT Devices} &
  \multicolumn{1}{c|}{Edge Server} &
  \multicolumn{1}{c|}{MEC Server} &
  \multicolumn{1}{l|}{\begin{tabular}[c]{@{}l@{}}- Offload Training Costs from IoT nodes\\ - Defend against Model Attack\\ - Enhance the FL Model Privacy\end{tabular}} \\ \hline
 &
  ~\cite{brik2020predicting} &
  \begin{tabular}[c]{@{}c@{}}Predicting Service-oriented \\ Network Slices Performances\end{tabular} &
  \begin{tabular}[c]{@{}c@{}}Federated\\  Learning\end{tabular} &
  ANN &
  \begin{tabular}[c]{@{}l@{}}- MSE / MAE\\ - Real and Predicted Delay\\ - Communication Overhead\end{tabular} &
  \begin{tabular}[c]{@{}l@{}}- Model Leakage\\ - Aggregator Server \\ Centralization\end{tabular} &
  \multicolumn{1}{c|}{\begin{tabular}[c]{@{}c@{}}EUs of\\ Network \\ Slices\end{tabular}} &
  \multicolumn{1}{c|}{MEC Server} &
  \multicolumn{1}{c|}{MEC Server} &
  \multicolumn{1}{l|}{\begin{tabular}[c]{@{}l@{}}- Enhance Prediction Accuracy\\ - Reduce the Communication\\ Overhead\end{tabular}} \\ \cline{2-11} 
 &
  ~\cite{luo2018channel} &
  \begin{tabular}[c]{@{}c@{}}OCEAN Framework:\\ Automated Prediction \\ of Channel State Information (CSI)\end{tabular} &
  \begin{tabular}[c]{@{}c@{}}Unsupervised\\ Learning\end{tabular} &
  \begin{tabular}[c]{@{}c@{}}CNN\\ $+$\\ LSTM\end{tabular} &
  \begin{tabular}[c]{@{}l@{}}- Accuracy\\ - Computing Time\end{tabular} &
  \begin{tabular}[c]{@{}l@{}}- Centralized CSI Prediction\\ - High Computational Resources\end{tabular} &
  \multicolumn{1}{c|}{\begin{tabular}[c]{@{}c@{}}- Cellular End\\ Devices\end{tabular}} &
  \multicolumn{1}{c|}{- gNB/gNB-CU} &
  \multicolumn{1}{c|}{\begin{tabular}[c]{@{}c@{}}- gNB/gNB-CU\\ -OAM\end{tabular}} &
  \multicolumn{1}{l|}{\begin{tabular}[c]{@{}l@{}}- Enhance both Accuracy\\ and  Estimation Computing Time\end{tabular}} \\ \cline{2-11} 
 \multirow{-3}{*}{\centering\begin{tabular}[c]{@{}c@{}}Zero \\ Touch \\ System \\ Management\end{tabular}} &
  ~\cite{thantharate2019deepslice} &
  \begin{tabular}[c]{@{}c@{}}DeepSlice\\ Automatic Network \\ Slicing Prediction\end{tabular} &
  \begin{tabular}[c]{@{}c@{}}Supervised\\  Learning\end{tabular} &
  \begin{tabular}[c]{@{}c@{}}DNN\\ $+$\\ RF\end{tabular} &
  \begin{tabular}[c]{@{}l@{}}- Accuracy / Loss\\ - Number of Active Users\end{tabular} &
  - Centralized Prediction &
  \multicolumn{1}{c|}{\begin{tabular}[c]{@{}c@{}}- Connected \\ Devices\end{tabular}} &
  \multicolumn{1}{l|}{\begin{tabular}[c]{@{}l@{}}- Base Station\\ - gNB/gNB-CU\end{tabular}} &
  \multicolumn{1}{c|}{\begin{tabular}[c]{@{}c@{}}- MEC Server\\ - OAM\end{tabular}} &
  \multicolumn{1}{l|}{\begin{tabular}[c]{@{}l@{}}- Distributed Prediction\\ - Load Balancing Training\end{tabular}} \\ \hline

\end{tabular}
}
\end{sidewaystable*}
%\end{table*}

\section{SFL for 6G Use Cases} 
\label{sec:SplitFed for 6G Use Case}

\subsection{Industry 5.0, Digital Twin, and Autonomous Robots}
\subsubsection{Motivation} 
Digital Twin technology has a great impetus in the development of governments and businesses in many fields such as healthcare, industry, education and sports. It consists in creating a connection between the virtual space and the physical world~\cite{pal2022digital} by developing a digital replica of a physical asset (animate or inanimate). It has innumerable added value applications, for instance creating an online meeting via avatars that would have the same features of physical persons (voice, behavior and intelligence). In the Qatar World Cup, the digital twin concept was applied by bridging the digital and physical stadiums. The latter ingests real-time data from millions of IoT devices that help in monitoring the situation at every stadium (climate, security, lights, etc.) and therefore take the appropriate action at the adequate time. As well, the introduction of automation and robotics have largely changed the way of working in several areas, wherein some jobs have completely disappeared and been replaced by machines. Examples include agriculture robots for fruit harvesting, plant irrigation and scanning, medical robots for assisted surgery and commercial robots for delivery and sale. Both aforementioned technologies have a profound impact in the development of the current industry that has passed through many phases from the water power, steam engine, electricity, oil and computers to the industry 5.0 that fuses all the emerging technologies. Digital Twin was used in the maritime industry for shipbuilding processes in order to comprehend the ship's behaviors under the different conditions by designing a cyber-physical system (CPS) and therefore improve the safety of marine transportation. Cobots or collaborative robots constitute one of the most remarkable supporting technologies in industry 5.0. Unlike traditional robots that work for humans, Cobots focus on the integration of the human in the manufacturing process by charging machines with dull tasks and entrust duties that demand critical and cognitive thinking to humans~\cite{maddikunta2022industry}. This collaboration and communication permit to leverage human intellect and enhance the industrial process. 

\subsubsection{How SFL can help}
AI and ML play a critical role in the future industry to concretise the use of Autonomous Robots and Digital Twins. For instance, to create a digital model for predictive maintenance systems~\cite{lee2020integration}~\cite{mihai2021digital}, a pool of real-time data generated by the sensors embedded in the physical system is collected, then processed and analyzed by a central unit to be leveraged by the DT model (for potential enhancements to the real asset). However, the recurring transfer of complex system data to a central entity may result in massive network traffic and data privacy leaks. This was the basic cause for the adoption of new advanced distributed solutions. For instance, in~\cite{sun2020adaptive} authors propose a new Federated Learning framework for Industrial IoT where industrial devices (robots, excavators, construction machinery, etc.) perform their manufacturing tasks locally (e.g, training a defective detection model) without sharing their own dataset. This will increase the data privacy and decrease the communication costs. However, the authors do not consider some network constraints, such as the bandwidth utilization and potential vulnerabilities that could be caused by the transfer of the whole global model towards the industrial devices. SFL, as a new technology, could be integrated along with Industry 5.0, Digital Twin and Autonomous Robots as well to deal with issues not previously considered for the development of smart manufacturing models.
For instance, for Digital Twin-enabled 6G networks, SFL can improve the efficiency of machine learning models dedicated to various twin sources to manage the real system and predict its future states by ensuring cost efficiency, resource-optimized operation and instant wireless connectivity that keeps a synchronized digital plane with the physical system. As well, SFL models embedded on robots and Cobots could improve the learning process performance by providing a distributed and parallel training of complex systems over shared learning models (client- and serve-side) and various data islands.

\subsection{Connected and Autonomous Vehicles}
\subsubsection{Motivation}
Preventing car crashes and saving human lives are among the root reasons for smart traffic systems. Recent developments in wireless communication technologies have brought rapid and massive changes in the automotive industry world. A new era of safer and smarter transportation is dawning, namely via connected autonomous vehicles (CAV)~\cite{he2019cooperative}. The concept of a connected vehicle (CV) means that the vehicle is equipped with sophisticated communication and sensing modules (GPS, radars, cameras, sensors, etc.) that allow it to exchange with its surrounding neighbors (other vehicles, infrastructure and personal devices) giving birth to many applications summarized under the term V2X (Vehicle-To-Everything). An autonomous vehicle (AV) refers to a vehicle that has the power to react by itself to any road event without the need of a driver (braking, steering, obstacle avoidance, etc.). A vehicle that possesses the potential to carry out both actions is termed as a CAV. To enable these activities, two sorts of technologies are used. The first one pertains to connectivity, and features many developed standards such as DSRC and C-V2X~\cite{jin2022dsrc}. Secondly, we find AI technology and its branches (see Section~\ref{sec:Background}) dedicated to the automation part. In view of stringent and diverse Quality-of-Service (QoS) needs imposed by smart-transportation applications, that are data-intensive and delay-sensitive, 6G is expected to be the foundation of ITS deployment because of its uniqueness in terms of reliability, latency and massive connectivity. Artificial Intelligence as well will be one of the prime pillars of the future 6G-supported ITS, and research in this field begins to have contributors that discuss a diversity of issues and viewpoints around the topic. In~\cite{yuan2022machine}, an in-depth review on how machine learning can enhance the next-generation ITS main tasks in terms of perception, prediction and management is deliberated. In~\cite{noor20226g}, the authors investigate the key enabling technologies for 6G-V2X and their effect on the different 6G vehicular network aspects, namely communication, computing, and security. The authors subdivide these technologies into two categories: revolutionary and evolutionary V2X technologies. Besides tactile communications, quantum computing, brain-controlled vehicles and blockchain-aided V2X, one of the appealing revolutionized technologies discussed in the paper that can improve the 6G-assisted ITS systems is intelligent reflecting surfaces (IRS)~\cite{song2022deep}. After that, the authors expose a variety of evolutionary technologies that need some changes to become more suited to 6G-V2X, such as advanced resource allocation.

\subsubsection{How SFL can help} AI-aided solutions present a key part of a smart transportation system strategy. However, the majority of existing research papers for next generation ITS share the same basic patterns of training. One pattern follows the traditional design in which transport entities (smart car, smart road, smart traffic light, etc.) forward, in a wireless and continuous manner, information observed from embedded equipment (e.g., smart-car velocity, trajectory, direction and geographical coordinates) to be processed in a central location~\cite{hecker2018failure}. The transmission of these details menaces the driver's privacy. For instance, if attackers gain access to the future location of a driver, they can easily determine the latter's traveled routes and therefore deduce sensitive information about the driver, such as residence, job, health state, religious beliefs and other. The second pattern consists in applying the reverse operation, i.e., transmitting the whole model to the implicated entities as in federated learning paradigm~\cite{lu2019collaborative}~\cite{yu2020mobility}. Model privacy leakage is the common threat with both techniques. Future research directions are expected to strive not only for solutions that consider data protection, but also that propose model privacy-preserving mechanisms. The implementation of SFL in the transportation and mobility industry would improve on this aspect. SplitFed, as a promising technology, will allow connected and autonomous vehicles to share the training of a complex model without violating its privacy. The model would be decoupled among the transport entities, one part at the client's side (vehicular nodes) and the second part at the aggregator's side (main server). In addition, the interplay between SFL and edge computing can be considered to achieve better, more timely, and safer decisions via increasing the proximity among the involved servers by hosting them at the edge. However, the high speed and frequency changing of vehicular nodes may lead to intermittent connection between nodes and servers (Fed server and Main server) in the transmission of gradients and smashed data. The duplication of servers could be a way to address this issue.

\subsection{Intelligent eHealth and Body Area Networks}
\subsubsection{Motivation} 

Healthcare is one of the areas that have witnessed vigorous advances and improvements in both hardware and software platforms through eHealth services and applications. For instance, telemedicine has changed the standard practices in healthcare centers, mainly during the COVID-19 pandemic that hastened the use of teleconsultation, telediagnosis and telemonitoring. These techniques provide patients a personalized and easy access to health services irrespective of their geographical positions especially in rural areas where the healthcare system is unavailable or underdeveloped. A significant development is the Internet of Medical Things (IoMT)~\cite{vishnu2020internet}, also known as H-IoT (Healthcare-IoT). It consists of all implantable sensors and wearable devices (e.g., diabetic pump, smart watches, fitness bands) that record constantly vital parameters such as blood pressure, body temperature, oxygen saturation, glucose level and heart rates, allowing users to track their health state, and detect any abnormal signs. The interconnection among all devices forms a new extension of sensor networks, called Wireless Body Area Networks (WBANs)~\cite{salayma2017wireless}. Through AI-assisted analysis of big health-related data, produced by distinct types of H-IoT devices, several  medical applications of diverse benefits are possible~\cite{yu2018artificial},~\cite{bohr2020rise}, like lung cancer detection, Alzheimer's and Parkinson's disease prediction and diabetic retinopathy recognition. Because medical imaging is the most used clinical examination for disease diagnosis, CNN architectures have taken the lead in the wellness research space and demonstrated high performance for various fundamental tasks. The deployment of WBANs must consider many QoS features such as latency, power-consumption, stable communication (even if the person is moving) and interference mitigation~\cite{taleb2021wireless}. Therefore, the selection of the most appropriate communication technology is of great importance. 6G is envisioned to streamline many aspects in the smart healthcare systems by providing efficient and economic remote services (e.g., surgical intervention via video streaming, out of hospital care using holographic communication) that would support healthcare practitioners in their daily tasks~\cite{9083916}.

\subsubsection{How SFL can help} In 2020, Rieke et al~\cite{rieke2020future} introduced a paper that explores the future of digital health with federated learning (FL) and its potential impact on the various healthcare stakeholders. The study shows how FL overcomes data security and privacy by sharing the updated weights of trained local models instead of the raw data from edge users. However, some points make FL-based approaches not always efficient for healthcare use cases. First, as it is known, images are the type of data highly used in the e-health domain. In general, they are characterized by a large size that requires models with huge numbers of parameters. The frequent transfer of the whole model over unreliable channels and limited bandwidth is a big issue. Second, in contrast to some medical equipment such as Computed Tomography (CT) and Magnetic Resonance Imaging (MRI) scanners, clients are not always powerful; they could be low-power electronic devices. In this case, the training over the high-dimension model is not feasible. Third, the proportional relation between the size of the model and attack success rate (the higher the dimensionality of a model, the higher its probability of being perturbed by an attacker)~\cite{khan2022security} render the model less secure. Using SplitFed, medical devices would only have a lower dimensionality part of the complete model (SplitFed has the ability to vary/decrease the model portion of clients). 
This would save more energy and make the model more robust to poisoning attacks. Besides, given that the computation is performed in parallel, a fast model training is envisioned. In sum, SFL would be favorable for many future healthcare applications that are based on resource-constrained devices and require real-time services. 

\subsection{Multisensory XR Applications and Holographic Telepresence}
\subsubsection{Motivation}
Developments at the intersection of diverse research fields, including computer vision, sensing technology, wearables, holographic display technology, edge computing, specialized AI hardware technology, and high-capacity/low-latency wireless communication have made it possible to offer users new experiences via e\textbf{X}tended \textbf{R}ealities (XR): \textbf{V}irtual \textbf{R}eality (VR), \textbf{A}ugmented \textbf{R}eality (AR) and \textbf{M}ixed \textbf{R}eality (MR). These experiences are difficult, even impossible, to engage with in real life, such as visiting the moon's surface and climbing the deadliest mountain.
VR consists in creating an entire 3D virtual environment that highly simulates the real world. The basic idea behind VR systems is to display computer-generated images of the virtual space in three forms: visual, aural and haptic~\cite{podobnik2012haptics} taking into consideration the person's attitude (position, orientation, eye motion, etc). This representation immerses the user (physically and mentally) into the virtual universe. The AR concept, as its name indicates, permits to augment the real environment. Unlike VR that isolates users totally from their existing world, AR enhances it by adding more virtual objects (digital contents) through digital devices like smartphones, tablets, or AR glasses~\cite{devagiri2022augmented}. 
In effect, the definition of the term MR is debatable, even among experts~\cite{speicher2019mixed}. 
E.g., in~\cite{flavian2019impact}, the authors positioned MR in the middle of AR and AV (Augmented Virtuality), while in~\cite{hoyer2020transforming}, authors categorize MR as an extension of AR, although in~\cite{wedel2020virtual}, MR is defined as an amalgamation of both VR and AR. The latter is the most common among academic researchers. Numerous XR tools, such as VR headsets, VR gloves, AR glasses, Teslasuit and Holosuit, are used to support a wide range of XR applications in various fields, including tourism, education, marketing, agriculture, and medicine~\cite{minopoulos2022opportunities}. Hologram is another immersive media technology that would surmount the distance barrier and provide a real-time presence. It permits people to collaborate and connect to each other by offering a natural conversation experience to such an extent they feel like they are inside the same room~\cite{Clemm20}.
Unlike AR and VR, holographic telepresence does not require wearable devices. At a remote environment, images of humans and their surrounding items are compressed and optimized before being sent over a high bandwidth network connection. Afterwards, these images will be reconstructed (decompressed and laser-projected) at the users’ site. All the aforementioned technologies require QoS guarantees (high processing, sufficient computation power and extra-reliability connections) that surpass the limits of the 5G network. Due to its specific technological and technical aspects, the future 6G is the suitable candidate to fulfill the requirements of XR/Holographic Telepresence systems. 

\subsubsection{How SFL can help} AI/ML aided solutions are key in automating the complex decision-making processes needed to realize future XR/Holographic Telepresence systems. However, the huge quantity of data generated by VR and AR users make the use of a centralized learning paradigm almost impossible because of the high network resource utilization. SFL, as a distributed learning algorithm, is an effective way to reduce network load by just forwarding the model updates that are smaller than user data. In addition, if 6G communication is coupled with SFL algorithms, the performance of XR classification/prediction models could be enhanced. For instance, data confidentiality and privacy are important in all XR applications since sensitive information is implicated to control the XR contents (e.g., eye motion and fingers' position for the estimation of 3D human body pose). In reference to this, SFL can offer several benefits. First, the model is subdivided into multiple sections (sub-models) between XR devices and the main server. Clients start processing on their local data before forwarding their intermediate outputs to the main server to proceed with the training. This will reduce the computational load of individual devices, ensure both user and model privacy, capture context-aware information (e.g., user preferences) leading to customized XR content and interactions. As well, 6G will guarantee a high bandwidth and reliable communication for smashed data and gradients transfer. Similarly, this also holds at inference time for an object detection model that needs an accurate and rapid identification (size, shape, color, location, motion) in order to mix both virtual and physical environments instantaneously. With SFL, the client first extracts low-level features locally (e.g., color or motion information) from the data collected in the physical environment. Then, it sends the intermediate results to the server for extended analysis (such as object localization). Accordingly, this will reduce privacy concerns since sensor data from the physical environment remain on the XR devices, minimize data transmission requirements (only smashed data are shared with the server) and optimize bandwidth usage. In the same way, with Holographic Telepresence where the assimilation of the real environment depends on well understanding all its components, a high throughput and deep analysis of the data collected from the real world are needed for a perfect virtual human training. The combination of SFL and 6G will accelerate the understanding of the real environment and prompt the customers to live an interactive and convincing experience. 

\subsection{Smart Grid 2.0}
\subsubsection{Motivation} Most electrical power distribution systems rely on fossil-fuel generators, which are detrimental to the environment due to the high air pollutants induced by this technique. According to~\cite{you2022impact}, almost 40\% of the $CO_{2}$ emissions are due to power generation. Furthermore, producing a large amount of power in one site increases the delivery cost, particularly for distant consumers where long transmission cables are required. Likewise, a centralized grid topology is more prone to reliability issues, as a sudden fault involves a full blackout. Introducing new information and communication technologies is a good solution to deal with energy issues~\cite{Zakii}. Smart Grid (SG) is an intelligent and distributed digital power system designed to effectively utilize the electricity network. In conventional power networks, there is only one single source and only one way to feed end-users. With SG, the power is emanated from various sources (e.g., solar farm, wind farm) using multi-way communication~\cite{ege2021communication}. In order to build a flawless system, incorporating intelligent and high performance technologies into the diverse smart grid subsystems (generation, storage, transmission, monitoring and distribution) is essential~\cite{dileep2020survey}.  
Indeed, Massive Internet of Things (MIoT) is one of the cutting-edge technologies in supporting smart power grids, including, for instance, smart meters, automated meter reading, vehicle-to-grid systems, smart sensor and actuator networks~\cite{8642293} to mention a few.
All these smart devices participate in the accurate and automated measurement, extraction and transfer of parameter values from the different part of the smart grid system. With the help of AI/ML techniques, the analysis of the collected data would be beneficial to estimate the state of the grid network, boost the quality of experience (QoE),  deploy dynamic pricing and personalized energy services.  
Several ML/DL approaches have been applied for smart grid networks, including, but not limited to: CNN for load forecasting~\cite{khan2019electricity}, LSTM-RNN for photovoltaic power prediction~\cite{abdel2019accurate}, KNN for load and price prediction~\cite{tahir2019load}, SAE for detection and classification of transmission line faults~\cite{7539352}, SVM for cyberattack detection (covert cyber deception assault)~\cite{ahmed2018feature}, and, lastly, random forests  combined with CNN for energy theft detection~\cite{li2019electricity}. 
\subsubsection{How SFL can help} 
\begin{figure*}[!t]
\centering
  \includegraphics[height=11cm,width=18cm]{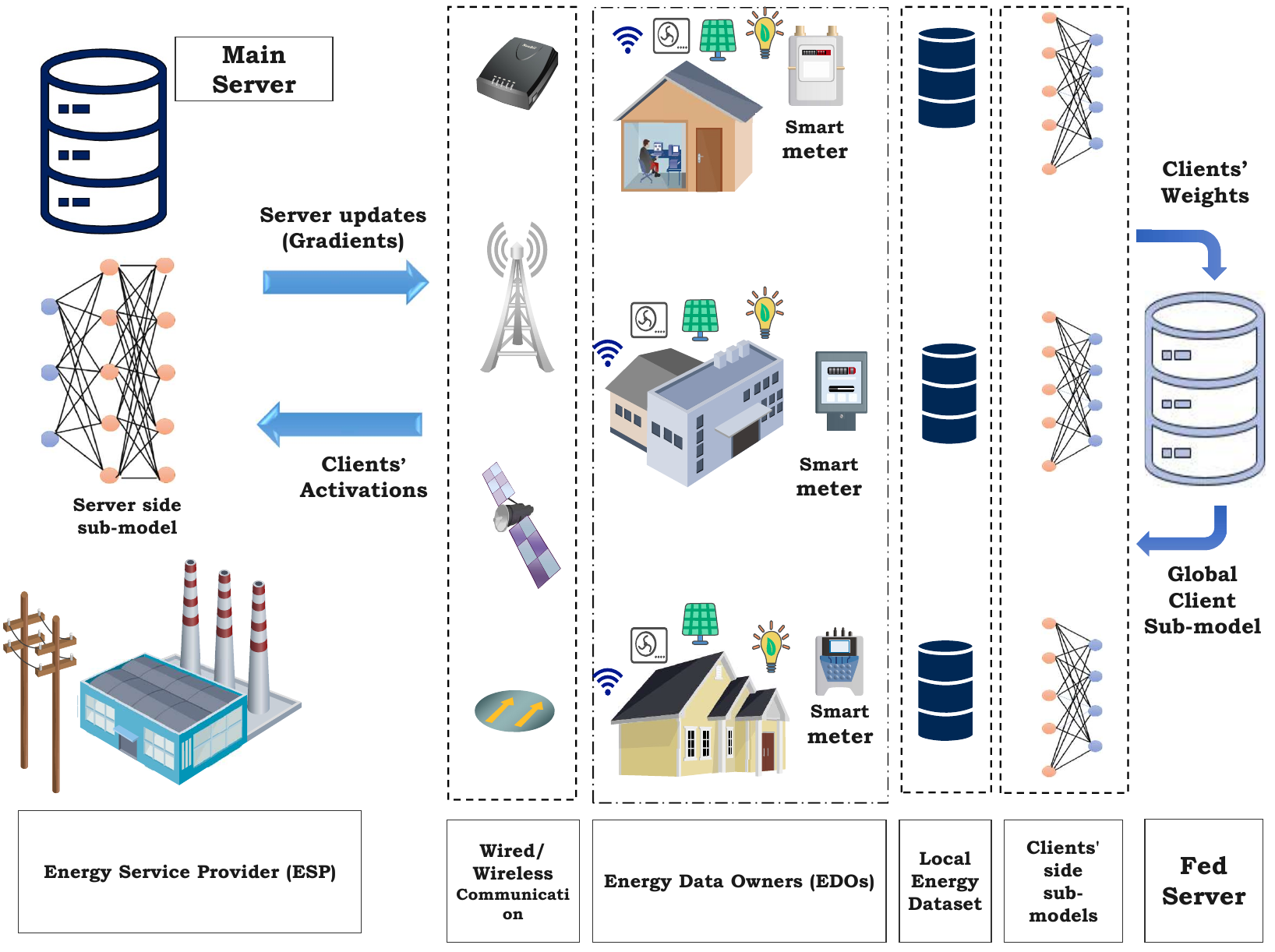}
  \caption{Split Federated Learning in Smart Grid.}  
   \label{fig:SFL-smart grid}
\end{figure*}

For better and faster grid management, energy data collected from smart components should be analyzed quickly and securely. However, it is difficult to address these challenges with centralized learning schemes, which are the most commonly encountered in literature. In effect, broadcasting all the grid information towards a central location (e.g., cloud) extends the transmission time and augments security and privacy threats since customer load profiles reveal a lot of sensitive data (e.g., daily routine, time spent home). Beyond that, the abuse of this information could involve social issues, such as increased burglary threats when residences are unoccupied. Applying distributed learning algorithms in power and energy domains permits not just understanding grid activities but protecting system data, too. Many interesting works have used the federated learning paradigm to tackle transmission delay and data privacy concerns. For instance, in~\cite{wang2021electricity}, the authors proffer a federated framework for electricity consumer characteristics identification where smart meter data are kept locally within retailers, and only local weights are sent to a computational center. Similarly, in~\cite{liu2021federated} and ~\cite{wen2021feddetect}, the authors expose collaborative FL architectures for learning power consumption patterns and energy theft detection, respectively. In~\cite{taik2020electrical}, a new approach for electrical load prediction based on edge computing and FL using residents’ behavior data is proposed. Certainly, all smart-grid FL-based studies ensure data privacy. Nevertheless, all neglect the model privacy aspect. SFL provides a direct answer to this challenge. For instance, to develop a SFL-based approach for faults location and detection in a smart-grid system, the model would not be sent over the network but partitioned into client and server sides. All the grid clients, also called energy data owners (EDOs), e.g., substations, train in parallel their sub-models using their local data gathered from sensors and smart meters installed in the smart grid. Then, the obtained parameters are exchanged with the main server that resumes the training, aggregates the EDOs shared models, and forwards the outcome back to EDOs. After some rounds, the EDOs upload their local models to the Fed server for aggregation. The process is repeated until a desired accuracy is achieved. As a result, less processing power from the grid network nodes is required, while the split and parallelism features of SFL help to protect the model against inference attacks and supply service providers (SPs) with energy-related knowledge in a brief delay. \figurename~\ref{fig:SFL-smart grid} illustrates how smart-grid systems could benefit from the SFL paradigm.

Table~\ref{tab:sum-use-cases} analyses a selected group of studies related to 6G use cases. In addition, it shows how SFL can be applied and the advantages that this new algorithm presents.

% \usepackage{multirow}
% \usepackage{graphicx}
\begin{table*}[]
\caption{SFL Contribution on a Selected Group of Studies Related to 6G Use Cases.}
\label{tab:sum-use-cases}
\centering
\resizebox{\textwidth}{!}{%
\begin{tabular}{|c|l|c|c|l|l|l|cccl|}
\hline
\cellcolor[HTML]{C0C0C0} &
  \multicolumn{1}{c|}{\cellcolor[HTML]{C0C0C0}} &
  \cellcolor[HTML]{C0C0C0} &
  \cellcolor[HTML]{C0C0C0} &
  \multicolumn{1}{c|}{\cellcolor[HTML]{C0C0C0}} &
  \cellcolor[HTML]{C0C0C0} &
  \cellcolor[HTML]{C0C0C0} &
  \multicolumn{4}{c|}{\cellcolor[HTML]{C0C0C0}\textbf{SFL Contribution}} \\ \cline{8-11} 
\cellcolor[HTML]{C0C0C0} &
  \multicolumn{1}{c|}{\cellcolor[HTML]{C0C0C0}} &
  \cellcolor[HTML]{C0C0C0} &
  \cellcolor[HTML]{C0C0C0} &
  \multicolumn{1}{c|}{\cellcolor[HTML]{C0C0C0}} &
  \cellcolor[HTML]{C0C0C0} &
  \cellcolor[HTML]{C0C0C0} &
  \multicolumn{3}{c|}{\cellcolor[HTML]{C0C0C0}\textbf{SFL Agents}} &
  \multicolumn{1}{c|}{\cellcolor[HTML]{C0C0C0}} \\ \cline{8-10}
\multirow{-3}{*}{\cellcolor[HTML]{C0C0C0}\textbf{Use Case}} &
  \multicolumn{1}{c|}{\multirow{-3}{*}{\cellcolor[HTML]{C0C0C0}\textbf{Ref}}} &
  \multirow{-3}{*}{\cellcolor[HTML]{C0C0C0}\textbf{\begin{tabular}[c]{@{}c@{}}Focus\\ Point\end{tabular}}} &
  \multirow{-3}{*}{\cellcolor[HTML]{C0C0C0}\textbf{\begin{tabular}[c]{@{}c@{}}DL \\ Technique\end{tabular}}} &
  \multicolumn{1}{c|}{\multirow{-3}{*}{\cellcolor[HTML]{C0C0C0}\textbf{\begin{tabular}[c]{@{}c@{}}Evaluated\\ Metrics\end{tabular}}}} &
  \multirow{-3}{*}{\cellcolor[HTML]{C0C0C0}\textbf{\begin{tabular}[c]{@{}l@{}}Data\\ Privacy\end{tabular}}} &
  \multirow{-3}{*}{\cellcolor[HTML]{C0C0C0}\textbf{\begin{tabular}[c]{@{}l@{}}Model\\ Privacy\end{tabular}}} &
  \multicolumn{1}{c|}{\textbf{Clients}} &
  \multicolumn{1}{c|}{\textbf{\begin{tabular}[c]{@{}c@{}}Fed\\ Server\end{tabular}}} &
  \multicolumn{1}{c|}{\textbf{\begin{tabular}[c]{@{}c@{}}Main\\ Server\end{tabular}}} &
  \multicolumn{1}{c|}{\multirow{-2}{*}{\cellcolor[HTML]{C0C0C0}\textbf{SFL Benefits}}} \\ \hline
 &
  ~\cite{sun2020adaptive} &
  \begin{tabular}[c]{@{}c@{}}Digital Twin for\\ Industrial IoTs\end{tabular} &
  \begin{tabular}[c]{@{}c@{}}FL\\ (DQN)\end{tabular} &
  \begin{tabular}[c]{@{}l@{}}- Energy Consumption\\ - Accuracy, Loss\\ - Time Consumed by FL\\ - Total Number of Aggregation\end{tabular} &
  \cmark &
  \xmark &
  \multicolumn{1}{c|}{\begin{tabular}[c]{@{}c@{}}Industrial IoT\\  Devices\\ (Sensors, Machinery, \\ Industrial Robots, etc)\end{tabular}} &
  \multicolumn{1}{c|}{} &
  \multicolumn{1}{c|}{} &
  \begin{tabular}[c]{@{}l@{}}- Accelerate the convergence time\\ - Lightweight models for IIoT devices\\ - Enhance the energy consumed during\\   FL training under different channel state\end{tabular} \\ \cline{2-8} \cline{11-11} 
 &
  ~\cite{imteaj2020fedar} &
  \begin{tabular}[c]{@{}c@{}}Learning Model for \\ Distributed Mobile Robots\end{tabular} &
  FL &
  \begin{tabular}[c]{@{}l@{}}- FL Accuracy\\ - Trust Score\end{tabular} &
  \cmark &
  \xmark &
  \multicolumn{1}{c|}{\begin{tabular}[c]{@{}c@{}}Mobile\\ Robots\end{tabular}} &
  \multicolumn{1}{c|}{} &
  \multicolumn{1}{c|}{} &
  \begin{tabular}[c]{@{}l@{}}- Assure minimum Mobile Robots \\ Resources than FL tasks.\end{tabular} \\ \cline{2-8} \cline{11-11} 
 &
  ~\cite{pang2021collaborative} &
  \begin{tabular}[c]{@{}c@{}}Collaborative City\\  Digital Twin\end{tabular} &
  \begin{tabular}[c]{@{}c@{}}FL\\ (TCN)\end{tabular} &
  \begin{tabular}[c]{@{}l@{}}- MAE\\ - MAPE\end{tabular} &
  \cmark &
  \xmark &
  \multicolumn{1}{c|}{City DT} &
  \multicolumn{1}{c|}{} &
  \multicolumn{1}{c|}{} &
  \begin{tabular}[c]{@{}l@{}}- Lower FL Model Security Risks\\ - Optimized Learning Process\\ - Preserve the local data of each city DT\end{tabular} \\ \cline{2-8} \cline{11-11} 
\multirow{-4}{*}{\begin{tabular}[c]{@{}c@{}}Industry 5.0,\\  Digital Twin, \\ Autonomous\\  Robots\end{tabular}} &
  ~\cite{heo2019collision} &
  \begin{tabular}[c]{@{}c@{}}CollisionNet:\\ Robot Collision Detector\end{tabular} &
  DNN &
  \begin{tabular}[c]{@{}l@{}}- FP, FN\\ - Detection Delay\\ - Detection Accuracy\end{tabular} &
  \xmark &
  \xmark &
  \multicolumn{1}{c|}{\begin{tabular}[c]{@{}c@{}}Industrial \\ Robots\end{tabular}} &
  \multicolumn{1}{c|}{\multirow{-4}{*}{\begin{tabular}[c]{@{}c@{}}Edge\\ Server\end{tabular}}} &
  \multicolumn{1}{c|}{\multirow{-4}{*}{\begin{tabular}[c]{@{}c@{}}Cloud/Edge\\ Aggregator\\ Server\end{tabular}}} &
  - Enhance the Detection Latency \\ \hline
 &
  ~\cite{hecker2018failure} &
  \begin{tabular}[c]{@{}c@{}}Failure Prediction\\ For AC\end{tabular} &
  RCNet &
  \begin{tabular}[c]{@{}l@{}}- Percentage of Failures\\ - Drivability Score\end{tabular} &
  \xmark &
  \xmark &
  \multicolumn{1}{c|}{\begin{tabular}[c]{@{}c@{}}Autonomous \\ Cars\end{tabular}} &
  \multicolumn{1}{c|}{\begin{tabular}[c]{@{}c@{}}MEC \\ Server\end{tabular}} &
  \multicolumn{1}{c|}{\begin{tabular}[c]{@{}c@{}}MEC \\ Server\end{tabular}} &
  \begin{tabular}[c]{@{}l@{}}- Lower the Latency and Augment\\ the Privacy of the Failures Occurrence\\ Model Prediction\end{tabular} \\ \cline{2-11} 
 &
  ~\cite{yu2020mobility} &
  \begin{tabular}[c]{@{}c@{}}Proactive Edge Caching\\  for Connected Vehicles\end{tabular} &
  \begin{tabular}[c]{@{}c@{}}FL\\ (AE)\end{tabular} &
  - Cache Hit Ratio &
  \cmark &
  \xmark &
  \multicolumn{1}{c|}{\begin{tabular}[c]{@{}c@{}}Connected \\ Cars\end{tabular}} &
  \multicolumn{1}{c|}{RSU} &
  \multicolumn{1}{c|}{\begin{tabular}[c]{@{}c@{}}Edge Server\\ (BS or RSU)\end{tabular}} &
  \begin{tabular}[c]{@{}l@{}}- Enhance the privacy of the content \\ popularity prediction model\\ - Reducing Vehicle's Energy Consumption\end{tabular} \\ \cline{2-11} 
 &
  ~\cite{barbieri2022decentralized} &
  \begin{tabular}[c]{@{}c@{}}Extended sensing in 6G \\ connected vehicles\end{tabular} &
  \begin{tabular}[c]{@{}c@{}}Decentralized\\ FL\end{tabular} &
  - Accuracy, Loss &
  \cmark &
  \xmark &
  \multicolumn{1}{c|}{\begin{tabular}[c]{@{}c@{}}Connected\\ Vehicles\end{tabular}} &
  \multicolumn{1}{c|}{\begin{tabular}[c]{@{}c@{}}Base\\  Station\end{tabular}} &
  \multicolumn{1}{c|}{\begin{tabular}[c]{@{}c@{}}Edge \\ Server\end{tabular}} &
  - Optimize the Vehicles' Energy \\ \cline{2-11} 
\multirow{-4}{*}{\begin{tabular}[c]{@{}c@{}}Connected and\\  Autonomous \\ Vehicle\end{tabular}} &
  ~\cite{lu2019collaborative} &
  \begin{tabular}[c]{@{}c@{}}Collaborative Learning \\ Framework\\ for CAVs\end{tabular} &
  FL &
  \begin{tabular}[c]{@{}l@{}}- Precision, Recall\\ - Accuracy, F-measure\end{tabular} &
  \cmark &
  \xmark &
  \multicolumn{1}{c|}{\begin{tabular}[c]{@{}c@{}}Electric\\ Vehicles\end{tabular}} &
  \multicolumn{1}{c|}{\begin{tabular}[c]{@{}c@{}}Base\\  Station\end{tabular}} &
  \multicolumn{1}{c|}{\begin{tabular}[c]{@{}c@{}}Edge \\ Server\end{tabular}} &
  \begin{tabular}[c]{@{}l@{}}- Enhance the EV Battery\\ - Latency Reduction\\ - Model Privacy Protection\end{tabular} \\ \hline
 &
  ~\cite{fan2022covid} &
  \begin{tabular}[c]{@{}c@{}}COVID-19\\  Detection\end{tabular} &
  \begin{tabular}[c]{@{}c@{}}CNN\\ $+$\\ Transformer\end{tabular} &
  \begin{tabular}[c]{@{}l@{}}- Accuracy, Loss\\ - Specificity, sensitivity\\ - Precision, F1-Score\\ - Confusion Matrices\end{tabular} &
  \xmark &
  \xmark &
  \multicolumn{1}{c|}{\begin{tabular}[c]{@{}c@{}}Healthcare \\ Devices\end{tabular}} &
  \multicolumn{1}{c|}{} &
  \multicolumn{1}{c|}{} &
  \begin{tabular}[c]{@{}l@{}}- Guarantee both Patients' Data  \\ and COVID-19 Model Privacy\\ - Minimize Training Time\end{tabular} \\ \cline{2-8} \cline{11-11} 
 &
  ~\cite{chen2020fedhealth} &
  \begin{tabular}[c]{@{}c@{}}FedHealth Framework \\ for Wearable Healthcare\end{tabular} &
  FL &
  \begin{tabular}[c]{@{}l@{}}- Accuracy\\ - Precision\\ - Recall\\ - F1-score\end{tabular} &
  \cmark &
  \xmark &
  \multicolumn{1}{c|}{\begin{tabular}[c]{@{}c@{}}Wearable\\ Devices \\ (e.g; Smart Watches,\\ Glasses)\end{tabular}} &
  \multicolumn{1}{c|}{} &
  \multicolumn{1}{c|}{} &
  \begin{tabular}[c]{@{}l@{}}- Optimize Nodes Power Consumption\\ - Speed up the training\end{tabular} \\ \cline{2-8} \cline{11-11} 
 &
  ~\cite{li2019privacy} &
  \begin{tabular}[c]{@{}c@{}}Brain Tumor \\ Segmentation\end{tabular} &
  \begin{tabular}[c]{@{}c@{}}FL\\ (DNN)\end{tabular} &
  - Dice Score &
  \cmark &
  \xmark &
  \multicolumn{1}{c|}{\begin{tabular}[c]{@{}c@{}}Medical Sites\\ (Hospitals)\end{tabular}} &
  \multicolumn{1}{c|}{} &
  \multicolumn{1}{c|}{} &
  \begin{tabular}[c]{@{}l@{}}- Enhance the classification accuracy\\ - Enhance the Dice Score\\ - Preserve Model Privacy\end{tabular} \\ \cline{2-8} \cline{11-11} 
\multirow{-4}{*}{\begin{tabular}[c]{@{}c@{}}Intelligent eHealth\\  and Body Area\\  Networks\end{tabular}} &
  ~\cite{brisimi2018federated} &
  \begin{tabular}[c]{@{}c@{}}Hospitalization\\ Prediction\end{tabular} &
  \begin{tabular}[c]{@{}c@{}}FL\\ (Sparse SVM)\end{tabular} &
  \begin{tabular}[c]{@{}l@{}}- Communication Cost\\ - Computation Time\\ - AUC\end{tabular} &
  \cmark &
  \xmark &
  \multicolumn{1}{c|}{\begin{tabular}[c]{@{}c@{}}- Hospitals\\ - Patients\end{tabular}} &
  \multicolumn{1}{c|}{\multirow{-4}{*}{\begin{tabular}[c]{@{}c@{}}Edge \\ Server\end{tabular}}} &
  \multicolumn{1}{c|}{\multirow{-4}{*}{\begin{tabular}[c]{@{}c@{}}Cloud/Edge \\ Server\end{tabular}}} &
  \begin{tabular}[c]{@{}l@{}}- Enhance Cost Computation Time\\ - Enhance the Prediction Accuracy\end{tabular} \\ \hline
 &
  ~\cite{feng2019exploring} &
  \begin{tabular}[c]{@{}c@{}}User's Viewport \\ Prediction\end{tabular} &
  CNN &
  \begin{tabular}[c]{@{}l@{}}- Accuracy\\ - Timing Overhead\\ - CDF\end{tabular} &
  \xmark &
  \xmark &
  \multicolumn{1}{c|}{} &
  \multicolumn{1}{c|}{} &
  \multicolumn{1}{c|}{} &
  \begin{tabular}[c]{@{}l@{}}- Minimize The Bandwidth Utilization\\ - Enhance Timing Overhead\end{tabular} \\ \cline{2-7} \cline{11-11} 
 &
  ~\cite{kanade2021convolutional} &
  \begin{tabular}[c]{@{}c@{}}Eye-gaze\\ Tracking\end{tabular} &
  CNN &
  - n/a &
  \xmark &
  \xmark &
  \multicolumn{1}{c|}{} &
  \multicolumn{1}{c|}{} &
  \multicolumn{1}{c|}{} &
  - Enhance the Convergence Performance \\ \cline{2-7} \cline{11-11} 
\multirow{-3}{*}{\begin{tabular}[c]{@{}c@{}}XR Applications,\\  Holographic \\ Telepresence\end{tabular}} &
  ~\cite{8633843} &
  \begin{tabular}[c]{@{}c@{}}User's Head-Motion\\  Prediction\end{tabular} &
  GRU-CNN &
  \begin{tabular}[c]{@{}l@{}}- MAE\\ - RMSE\end{tabular} &
  \xmark &
  \xmark &
  \multicolumn{1}{c|}{\multirow{-3}{*}{\begin{tabular}[c]{@{}c@{}}Head-Mounted \\ Display (HDM)\end{tabular}}} &
  \multicolumn{1}{c|}{\multirow{-3}{*}{\begin{tabular}[c]{@{}c@{}}Edge \\ Server\end{tabular}}} &
  \multicolumn{1}{c|}{\multirow{-3}{*}{\begin{tabular}[c]{@{}c@{}}Cloud/ MEC\\ Server\end{tabular}}} &
  \begin{tabular}[c]{@{}l@{}}- Parallel and Decentralized Head Motion\\ Predictor\end{tabular} \\ \hline
\multicolumn{1}{|l|}{} &
  ~\cite{khan2019electricity} &
  Load Forecasting &
  DCNN &
  \begin{tabular}[c]{@{}l@{}}- Electricity Load\\ - MAPE, MAE\\ - RMSE\end{tabular} &
  \xmark &
  \xmark &
  \multicolumn{1}{c|}{\begin{tabular}[c]{@{}c@{}}- Energy \\ Consumers\end{tabular}} &
  \multicolumn{1}{c|}{} &
  \multicolumn{1}{c|}{} &
  \begin{tabular}[c]{@{}l@{}}- Ameliorate the Forecast Process\\ (Decentralized and Distributed Forecasting)\\ - Preserve the Privacy of Data's Consumers\\ as well as Load Energy Prediction Model\end{tabular} \\ \cline{2-8} \cline{11-11} 
\multicolumn{1}{|l|}{} &
  ~\cite{wang2021electricity} &
  \begin{tabular}[c]{@{}c@{}}Identification of Electricity\\  Consumer Characteristics\end{tabular} &
  \begin{tabular}[c]{@{}c@{}}FL\\ (ANN)\end{tabular} &
  \begin{tabular}[c]{@{}l@{}}- Accuracy\\ - Matthews Correlation \\      Coefficient (MCC)\end{tabular} &
  \cmark &
  \xmark &
  \multicolumn{1}{c|}{\begin{tabular}[c]{@{}c@{}}- Smart \\ Meter\end{tabular}} &
  \multicolumn{1}{c|}{} &
  \multicolumn{1}{c|}{} &
  \begin{tabular}[c]{@{}l@{}}- Preserve Consumers' Data\\ - Reduce the Identification Model \\ Security Risks\end{tabular} \\ \cline{2-8} \cline{11-11} 
\multicolumn{1}{|l|}{} &
  ~\cite{liu2021federated} &
  \begin{tabular}[c]{@{}c@{}}Power Consumption \\ Patterns\end{tabular} &
  FL &
  \begin{tabular}[c]{@{}l@{}}- Mean Squared Error (MSE)\end{tabular} &
  \cmark &
  \xmark &
  \multicolumn{1}{c|}{\begin{tabular}[c]{@{}c@{}}- Transformer \\ Stations\end{tabular}} &
  \multicolumn{1}{c|}{} &
  \multicolumn{1}{c|}{} &
  \begin{tabular}[c]{@{}l@{}}- Enhance Prediction Performance\\ (Training Time, Low MSE)\\ - Minimize Model Leakage\end{tabular} \\ \cline{2-8} \cline{11-11} 
\multicolumn{1}{|l|}{\multirow{-4}{*}{Smart Grid 2.0}} &
  ~\cite{wen2021feddetect} &
  \begin{tabular}[c]{@{}c@{}}FedDetect:\\ Energy Theft Detection\end{tabular} &
  \begin{tabular}[c]{@{}c@{}}FL\\ (TCN)\end{tabular} &
  \begin{tabular}[c]{@{}l@{}}- Accuracy\\ - Training Time\end{tabular} &
  \cmark &
  \xmark &
  \multicolumn{1}{c|}{\begin{tabular}[c]{@{}c@{}}Consumer Device or\\ Distributed Detection \\ Stations (DTSs)\end{tabular}} &
  \multicolumn{1}{c|}{\multirow{-4}{*}{\begin{tabular}[c]{@{}c@{}}Edge \\ Server\end{tabular}}} &
  \multicolumn{1}{c|}{\multirow{-4}{*}{\begin{tabular}[c]{@{}c@{}}Cloud/ MEC\\ Server\end{tabular}}} &
  \begin{tabular}[c]{@{}l@{}}- Optimize Training Duration\\ - Reduce Strain on the Consumer Device\\     (Train only a section of the FedDetect)\end{tabular} \\ \hline
\end{tabular}%
}
\end{table*}

\section{Datasets and Frameworks for Successful Implementation of SFL-driven 6G Networks} 
\label{sec:Datasets and implementation}
This section describes various tools that can support the development, evaluation, and validation of SFL-based solutions for 6G networks. We first list multiple existing datasets for different 6G technical aspects and use-cases. Then, we provide multiple existing frameworks related to each 6G aspect/use-case.

\subsection{Existing Datasets for 6G Networks} 

ML/DL model outcomes are greatly dependent on data quality (type, size, source, format, etc). As low-quality data leads to poor results, collecting and preparing appropriate data is the first hurdle in the AI domain. Actually, there is a lack of datasets related to the future 6G since it is still under development, despite some initiatives (e.g, DeepSense 6G). This scarcity is among the major reasons, that push researchers to adopt datasets from other relevant network technologies and configurations (e.g, 5G-specific datasets). We discuss in this section some publicly available datasets that could be used for solving both uses-cases and technical-related issues. In this context, we distinguish two types of datasets: (a) Technical datasets, where the data are used to enhance core technical aspects of 6G networks, (b) application datasets that can support 6G-enabled split federated learning applications like smart-grid, healthcare and smart-agriculture.

\subsubsection{Datasets for 6G technical aspects}

\begin{itemize}
    \item DeepSense 6G~\cite{deepsense6g}: In order to promote ML/DL research in cellular communication, a public, large-scale and real-world dataset has been recently published in~\cite{alkhateeb2022deepsense}. It contains data about multiple dynamic scenarios (34 scenarios as stated on the official website of the dataset). Each scenario emulates a use case (indoor communication, vehicle-to-infrastructure communication, millimeter wave drone communication, night/day time, rainy weather, and others). A scenario comprises a set of units (camera, 2D/3D LiDAR, radar, GPS, smart car, stationary base station, etc.), and collects a set of modalities (e.g., RGB images, position, beam power). These measurements could be used for many SFL-enabled applications such as beam prediction, user identification, positioning, object detection/classification and more.
    
    \item 5GMdata~\cite{Klautau18}: It is a public and simulated telecommunication dataset, that has been generated using traffic and ray-tracing simulators (SUMO/Remcom Wireless InSite). The simulations with SUMO and InSite represent the first phase before organizing the raw data into a 5GMdata database, based on the target application. Next, data post-processing is performed by converting the 5GMdata into a format utilizable by the ML/DL algorithm. Finally, the ML/DL experiment using the associated data is run. This dataset could be personalized to fit 6G cellular networks and develop new split federated learning applications such as channel estimation and beam selection.
    
    \item DeepMIMO~\cite{deepmimo}: It is a public and generated dataset for mmWave/massive MIMO channels introduced in~\cite{alkhateeb2019deepmimo}. It includes seven scenarios (according to the official dataset's website). For example, the first scenario mimics an outdoor environment of two streets, one intersection of $18$ base stations (the height of each BS is 6 $m$) and more than a million users distributed on three uniform grids. The dataset provides three versions (DeepMIMO v1, DeepMIMO v2, DeepMIMO 5G NR) and many applications in the mmWave area, including Intelligent Reflecting Surfaces (IRS), channel estimation, and blockage prediction, to mention just a few.
    
    \item Telecom Italia~\cite{barlacchi2015multi}: It is a rich, open and multi-source dataset largely used by academic researchers. It aggregates telecommunications activities, namely: SMSs, calls, and Internet usage data in the city of Milan and the Province of Trentino. The data are recorded every 10 minutes for two months (from 1/11/2013 to 1/1/2014). The dataset could be adopted for user traffic prediction that plays a major role in designing smart resource management solutions.~\cite{nan2022regional}~\cite{bolla2022ai} are recent works that have made use of Telecom Italia dataset for 6G networks.
    
    \item Cellular Traffic Analysis Data~\cite{azari2019cellular}: It is a real-world and labeled cellular dataset available at Github~\cite{CTAD2019}. The traffic has been captured on several android devices using virtual private network tunneling. The dataset could be utilized to predict user traffic, and therefore, ensure an adequate traffic-aware resource management.

    \item 5G dataset~\cite{5Gdataset}: This is a public dataset described in~\cite{raca2020beyond}. It is generated when a user is running two services: downloading files and streaming videos (Netflix and Amazon Prime applications) from two mobility patterns, separately, namely: driving and static. Various channel, context and cell-related details were measured, such as downlink/uplink rates, mobile device speed, GPS coordinates, SNR (Signal to Noise Ratio), RSPR (Reference Signal Received Power), RSPQ (Reference Signal Received Quality), etc. Based on its features, the dataset could be used to automate various network functions for 5G and beyond, such as intelligent predictive handover, intelligent resource management, and bandwidth prediction.

    \item UNSW-NB15~\cite{moustafa2015unsw}~\cite{unsw-nb15}: This is one of the most popular datasets used in ML/DL-aided network security. It was released in 2015. It has 44 features and about 2,540,044 data records distributed between normal and attack traffic. UNSW-NB15 uses contemporary methods to more reflect the real network traffic, and contains modern attacks divided into 9 types, namely Fuzzers, Backdoors, Analysis (e.g., port scanning), DoS, Exploits, Generic, Shellcode, Reconnaissance, and Worms.
    
    \item AWID (Aegean WiFi Intrusion Dataset)~\cite{9360747}: It implements real Wi-Fi network traces of both legitimate and illegitimate IEEE 802.11 WLAN traffic. Each record in the dataset comprises 155 attributes between numeric and nominal values. The last version AWID3 focuses on IEEE 802.11w, Wi-Fi5 and WPA2 enterprise attacks. The dataset is also considered for other wireless communication technologies such as IoT and 5G~\cite{rezvy2019efficient}.
    
    \item CIC-IDS2017~\cite{sharafaldin2018toward}: It is an intrusion detection labeled dataset published by the Canadian Institute for Cybersecurity in 2017. It contains benign and the most up-to-date common attacks such as DDoS, Brute Force, XSS, SQL Injection, Infiltration, Port Scan, and Botnet. The dataset contains 2,830,743 records split into 8 files with 78 features for each record.
    
    \item 5G-NIDD (5G - non-IP data delivery)~\cite{samarakoon20225g}: This is a fully labeled dataset generated from a functional 5G test network, that can be used to develop and test AI/ML solutions for identification, and detection of malicious content in network traffic.

    \item Microservices configurations dataset~\cite{mohamed_mekki_2022_6907619}: In order to verify if a cloud's tenant configuration (in terms of memory and CPU) is appropriate to its service requirements, authors in~\cite{mekki2022microservices} conducted an experiment based on the execution of three concurrent applications under diverse resource configurations, namely: Web servers, RabbitMQ broker and the OpenAirInterface 5G Core network AMF (Access and Mobility Management Function). The experimental results led to the generation of three different datasets (one for each deployed application). For instance, the webserver dataset has been produced from a parallel and increasing number of requests (between 100 and 1000) sent to each web server instance. The experiment left 16 features in the web server dataset, 15 features in the 5G AMF dataset, and 12 features in the RabbitMQ dataset. The features include the timestamp of metrics' collection, the memory and CPU allocated to the container, the memory and CPU used by the container, and other application-related features. The dataset could be adopted to train split federated learning models that manage automatically the configuration of services' resources for an optimal execution and efficient computing resources usage.

    \end{itemize}

\subsubsection{Datasets for 6G Use-Cases} 
This section describes datasets related with popular 6G use-cases that could be leveraged for SFL methods.

\begin{itemize}

    \item PlantVillage~\cite{hughes2015open}: It is a public plant disease dataset used in~\cite{angin2020agrilora} to develop a Digital Twin Framework for Smart Agriculture, which represents one of the main applications of 6G networks in industry 5.0. It contains 54305 leaf images from 14 crops, between healthy and diseased, divided into 38 classes. PlantVillage is the most cited among available plant disease datasets. The authors removed the leaves from the plants and photographed them with a single digital camera. PlantVillage is a real dataset that could be used in implementing SFL-based smart farming solutions. The idea is to partition the data across the different simulated collaborators (drones, cameras, etc.), potentially experimenting with different data distributions (uniform/non-uniform) and different forms of data partitioning (e.g., vertical vs. horizontal).

    \item Berkeley Deep Drive-X (eXplanation)~\cite{kim2018textual}: It is a real, public and large-scale dataset, that contains 77 hours of driving in 6,970 videos shot under various driving conditions (day/night, highway/urban/rural area, rainy/sunny weather, etc). Each video is approximately 40 seconds long and comprises 3-4 actions (accelerating, slowing down, stopping, turning left, moving into the right lane, and so on). All actions are annotated with a description and explanation.
    
    \item Dataset of Annotated Car Trajectories (DACT)~\cite{moosavi2017annotation}: It is a set of driving trajectories captured in Columbus, Ohio, where each trajectory registers over 10 minutes that can be divided into multiple segments annotated by the operating pattern (e.g., speed-up and slow-down). Furthermore, each trajectory is an ordered set of tuples and each tuple consists of 11 attributes, such as: Trip ID, vehicle's speed in mph (miles per hour), vehicle's acceleration, latitude, longitude, type of segment (exit, loop, turn, etc).
    
    \item MIMIC-III~\cite{johnson2016mimic}: This is the acronym for Medical Information Mart for Intensive Care III. It is a big, anonymized and freely accessible medical database. It covers data of forty thousand patients admitted in intensive care units in Boston, USA, hospitals between 2001 and 2012. It provides an important benchmark for evaluating health-related models based on split federated learning.

    \item COVID-19 image data collection~\cite{COVID19}: According to~\cite{cohen2020covid}, it is the largest public dataset for the diagnosis of coronavirus disease. In addition to the chest X-ray (CXRs) images, the dataset includes a list of metadata such as patient ID, sex, age, temperature, time since first symptoms, intensive care unit (ICU) status, incubation status, hospital location, etc. It is a suitable resource for building and evaluating several split federated learning based applications, for example automatic detection of COVID-19 cases, patient's severity and the need for mechanical ventilation prediction.
    
    \item VR streaming~\cite{wu2017dataset}: It is a publicly-available dataset that comprises head tracking of $48$ users fairly divided between males and females. The data have been recorded while users were watching $18$  spherical videos from $5$ categories. Taking advantage of its features (users' way of watching, their head movements, and directions they focus on), the data can serve as a powerful source to enhance user experience in VR applications by building SFL-based patterns for gazing prediction and user identification.
    
    \item Irish CER~\cite{IrishCER}: The dataset is provided by the Irish Commission for Energy Regulation (CER). It contains customers' electric load profiles from 6435 smart meters conducted on 536 days and half-hourly basis. Besides, through a questionnaire filled in by the experiment's contributors, the dataset is enriched by many variables on occupant socio-demographic factors, their consumption behavior, domestic properties, and home appliances. By segmenting the whole dataset into different parts, it could be used to build several SFL-based models to understand the electricity customer conduct, for instance predicting the future load in one or multiple nodes of a smart-grid network, forecasting the electricity demand, and detecting faults and attacks.
    
    \item RAE (Rainforest Automation Energy)~\cite{DVN/ZJW4LC_2017}: The dataset includes 1\,Hz energy readings (mains and sub-meters) from two residential dwellings. In addition to power data, environmental and sensor data from the house's thermostat are included. As well, pertinent sub-meter data for power utilities (heat pump and rental suite) are captured and incorporated. The dataset recordings could be adopted for various applications including, but not restricted to, energy saving, abnormal detection, occupancy pattern and energy demand prediction.

\end{itemize}

Table~\ref{tab:summary of datasets} presents a comparative view between the above existing datasets according to their properties (public or private), label class (labeled or unlabeled), data distribution (IID or Non-IID), generation procedure (real vs. simulated) and applicable area (core 6G technical aspects or use case-specific ones). In the last column, we provide some potential 6G applications for which the concerned dataset could be used to develop SFL models. Note that datasets are arranged in descending chronological order according to the publication year.

% \usepackage{booktabs}
\begin{table*}[]
\caption{Benchmark Datasets.}
\label{tab:summary of datasets}
\centering
\begin{tabular}{@{}cp{0.5cm}p{1cm}p{1cm}p{1cm}ccl@{}}
\toprule
\rowcolor[HTML]{C0C0C0} %FFFFFF
\textbf{\rotatebox{90}{Dataset}} &
  \multicolumn{1}{c}{\cellcolor[HTML]{C0C0C0}\textbf{\rotatebox{90}{Year}}} &
  \textbf{\rotatebox{90}{Public / Private}} &
  \textbf{\rotatebox{90}{Labeled / Unlabeled}} &
  \textbf{\rotatebox{90}{IID / Non- IID}} &
  \textbf{\rotatebox{90}{Real / Simulated}} &
  \textbf{\rotatebox{90}{Technical / Use case Dataset}} &
  \multicolumn{1}{c}{\cellcolor[HTML]{C0C0C0}\textbf{Potential 6G-enabled SFL Application}} \\ \midrule
DeepSense 6G &
  \multicolumn{1}{c}{2022} &
  Public &
  Labeled &
  Non-IID &
  Real &
  \begin{tabular}[c]{@{}c@{}}Technical \end{tabular} &
  \begin{tabular}[c]{@{}l@{}}- Beam Prediction\\ - Resource Management\\ - Interference Management\\ - User Scheduling\end{tabular} \\ \midrule
DeepMIMO &
  2022 &
  Public &
  Unlabeled &
  Non-IID &
  Simulated &
  Technical &
  \begin{tabular}[c]{@{}l@{}}- Intelligent Reflecting Surfaces (IRS)\\ - Channel Estimation\\ - Blockage Prediction\end{tabular} \\ \midrule
5G-NIDD &
  2022 &
  Public &
  Labeled &
  Non-IID &
  Real &
  Technical &
  - Cellular Network Intrusion Detection \\ \midrule
\multicolumn{1}{c}{\begin{tabular}[c]{@{}l@{}}Microservices\\ Configurations\\                 Data\end{tabular}} &
  2022 &
  Public &
  Unlabeled &
  Non-IID &
  Simulated &
  Technical &
  - Applications' Resources Configuration \\ \midrule
AWID3 &
  2021 &
  Public &
  Labeled &
  Non-IID &
  Real &
  Technical &
  - Security Issues (Network Intrusion Detection) \\ \midrule
COVID-19 image data &
  2020 &
  Public &
  Labeled &
  Non-IID &
  Real &
  Use case &
  \begin{tabular}[c]{@{}l@{}}- Coronavirus Automatic Detection\\ - Patient's Severity Prediction\\ - Mechanical Ventilation Need Prediction\end{tabular} \\ \midrule
5G dataset &
  2020 &
  Public &
  Unlabeled &
  Non-IID &
  Real &
  Technical &
  \begin{tabular}[c]{@{}l@{}}- Intelligent predictive handover\\ - Intelligent resource management\\ - Bandwidth prediction.\end{tabular} \\ \midrule
Cellular Traffic Analysis &
  2019 &
  Public &
  Labeled &
  Non-IID &
  Real &
  Technical &
  \begin{tabular}[c]{@{}l@{}}- Cellular Load Traffic Forecasting\\ - Resource Management\end{tabular} \\ \midrule
Berkeley Deep Drive-X &
  2018 &
  Public &
  Labeled &
  Non-IID &
  Real &
  Use case &
  \begin{tabular}[c]{@{}l@{}}- Explainable Model for Autonomous cars\\ - Driving Behavior Explanation\end{tabular} \\ \midrule
5GMdata &
  2018 &
  Public &
  Labeled &
  Non-IID &
  Simulated &
  Technical &
  \begin{tabular}[c]{@{}l@{}}- Beam Selection\\ - Channel Estimation\end{tabular} \\ \midrule
DACT &
  2017 &
  Public &
  Labeled &
  Non-IID &
  Real &
  Use case &
  - Driving Pattern \\ \midrule
CIC-IDS2017 &
  2017 &
  Public &
  Labeled &
  Non-IID &
  Simulated &
  Use case &
  \begin{tabular}[c]{@{}l@{}}- Network Intrusion Detection\\ - Attack Forecasting\end{tabular} \\ \midrule
RAE &
  2017 &
  Public &
  Unlabeled &
  Non-IID &
  Real &
  Use case &
  \begin{tabular}[c]{@{}l@{}}- Energy Saving / Anomalous Consumption Detection\\ - Occupancy Pattern / Energy Demand Prediction\end{tabular} \\ \midrule
VR Streaming &
  2017 &
  Public &
  Labeled &
  Non-IID &
  Real &
  Use case &
  \begin{tabular}[c]{@{}l@{}}- Patterns for gazing prediction \\ - User identification.\end{tabular} \\ \midrule
MIMIC-III &
  2016 &
  Public &
  Labeled &
  Non-IID &
  Real &
  Use case &
  \begin{tabular}[c]{@{}l@{}}- Predicting hospital length of stay\\ - Early Detection of Diseases (Sepsis, Pancreatitis, etc)\end{tabular} \\ \midrule
Irish CER &
  2015 &
  Public &
  Unlabeled &
  Non-IID &
  Real &
  Use case &
  \begin{tabular}[c]{@{}l@{}}- Electric Demand Prediction\\ - Electric Load Forecasting\\ - Faults and attacks Detection\end{tabular} \\ \midrule
PlantVillage &
  2015 &
  Public &
  Labeled &
  Non-IID &
  Real &
  Use case &
  \begin{tabular}[c]{@{}l@{}}- Digital Twin Smart-Farming Framework\\ - Plant Disease Detection\end{tabular} \\ \midrule
UNSW-NB15 &
  2015 &
  Public &
  Labeled &
  Non-IID &
  Simulated &
  Use case &
  \begin{tabular}[c]{@{}l@{}}- Network Anomaly Detection\\ - Identify Cyber-Attacks\end{tabular} \\ \midrule
Telecom Italia &
  2015 &
  Public &
  Labeled &
  Non-IID &
  Real &
  Technical &
  \begin{tabular}[c]{@{}l@{}}- Cellular Traffic Prediction\\ - Resource Management\end{tabular} \\ \bottomrule
\end{tabular}
\end{table*}

\subsection{Existing Implementation Frameworks}

To implement Split Federated Learning in 6G networks, we need two sorts of tools. Firstly, network simulators that play a significant role in modeling and analyzing the cellular system. Secondly, ML/DL platforms for training, testing and validating SFL models before being applied to the 6G network. For that, a plethora of network simulators and AI frameworks could be used. In the following, we present the most popular and widespread tools in both academia and industry. 

\subsubsection{Mobility and Network Simulators}

\begin{itemize} 

    \item SUMO (Simulation of Urban MObility)\footnote{\url{https://www.eclipse.org/sumo/}}: This is an open source traffic generator, developed in 2001 by the German Aerospace Center (DRL). It permits the simulation and analysis of realistic user mobility and traffic-related models. It offers many features; we briefly cite a few of them: building roads, considering streets, intersections, traffic lights, high-speed routes, lane and direction changing, etc. The extracted data from the trace file could be used by another network simulator such as OMNET++, NS2 and NS3.
      
    \item NS3\footnote{\url{https://www.nsnam.org/}}: This is a popular event-driven emulator/simulator designed specifically for research and educational purposes in computer communication networks. It is based on two programming languages: C++ and Python. The simulator core is developed entirely in C++ with optional python bindings, which gives users the ability to choose between C++ and Python to write simulation scripts. It supports diverse network technologies, including cellular networks such as 4G (LTE) and 5G (NR).
    
   \item OMNET++ (Objective Modular Network Testbed in C++)\footnote{\url{https://omnetpp.org/}}: It is an extensible, modular, discrete-event and free software simulator, targeted mainly for computer network simulation (wired and wireless). It is programmed exclusively in C++, and can be coupled with several external frameworks such as Tensorflow for ML/DL development. It is widely used for 4G and 5G networks simulation.

   \item NetSim\footnote{\url{https://www.tetcos.com/}}: NetSim is a C language-based network simulator that allows not only simulation but also emulation of real-time traffic from real devices. 
   In addition, it is available under three versions (Pro, Standard and Academic). Each version has different features, support options and pricing (no free usage). It provides an easy graphical user interface and a packet trace file with all information needed for further analysis and evaluation of performance metrics.

   \item Riverbed Modeler\footnote{\url{https://www.riverbed.com/}}: It is a commercial discrete event-simulation environment, formerly known as OPNET, used for the analysis of communication applications, protocols and networks. Its sophisticated graphical interface permits the user to build network topology (nodes and links), display results, adjust the different parameters, performing various experiments and scenarios visually and rapidly.

\end{itemize}

\subsubsection{ML/DL Frameworks}

\begin{itemize}
    \item PySyft\footnote{\url{https://github.com/OpenMined/PySyft}}: This is an open-source Python library, developed by OpenMined. It integrates secure and private deep learning algorithms such as Federated Learning. In also implements differential privacy, and encrypted computation. On Github, it has 8.6k stars and 1.9k forks.
    \item Federated AI Technology Enabler (FATE)\footnote{\url{https://github.com/FederatedAI/FATE}}: This software is an open source federated learning framework in Linux. It implements secure computation protocols based on homomorphic encryption and multi-party computation (MPC). It has been applied in many domains such as the finance and medical ones, and acquired 4.8k stars and 1.4k forks on its Git repository.
    \item FedML\footnote{\url{https://github.com/FedML-AI/FedML}}: This framework is an open research library enabling collaborative machine learning on decentralized data. It was developed at the University of Southern California based on PyTorch. It has 2.4k stars and 572 forks on Github. It assures three computing paradigms: on-device training for edge devices, distributed computing, and single-machine simulation. Further, it encourages diverse algorithmic research through the design of flexible APIs and comprehensive reference implementations (optimizer, models, and datasets).
    \item TensorFlow Federated (TFF)\footnote{\url{https://www.tensorflow.org/federated}}: It is a free and open-source TensorFlow-based framework. It is developed by Google for federated learning. It has attracted about 2k stars and 532 forks on GitHub. TFF proposes two APIs, namely: Federated Learning (FL) API for high-level interfaces (training and evaluation of users' models) and Federated Core (FC) API for low level interfaces (e.g., developing new FL algorithms).
    \item OpenFL\footnote{\url{https://github.com/intel/openfl}}: It is a python-based, open-source framework, developed by Intel. OpenFL is a versatile tool that was initially deployed for medical imaging usage (training brain tumor segmentation models). It provides an efficient and reproducible method for developing and evaluating FL algorithms. It gained 1.7k stars and 405 forks on Github.
    \item Flower\footnote{\url{https://flower.dev/}}: It is an open-source federated learning framework created by Adaptech Research. It provides a high-level API that enables researchers to experiment and build various FL use cases. It is compatible with both PyTorch and TensorFlow frameworks and supports a large number of clients. It gained 470 forks and 2.2k stars on the github repository.  
\end{itemize}

\section{Open Challenges and Future Directions}  

\label{sec:Future Directions and Discussions}
While SplitFed is a promising technique for collaborative machine learning in decentralized 6G systems, there are still several open challenges that need to be addressed for its effective implementation in 6G networks. We discuss these challenges along two dimensions: (i) SFL-specific issues, which are related with its space of available architectural and algorithmic configuration options, and have a scope that goes beyond its application in/for 6G, and (ii) 6G-specific challenges that stem from the expected characteristics of this communication technology and their interplay with SFL.

\subsection{Open Challenges in SFL}
\label{sec:Open Challenges in SFL}
 
 \noindent\textbf{Splitting strategy:} Sharing the model among all the involved learners is the first stage of the SplitFed algorithm. If we assume a SplitFed Model $\mathcal{M}$ with $\boldsymbol\ell$ layers, the total number of possible splitting combinations is then $(\boldsymbol\ell-1)$. For each possibility $\mathrm{P_{i}}$, $i\in \left\{1, \dots (\boldsymbol\ell-1)  \right\}$, the client and server sides would have $i$ and $(\boldsymbol\ell-i$) layers, respectively. Based on these alternatives, the following questions should be answered: Which combination would be suitable to split the model? Is dividing the model in a random way a good strategy? Should the splitting strategy take into consideration certain criteria, such as the number of collaborators and clients' computing resources?

\noindent\textbf{Computation requirements at the server side:}
An aspect that is relatively downplayed in SFL research is the fact that the desired parallelization in processing client data is achieved at the expense of increased compute requirements at the server side, as the server needs to maintain a copy of the server-side model portion per client participating in a training round. For settings with massive numbers of clients, this could put significant strain on the server and increase costs. At the same time, this observation reveals interesting resource allocation and orchestration problems for future research. For example, the size of the set of recruited clients per training round can be dynamically tuned based on the available resources on the server side, while multiple main server instances, each responsible for a different client subset, can be introduced -- at the same time dealing with synchronization and consistency issues towards building a global shared model.

\noindent\textbf{Data fairness (imbalanced and non-IID data):} In SplitFed, each participating host collects its local data from various heterogeneous sources and with different features (source, location, period, etc.). Therefore, in many settings, is unrealistic to assume that all SFL clients will have iid-distributed local data. On the contrary, various and non-stationary data are expected. Training a SFL model under highly skewed data will cause a high weight divergence and thus a model accuracy degradation~\cite{duan2020self}. 
Hence, new techniques and algorithms have to be proposed to handle data heterogeneity and improve the learning process on non-IID data. Furthermore, some parties may represent less or more samples than others leading to uneven data distribution that will affect the learning model convergence as well as the performance of SplitFed training process. To overcome this, each client can apply dataset enhancement strategies locally, for instance, using GAN algorithms or data augmentation techniques (e.g., rotation, shearing, and flipping for images)~\cite{shorten2019survey}. 
An alternative option is to support collaborators by supplementary data from the main or Fed server.

\noindent\textbf{Dataset labeling:} 
The labeling procedure is an integral part of data preparation for Supervised Split Federated Learning (SSFL). It consists of appending informative tags to data (text, image, audio, and/or video) to help the SFL model identify the class of an unlabeled object. However, the distributed nature of client data may lead to divergences in the labeling results (owing to different annotators' expertise level, biases or even malicious falsification). This can cause noisy labels that severely degrade the performance of the learned model. Ensuring uniform and accurate labeling among all the SFL clients is a challenging task that must be considered to enhance the quality of local datasets and maximize the reliability and accuracy of models. In this context, some techniques could be utilized such as meta learning, label correction, and knowledge distillation~\cite{9729424}.
  
\noindent\textbf{Aggregation technique:} The aggregation algorithm plays a crucial role for achieving a good performance in a SFL design. It permits combining the local sub-model updates from all the SFL nodes participating in the training round. A robust aggregation technique should be able to maximize the accuracy of the global model, enhance the privacy of local updates, optimize the communication bandwidth and identify suspicious clients. In this respect, multiple questions might be usefully discussed: is it accurate to adopt the aggregation mechanisms developed for Federated Learning~\cite{9530694} or new aggregation techniques should be designed since model training in FL and SplitFed are different? Furthermore, is it adequate to apply the same aggregation algorithm for both the client and the server side or should each part have its aggregation method that aligns with its specific objectives? 

\subsection{Open Challenges in 6G}

\noindent\textbf{Wireless channel constraints:} 
Higher frequency bands and Terahertz communication are among the main features that set 6G networks apart from other wireless technologies. These aspects have the potential to provide faster data rates and lower latency than current networks. However, they pose some challenges including high path loss, signal attenuation, interference and fading channels. Given these issues, a user in SplitFed may disconnect from the main or Fed server during the transmission of smashed data and/or clients' weights. This will interrupt the training process and call for higher time to complete it. To enhance the reliability of the global model learning, deploying more than one main/fed server is recommended (server redundancy). However, starting training on a different main/fed server from scratch is not a good idea. For instance, if a client moves in the last stages of its sub-model training, the required time to finish the new training would be very lengthy. For that, it is very important to develop effective data/service migration techniques to resume users' training after being interrupted rather than starting over.

\noindent\textbf{Disproportionate and heterogeneous 6G users:} In effect, the number of SFL contributors depends greatly on the 6G target application. Some scenarios have few participants where the failure of any party impacts the whole communication system. In contrast, in some cases with a great number of participants (IoT devices, mobile phones, etc.), the disconnection of a 6G user does not affect widely the performance of the SFL learning process as a whole. Moreover, 
6G clients can have multiple and heterogeneous computing resources (CPU, storage, etc). This will engender different sub-model training times that penalize the generation of the global model. In this context, an efficient strategy to choose the suitable number of clients with respect to the target application and clients' resource constraints should be applied to allow the participation of as many clients as possible and accelerate the performance improvement of the SFL model.
    
\noindent\textbf{Irrelevant and heterogeneous 6G features:} 6G clients may sense irrelevant features of their private data (lack of domain knowledge, measurements in imperfect network conditions such as interference, noise, collision, etc) that can effectively increase the computational and time complexity of the data preparation phase. 
Extracting only representative information and filtering out inessential details, that do not affect the decision-making process is of paramount importance. In this context, a semantic information extraction model trained through the SplitFed algorithm could be implemented. The model will learn from diverse SFL collaborators that integrate multiple features. The parameters of the local sub-models are then transmitted to the main server for aggregation. The obtained model is expected to improve the accuracy of extracted information, clients' energy efficiency and model training time overhead. 

\noindent\textbf{Black-box and complex deep learning:} One of the main challenge of DL-based models, including SFL, is that they do not provide any details about how and why their decisions are made, and thus such decisions cannot be properly understood and trusted by the different 6G stakeholders such as managers and executive staff. Therefore, the 6G stakeholders may not perform/execute the SFL-based decisions. To deal with this issue, eXplainable Artificial Intelligence (XAI) is an emerging paradigm that provides a set of techniques, e.g., ante-hoc, post-hoc, visualization, model-agnostic, etc., and aims to improve the transparency of black-box DL decision-making processes~\cite{exp2}\cite{exp3}\cite{Exp4}. In other words, XAI helps to explain the SFL-based decisions to make them trustable and interpretable by the different 6G actors~\cite{exp1}.
     
\noindent\textbf{Security Issues:} 
There is no doubt that the SplitFed reduces the risk of data and model disclosures. However, this does not mean that SplitFed is entirely proof against all attacks. Indeed, it is still prone to security and privacy risks from the client level to the server level. Diverse data-oriented attacks such as data tampering and data poisoning may target the SFL process causing a significant loss in the global model's accuracy as demonstrated by a recent study~\cite{SFLPoisoningAttack}~\cite{Sabra}. As well, other threats and vulnerabilities could impact the success of the SplitFed models: compromised/malicious Fed/main servers, unsecured communication channel, clients dropout and free-riding attacks. Hence, new defensive techniques for protecting SFL entities (Fed server, clients local data, cut layer activations, main server, etc.) during model training, aggregation, and transmission are mandatory to build a risk-free split federated learning ecosystem. In this regard, Blockchain technology and Secure Multi-Party Computation (MPC) can highly benefit SFL.

\noindent\textbf{Performance degradation of SFL-based models:} As mentioned before, SFL can be used to optimize different functions/operations related to 6G systems. However, a critical challenge is how to train and deploy SFL-based models, while providing stable life-cycle performance. In fact, data profiles evolving may cause performance degradation of the AI learning models~\cite{MLOP}. Thus, both models' performance degradation and new data profiles should be studied, to ensure a stable performance of the intelligent 6G functions/operations over time. 
Therefore, it is required to not only perform continuous monitoring of both data and model profiles, but also automate the whole development process of SFL learning models, including data collection/extraction, model training, validation, and deployment~\cite{kreuzberger23}\cite{ORAN-Brik}. In this context, the DevOps paradigm can be leveraged. DevOps includes a set of practices that combine software development (Dev) and IT operations (Ops). DevOps aims not only to reduce the systems' development life cycle, but also to provide a continuous software delivery with high quality, by leveraging paradigms and concepts like Continuous Integration and Delivery (CI/CD). When dealing with machine learning operations, and automation of the learning process, the paradigm is also called MLOps~\cite{kreuzberger23}. 
    
\noindent\textbf{SFL scalability:} The number and stability of SFL collaborators are key factors in the success of SFL-enabled schemes. Consequently, frequent client drop-outs, whatever the reason (intermittent connection, selfish client, malicious client, low battery, mobility, etc.) would have a negative effect on the model convergence time and accuracy. Therefore, proposing a novel strategy to make the SFL system more robust to this issue would be of great value. One solution could be predicting the device disconnection, based on its mobility or resources capacity. Moreover, the incorporation of incentive mechanisms could be beneficial. For instance, in a reputation-based incentive scheme, each client in the network would have a reputation rank based on its participation rate. The SFL clients performing the training in an efficient way will be rewarded. The goal is to encourage the participation of qualified nodes in the SFL training process~\cite{9369019}. 

Overall, addressing the open challenges of split federated learning in 6G networks will require a combination of novel algorithms, optimization techniques, hardware architectures, and security and privacy mechanisms. An interdisciplinary approach that draws on expertise from computer science, electrical engineering, mathematics, and statistics will be necessary to fully realize the potential of split federated learning in 6G networks.

\section{Limitations of this survey and obstacles faced}
\label{sec:Limitations of the Survey}

In this section, we address the potential gaps that could be attributed to our survey and should be considered as an opening window for further research. 

\begin{itemize}
    \item Lack of existing works: In fact, there are very few studies that focus on the implementation of Split Federated Learning, and specifically for 6G systems. Most works deal with FL or SL separately, of which a large number are dedicated to federated learning techniques. To identify previous works related to the SplitFed algorithm, we conducted a literature search on ACM Digital Library, Springer Link, IEEE Xplore, ScienceDirect, Wiley, Taylor \& Francis Online, MDPI, and arXiv databases from 2020 (first appearance of SFL) to 2023. To the best of our knowledge, we found only seven articles that are published in IEEE~\cite{SFL1}~\cite{SFL2}~\cite{SFL3}~\cite{SFL4}~\cite{SFL5}~\cite{SFL6}~\cite{SFL7}, three articles in arXiv and MDPI~\cite{SL_6G}~\cite{SFLArxic1}~\cite{SFLArxic2}, while no works are published in the other databases~\footnote{We have used the advanced search filter to find works with titles containing the term ``Split Federating Learning.''}. These results confirm the few existing studies on SFL for 6G networks which was one of the main difficulties we faced in our study. This also shows the need for further development in this emerging scope to enrich the literature review with relevant papers.

    \item Other 6G aspects and use cases: Our research work covers the most important and representative technical aspects and use cases foreseen for 6G systems. However, the list is not exhaustive and other new technical aspects and applications can be envisioned, driven by 6G requirements and users demands, such as network slicing, smart-governance, unmanned mobility, education, online advertising, sustainable development, etc.
    
    \item Experimental studies and results: Another potential limitation is related to the shortage of experiments and results on the implementation of SFL either for technical aspects or use cases. The responding argument is the scarceness of previous research enabling the connection between 6G systems and Split Federated Learning. To overcome this limitation, we encourage researchers to combine both technologies by implementing new models so that the results of future research can improve on this aspect.

\end{itemize}
 
\section{Conclusion}
\label{sec:Conclusion}
Distributed and collaborative deep learning have taken great strides over recent years and have been applied to many applications. Split Federated Learning, as a nascent technique, provides a secure and faster model training strategy. The core idea is to parallelize the training phase by dividing the global model amongst the participating agents and performing a local training process based on their private and local data. This new method dramatically reduces the training time and, in addition to data privacy, preserves model privacy. As we have seen, works on SFL are still in their growth phase and it is therefore necessary to explore new research avenues about the topic. The current study takes a deep dive on the potential of using the SplitFed Learning algorithm to improve the reliability of the future SFL-based 6G systems. At the beginning, we provide the reader with the existing AI ideas for 6G networks. To the best of our knowledge, our survey is the first to present a comprehensive view of the application of Split Federated Learning in the 6G networks. Afterwards, we outline the primary contributions and organization of the paper along with an exhaustive background on artificial intelligence algorithms, collaborative deep learning and 6G Mobile Networks as foundations and cornerstones. Following that, several 6G technical aspects are thoroughly examined with a representative realistic scenario for each aspect. Furthermore, the applicable 6G use cases which would benefit from Split Federated Learning are analyzed. We believe that the synergy between Split Federated Learning and Edge Computing would enable a significant improvement of both 6G applications and technical aspects. Moreover, a series of datasets and development frameworks that can support the implementation of SFL within 6G networks are summarized. In this context, we realized that only few 6G-related datasets exist, which explains the reason behind the use of other non-6G inputs. Next, we draw researchers and practitioners attention to the fact that SplitFed cannot deal with all issues, through an overview of the relevant limitations and challenges. Therefore, we discuss how to overcome these challenges by giving new research hints. To conclude, we expect this article to stimulate researchers to design, test, and deploy innovative SFL-based solutions for 6G technologies.

\bibliographystyle{IEEEtran} 
\bibliography{Ref}

\end{document}